\documentclass[12pt]{article}
\usepackage[letterpaper,top=1in,bottom=1in,left=1in,right=1in]{geometry}
\usepackage{setspace}
\usepackage{amsfonts}       
\usepackage{nicefrac}       
\usepackage{microtype}      
\usepackage{lipsum}
\usepackage[authordate, uniquename=false]{biblatex-chicago}
\usepackage{comment}

\usepackage[english]{babel}
\usepackage{graphicx}
\usepackage{multirow}
\usepackage{amsthm}
\usepackage{amssymb}
\usepackage{bbm}
\usepackage[colorlinks=true, allcolors=blue]{hyperref}
\usepackage{caption}
\usepackage{float}
\usepackage{subcaption}
\usepackage{makecell}
\usepackage{mathtools}
\usepackage{bm}
\usepackage{physics}
\usepackage{appendix}
\usepackage{enumitem}
\usepackage{csquotes}
\usepackage{cancel}
\usepackage{algorithm}
\usepackage{algpseudocode}
\usepackage{authblk} 
\usepackage{placeins}
\usepackage{booktabs}

\renewcommand{\v}[1]{\ensuremath{\bm{#1}}}
\newenvironment{significance}
{
  \small                 
  \begin{center}
    \bfseries Statement of Significance 
  \end{center}
  \quotation             
}
{\endquotation}

\title{Calibrating hierarchical Bayesian domain inference for a proportion}

\author[1]{%
  \large Rayleigh Lei$^*$  \\ 
  \vspace{-0.5cm}
  \small Continuing Academic Specialist \\ 
  \small Center for Statistical Training and Consulting, Michigan State University, United States of America\\
  \small \texttt{rayleigh@umich.edu}\\ 
\vspace{0.5cm}
  \large Yajuan Si \\
  \small Research Associate Professor \\ 
  \small Institute for Social Research, University of Michigan, United States of America\\
  \small \texttt{yajuan@umich.edu}
}

\date{}

\begin{document}
\maketitle
\thispagestyle{empty}

\subsection*{Acknowledgments}

This work is supported by the National Institutes of Health grant (U01MD017867). 

\subsection*{Word count}

The total word count is 8,811.

\newpage
\pagenumbering{arabic}

\begin{abstract}
Small area estimation (SAE) improves estimates for local communities or groups, such as counties, neighborhoods, or demographic subgroups, when data are insufficient for each area. This is important for targeting local resources and policies, especially when national-level or large-area data mask variation at a more granular level. Researchers often fit hierarchical Bayesian models to stabilize SAE when data are sparse. Ideally, Bayesian procedures also exhibit good frequentist properties, as demonstrated by calibrated Bayes metrics. However, hierarchical Bayesian models tend to shrink domain estimates toward the overall mean and may produce credible intervals that do not maintain nominal coverage. Hoff et al. developed the Frequentist, but Assisted by Bayes (FAB) intervals for subgroup estimates with normally distributed outcomes. However, non-normally distributed data present new challenges, and multiple types of intervals have been proposed for estimating proportions. We examine domain inference with binary outcomes and extend FAB intervals to improve nominal coverage. We describe how to numerically compute FAB intervals for a proportion and evaluate their performance through repeated simulation studies. Leveraging multilevel regression and poststratification (MRP), we further refine SAE to correct for sample selection bias, construct FAB intervals for MRP estimates, and assess their repeated sampling properties. Finally, we apply the proposed inference methods to estimate COVID-19 infection rates across geographic and demographic subgroups. We find that the FAB intervals improve nominal coverage, at the cost of wider intervals.

{\bf Keywords: Hierarchical Bayes, calibrated Bayes, small domain inference, proportion estimate, coverage}
\end{abstract}

\begin{significance}
Accurate estimation of health and socioeconomic measures for small communities is essential for informed public policy and effective resource allocation. However, these efforts are often hampered by insufficient local data. Small area estimation (SAE) using hierarchical Bayesian models helps stabilize estimates by borrowing information across domains, yet it frequently yields credible intervals that fail to maintain nominal coverage for individual areas. This paper addresses the challenge of achieving reliable interval coverage for SAE with binary outcomes by extending the Frequentist, but Assisted by Bayes (FAB) interval methodology. We develop computational tools to construct FAB intervals for three widely used confidence intervals for proportions. In addition, we integrate this approach with multilevel regression and poststratification to produce generalizable estimates aggregated to higher-level geographic units. Through simulation studies and an application estimating COVID-19 infection rates across geographic and demographic subgroups, we show that the proposed intervals achieve more reliable coverage, albeit at the cost of slightly wider intervals. These advances provide a practical solution for calibrating confidence intervals for proportions estimated in small areas.
\end{significance}

\section{Introduction}

Researchers are often interested in estimating population quantities for the overall group and subgroups and in examining how well methods perform among groups. Domain inference moves beyond population averages and can characterize heterogeneity across subgroups, such as geographic and demographic variation. Policy makers rely on subgroup or domain estimates to tailor resource allocations and policy interventions in economics (e.g., \textcite{BugalloEtAlSmallAreaEstimation2024}), public health (e.g., \textcite{WakefieldEtAlSmallAreaEstimation2020, AllorantEtAlSmallAreaModel2023}), and other fields. However, in practice, datasets available for analysis may not have sufficient observations across all domains, some of which have sample sizes that are too small to allow stable inferences. Small area estimation (SAE) introduces models to directly collect group summaries and borrow information across groups to stabilize domain inference~\parencite{rao15}. In this paper, we use the terms `groups', `subgroups', and `domains' interchangeably, which are equivalent to areas defined in the SAE. 

The well-known Fay-Herriot model \parencite{Fay:herriot1979} for a continuous outcome variable with the summary statistics $y_i$ measured for domain $i$ (e.g., the domain average of individual measure values) is specified as following:
\begin{align*}
    y_i \sim \textrm{N}(\cdot \mid \theta_i, \sigma^2_i), \,\,\,\,\,\, \theta_i \sim \textrm{N}(\cdot \mid \mu_i, \tau^2),
\end{align*}
where the sampling model for $y_i$ is a normal distribution with the mean $\theta_i$ and within-domain standard deviation $\sigma_i$, and the linking model for $\theta_i$ is another normal distribution with the mean $\mu_i$ and between-domain standard deviation $\tau$, with $y_i, \theta_i, \mu_i \in \mathbbm{R}$ and $\sigma_i, \tau \in \mathbbm{R}^+$. The linking model often uses domain-level covariates $x_i$ by specifying $\mu_i = x^T_i\beta$. The SAE for $\theta_i$ combines both the direct summaries $y_i$ and the linking model, using various estimation methods such as empirical Bayes and hierarchical Bayes. With a mis-specified linking model, over-shrinkage may result in bias while increasing precision. Most work in the literature on SAE assesses the mean squared error to strike a balance between the bias and variance for the point estimates $\hat{\theta}_i$ when comparing different methods~\parencite{rao15}.

We focus on the hierarchical Bayesian estimation methods. The Bayesian credible interval for $\hat{\theta}_i$ can outperform classical confidence intervals, providing greater precision and narrower ranges, since hierarchical Bayesian estimates tend to shrink small group estimates with large variation toward the hierarchical or overall mean with smaller variances~\parencite{GhoshAdjustedBayesEstimators1999, JudkinsLiuCorrectingBiasRange2000}. One popular hierarchical Bayesian method for SAE is multilevel regression and poststratification (MRP, \textcite{gelman:little:97, mrp-si20}), which fits a multilevel regression by setting up a predictive model for the outcome with a large number of covariates and regularizing with Bayesian prior specifications and then poststratifies the SAE to correct for imbalances in the sample composition from the target population. The flexible modeling of survey outcomes in MRP can capture complex data structures conditional on poststratification cells, which are determined by the cross-tabulation of categorical auxiliary variables, and calibrate the sample discrepancy with population control information. MRP accounts for the sample survey design in Bayesian modeling, as a design-adjusted, model-based approach, and is expected to yield valid inference about the finite population quantities~\parencite{BNFP:SI15, Makela-sm18, prior-si2018, BayesRake18}.

Bayesian procedures are desired to exhibit good frequentist properties, a central goal of calibrated Bayes ~\parencite{CalibratedBayes:Little11,rubin2019conditional}. Therefore, we desire Bayesian credible intervals to achieve nominal coverage, e.g., the coverage rate of 95\% credible intervals is around 0.95. However, the hierarchical Bayesian credible interval achieves population-level coverage, but not domain-specific coverage~\parencite{YuHoffAdaptiveMultigroupConfidence2018}. Given $\theta_i$, ~\textcite{BurrisHoffExactAdaptiveConfidence2020} show that Bayesian credible intervals for domain inference are centered around the shrinkage estimator that could be a potentially biased estimator of $\theta_i$, can yield a coverage probability higher than the nominal level when $\theta_i$ is close to $\mu_i$, and can fail to achieve nominal coverage when $\theta_i$ is far away from $\mu_i$.

\textcite{YuHoffAdaptiveMultigroupConfidence2018} have proposed Frequentist, Assisted by Bayes (FAB) intervals for domain estimates of $\theta_i$ in SAE with a normally distributed outcome, which maintain exact domain-specific coverage while allowing for hierarchical shrinkage via a linking model, i.e., a prior distribution specification. Because of information sharing across domains, the FAB intervals are narrower on average than classical direct confidence intervals. With exact domain-specific coverage, the FAB interval procedure is calibrated for domain-level inferences, an improvement over hierarchical Bayesian credible intervals. \textcite{YuHoffAdaptiveMultigroupConfidence2018} introduced an exchangeable prior distribution across domains, and \textcite{BurrisHoffExactAdaptiveConfidence2020} included domain-specific covariates and spatio-temporal correlation across domain means. However, existing work is restricted to a normally distributed outcome. In this paper, we extend FAB intervals to binary outcomes and make domain inference for proportions. 

There have been historical arguments on the appropriate choice of intervals for a proportion, e.g., the standard Wald interval, the Wilson interval~\parencite{WilsonProbableInferenceLaw1927}, and the Agresti-Coull interval~\parencite{AgrestiCoullApproximateBetterExact1998}. \textcite{BrownEtAlIntervalEstimationBinomial2001a} examined different intervals for a proportion in terms of coverage probabilities and widths, and \textcite{FrancoEtAlComparativeStudyConfidence2019} conducted a comparison based on simulation studies with complex sample survey data, both recommending the Wilson interval. We are not aware of any literature work on the reasonable interval choice for domain inference with a proportion, which presents new challenges in  methodology, computation, and practical application studies. We aim to address these challenges by applying the FAB framework to the Wald, Agresti-Coull, and Wilson intervals. Primarily, this entails defining $\alpha$-level intervals for each $\theta_i \in [0, 1]$ that minimizes the Bayes risk and interval length. The FAB interval is the largest continuous subset of $\theta_i$ for which the $\alpha$-level intervals cover our estimate of $\theta_i$.

Meanwhile, the motivating application study of our investigation is the COVID-19 infection rate estimation. Researchers are interested in tracking COVID-19 transmission across geographic areas and demographic groups, but the publicly available data sources are flawed. \textcite{mrp-covid21, mrp-covid22} applied MRP to routine hospital COVID-19 test records and generated a synthetic random proxy for COVID-19 infection surveillance. It is important to monitor the infection incidence in local communities and detect group heterogeneity by geography and demography. However, obtaining reliable subgroup estimates or generating valid domain-specific inference with sparse data is challenging. 

We aim to calibrate Bayesian domain inference for a proportion, where traditional Bayesian and Frequentist intervals often struggle with poor coverage. We extend the FAB interval framework to proportion estimation and develop computational algorithms under hierarchical Bayesian models, addressing the lack of closed-form results for binomial data. We introduce an ``all-in penalty" in the risk criterion and correct FAB intervals near the boundaries of the probability space. We use simulation and empirical studies to demonstrate the methodological innovations that yield intervals with improved coverage for domain estimates and functions of domain estimates (via MRP estimates). Furthermore, our contributions provide a broadly applicable framework for calibrated interval estimation in hierarchical Bayesian and SAE models, supported by reproducible code.

Our paper is organized as following. First, we describe the methodology and how to numerically adjust these intervals in Section \ref{sec:method}. We then assess their performances with simulated datasets to understand the coverage properties and interval lengths in Section \ref{sec:sim_results}. Finally, we apply these intervals to estimate county-level COVID-19 viral infection rates in Section~\ref{sec:real_data_results}. Section~\ref{conclusion} summarizes our findings, recommendations, and future investigations.

\section{Method}
\label{sec:method}

\subsection{Inference for a proportion}
\label{original}

We consider the following binomial model. For domain $i$, assume that the sample size is $n_i$ and the sum of observed binary outcomes is $y_i$. We use $\theta_i$ to denote the true proportion for domain $i$, i.e. the positive rate of the binary outcome, $\theta$ to denote a generic parameter, and $\widehat{\theta}_i$ to denote an estimator of $\theta_i$. The originally defined confidence intervals use the sample mean, $\widehat{\theta}_i = y_i/n_i$, as an estimate of the proportion $\theta_i$. For Bayesian credible intervals, we use the posterior mean as the point estimate $\widehat{\theta}_i$.


We introduce a logit link for the proportion of a positive outcome, $\textrm{logit}(\theta_i) = \eta_i$, which is assigned a normal prior. The hierarchical Bayes model is summarized as
\begin{align}
\label{bin}
    y_i \mid \theta_i \sim \textrm{Binomial}(\cdot \mid \theta_i, n_i);\,\,\,   \theta_i = \textrm{inv\_logit}(\eta_i);\,\,\,  
    \eta_i \sim \textrm{N}(\cdot \mid \mu_i, \tau_i^2).
\end{align}
We can perform posterior inference for $\theta_i$ given the data $\{n_i, y_i\}$ and the prior, $\textrm{logit}(\theta_i)\sim \textrm{N}(\cdot \mid \mu_i, \tau_i^2)$, where in the normal prior we introduce domain-specific mean and variance. We can simply use fixed values, such as $\mu_i=0, \tau_i^2=1$, or assign hyperprior distributions for the mean and variance through the hierarchical Bayes framework; both cases will be examined in this paper. For each fixed value of $n_i$ and each possible outcome $y_i \in \{0,1,\ldots,n_i\}$, we use the binomial likelihood, $\text{Binomial}(y_i \mid \theta_i, n_i)$, under the prior $\theta_i = \textrm{inv\_logit}(\eta_i)$, $\eta_i \sim N(0,1)$, and draw 10,000 posterior samples of $\theta_i$ via Markov chain Monte Carlo (MCMC). We form a 95\% Bayesian credible interval for $\theta_i$ using the 2.5th and 97.5th percentiles of the posterior draws. For each $(y_i, n_i)$, we repeat this process 1000 times and let $\mathcal{C}^j_B(y_i, n_i)$ denote the $j$th Bayesian credible interval for $j = 1, 2, \ldots, 1000$. We then evaluate frequentist coverage over a grid of 101 evenly spaced values of the true parameter $\theta_i\in [0,1]$, including values near 0 and 1. For each grid point $\theta_i$, the coverage probability of $\theta_i$, $p_{C}(\theta_i \mid n_i)$, is computed by averaging the probability that the credible interval contains $\theta_i$ over the sampling distribution of $y_i \sim \text{Binomial}(\cdot \mid \theta_i, n_i)$ across all values of $y_i$, i.e.
\[
p_{C}(\theta_i \mid n_i) = \sum_{y_i=0}^{n_i}\left[ \sum_{j = 1}^{1000} \frac{1}{1000}\mathbbm{1}\!\bigl\{\theta_i \in \mathcal{C}^j_B(y_i, n_i)\bigr\}\right]\,\text{Binomial}(y_i \mid \theta_i, n_i).
\]
If $\theta_i = 0$, $y_i = 0$ with probability 1 and if $\theta_i = 1$, $y_i = n_i$ with probability 1. Further, we slightly abuse notation by using $y_i$ again in the summation index to emphasize that we sum across all outcomes and to avoid introducing extra notation.

\begin{figure}[tp]
    \centering
        \centering
        \includegraphics[width=0.65\textwidth]{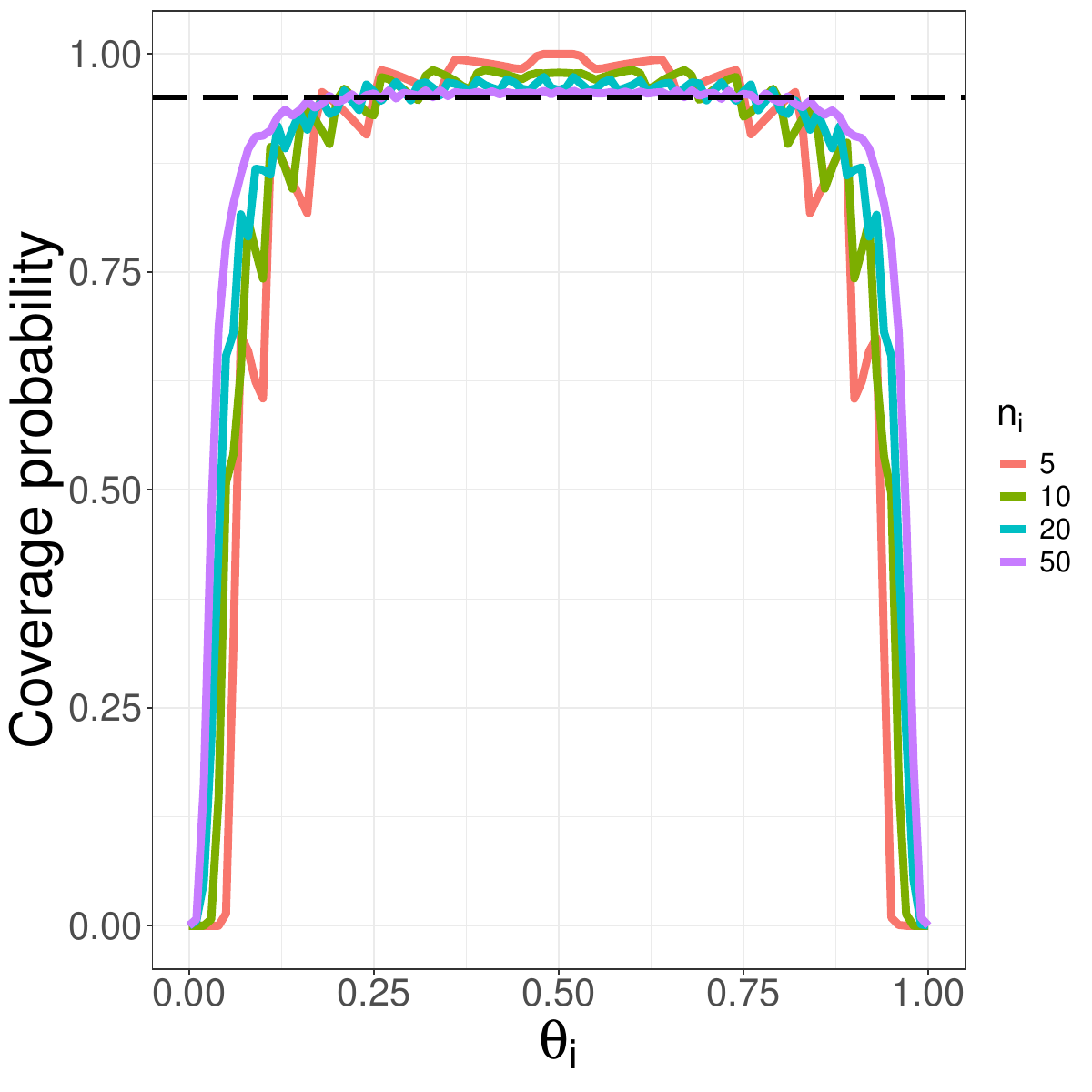}
    \caption{The coverage probability of Bayesian 95\% credible intervals of $\theta_i$ as a function of $\theta_i$ found by integrating over all responses under a hierarchical Bayes model: $y_i \sim \textrm{Binomial}(\cdot \mid \theta_i, n_i);\,\,\,   \theta_i = \textrm{inv\_logit}(\eta_i);\,\,\,  
    \eta_i \sim \textrm{N}(\cdot \mid 0, 1)$ for different $n_i$ values. The dashed horizontal line indicates the nominal coverage level of 0.95.}
    \label{fig:ci_coverage_prob}
\end{figure}

Figure \ref{fig:ci_coverage_prob} shows the coverage probabilities as a function of $\theta_i$ when we set $\mu_i = 0$ and $\tau_i^2 = 1$. The coverage rates of the credible intervals depend on the domain size $n_i$, the true parameter value $\theta_i$, and the difference between the truth and the prior value of 0.5. From the plot, the region that achieves nominal coverage is roughly the region of (0.23-0.25, 0.74-0.76) for $\theta_i$. The regions are slightly different across different $n_i$ values. However, when $\theta_i$ is near $0.5$, the coverage probability is larger for smaller $n_i$. On the other hand, the coverage probability for $\theta_i$ on the tails, close to 0 or 1, is greater for larger $n_i$, though lower than 0.95. As a result of this, if we integrate the coverage probability over $\theta_i$, the overall coverage probability increases from 0.81 for $n_i = 5$ to 0.88 for $n_i = 50$, still failing to achieve nominal coverage for the domain inference. 

This phenomenon also occurs for frequentist intervals \parencite{BrownEtAlIntervalEstimationBinomial2001a}. There are three common intervals. For a $1 - \alpha$ confidence interval with $\alpha \in [0,1]$, the Wald interval for a proportion $\widehat{\theta}_i$ is the following:
\begin{align}
\label{eq:wald_ci_deriv}   
\nonumber \mathcal{C}_N(\widehat{\theta}_i, n_i, \alpha) &= \left\{\theta: z_{\frac{\alpha}{2}} \leq \frac{ \widehat{\theta}_i - \theta}{\sqrt{\frac{1}{n_i}\widehat{\theta}_i (1 - \widehat{\theta}_i)}} \leq z_{1 - \frac{\alpha}{2}}\right\}\\
 &= \left(\widehat{\theta}_i + z_{\frac{\alpha}{2}} \sqrt{\frac{\widehat{\theta}_i(1 - \widehat{\theta}_i)}{n_i}}, \,\,\,  \widehat{\theta}_i +z_{1 - \frac{\alpha}{2}} \sqrt{\frac{\widehat{\theta}_i(1 - \widehat{\theta}_i)}{n_i}}\right).
\end{align}
Here $z_{\alpha} = \Phi^{-1}(\alpha)$, where $\Phi(\cdot)$ represents the cumulative function of the standard normal distribution. 

A slight change to this interval is ``adding two to the successes and four to the total" \parencite{AgrestiCoullApproximateBetterExact1998}, i.e. setting $\widetilde{y}_i = n_i\widehat{\theta}_i + 2$ and $\widetilde{n}_i = n_i + 4$, leading to the following definition of the Agresti-Coull (AC) interval:
\begin{align}
\mathcal{C}_{AC}(\widehat{\theta}_i, n_i, \alpha) = \mathcal{C}_N\left(\frac{\widetilde{y}_i}{\widetilde{n}_i}, \widetilde{n}_i, \alpha\right).    
\label{eq:ac_ci}
\end{align}

The Wilson interval can be derived from a similar starting point as the Wald confidence interval, except that the Wilson interval is an inversion of the score test, where the standard error term needs to be based on the parameter $\theta$ rather than the estimate $\widehat{\theta}_i$. As a result, the Wilson interval is defined as the following: 
\begin{align}
\label{eq:wilson_ci_deriv}   
\nonumber    \mathcal{C}_{W}(\widehat{\theta}_i, n_i, \alpha) &= \left\{\theta: z_{\frac{\alpha}{2}} \leq \frac{\widehat{\theta}_i - \theta }{\sqrt{\frac{1}{n_i}\theta (1 - \theta)}} \leq z_{1 - \frac{\alpha}{2}}\right\}\\     
&=\frac{n_i}{n_i + z^2_{1 - \frac{\alpha}{2}}}\left(\widehat{\mu}_{w, i} +  z_{\frac{\alpha}{2}}\widehat{s}_{w, i}, \,\,\, \widehat{\mu}_{w,i} + z_{1 - \frac{\alpha}{2}}\widehat{s}_{w, i}\right),
\end{align}
with
\begin{align*}
    \widehat{\mu}_{w, i} = \widehat{\theta}_i + \frac{z^2_{1 - \frac{\alpha}{2}}}{2n_i}, \,\,\,\,\,
    \widehat{s}_{w, i} = \sqrt{\frac{\widehat{\theta_i}(1 - \widehat{\theta_i})}{n_i} + \frac{z^2_{1 - \frac{\alpha}{2}}}{4 n_i^2}}.
\end{align*}

\subsection{FAB intervals for a proportion}
\label{fab}

We aim to improve the inference for a proportion estimate $\widehat{\theta}_i \in [0,1]$ based on the model in \eqref{bin}. We would like to calibrate the Bayesian intervals to achieve nominal coverage and ensure that they exhibit good frequentist properties. 

We describe the FAB framework to modify different confidence intervals: the Wald, the Agresti-Coull, and the Wilson intervals, respectively. We assume that the sampling model using the binomial distribution for the domain sums is correct. The statistics used to derive the confidence intervals in \eqref{eq:wald_ci_deriv} and \eqref{eq:wilson_ci_deriv} lie between $z_{\frac{\alpha}{2}}$ and $z_{1 - \frac{\alpha}{2}}$ for all $\theta$. The key idea behind the FAB intervals is to introduce a function $s_i(\cdot): (0, 1) \rightarrow (0, 1)$ such that the statistics lie between $z_{(1 - s_i(\theta))\alpha}$ and $z_{1 - s_i(\theta)\alpha}$ for each $\theta_i$~\parencite{BurrisHoffExactAdaptiveConfidence2020}, where the interval maintains the $1 - \alpha$ coverage probability for each $\theta$ because $1 - s_i(\theta)\alpha - (1 - s_i(\theta))\alpha = 1 - \alpha$. The resulting FAB intervals are defined below with the function $s_i(\cdot)$: 
\begin{align*}
    \mathcal{C}^F_N(\widehat{\theta}, n_i, \alpha, s_i) &= \left\{\theta: z_{(1 - s_i(\theta))\alpha} \leq \frac{\widehat{\theta}_i - \theta}{\sqrt{\frac{1}{n_i}\widehat{\theta}_i (1 - \widehat{\theta}_i)}} \leq z_{1 - s_i(\theta)\alpha}\right\}\\
    \mathcal{C}^F_{AC}(\widehat{\theta}_i, n_i, \alpha, s_i) &= \mathcal{C}^F_{N}\left(\frac{n_i \widehat{\theta}_i + 2}{n_i + 4}, n_i, \alpha, s_i\right)\\
    \mathcal{C}^{F}_{W}(\widehat{\theta}_i, n_i, \alpha, s_i) &= \left\{\theta: z_{(1 - s_i(\theta))\alpha} \leq \frac{\widehat{\theta}_i - \theta}{\sqrt{\frac{1}{n_i}\theta (1 - \theta)}} \leq z_{1 - s_i(\theta)\alpha}\right\}.
\end{align*}

To find $s_i(\theta)$, we aim to minimize the Bayes risk, which is the expected loss with respect to the prior distribution, taken over both the randomness in the observations $\{y_i\}$ and the uncertainty in the parameter $\theta_i$ as described by the prior. This optimization can be interpreted as minimizing the expected interval length, with the prior model determining how the cost is weighted across possible values of $\theta_i$. The motivation is to obtain shorter intervals when $\theta_i$ is close to $\mu_i$, where the prior assigns higher probability, at the expense of longer intervals when $\theta_i$ is farther from $\mu_i$ \autocite{YuHoffAdaptiveMultigroupConfidence2018}. This tradeoff is reasonable because values near $\mu_i$ are more likely under the prior model and therefore receive greater emphasis in the risk minimization.

Following \textcite{BurrisHoffExactAdaptiveConfidence2020}, $s_i(\cdot)$ minimizes:
\begin{align*}
R(s_i \mid \varsigma) = \int_{\theta \in [0,1]} \sum_{y'_i \in \{0, 1, \ldots, n_i\}} p(y'_i) \mathbbm{1}\left(\frac{y'_i}{n_i} \in \mathcal{I}^{F}(\theta, s_i, \varsigma)\right) d \theta.
\end{align*}
Here, we define 
\begin{align*}
    \mathcal{I}^{F}(\theta, s_i, \varsigma) = \left[\max\left(0, \theta + \varsigma z_{(1 - s_i(\theta))\alpha}\right),\,\,\, \min\left(1, \theta + \varsigma z_{1 - s_i(\theta)\alpha}\right)\right],
\end{align*}
where $\varsigma \in \mathbbm{R}^+$ represents the standard deviation associated with the interval, which is interval-specific (we use a generic notation for simplicity), and the marginal distribution, $p(y'_i)$, is given by
\begin{align}
    \label{eq:risk}
    p(y'_i) =  \int_{-\infty}^{\infty} p(y'_i \mid \textrm{inv\_logit}(\eta)) p(\eta) d \eta = \int_{-\infty}^{\infty} \textrm{Binomial}(y'_i \mid \textrm{inv\_logit}(\eta), n_i) \textrm{N}(\eta \mid \mu_i, \tau_i^2) d \eta.  
\end{align}

The risk for a given $\theta$ is calculated based on all possible data realizations $y_i'/n_i \in \mathcal{I}^{F}(\theta, s_i, \varsigma)$, but not the observed $y_i$. For the Agresti-Coull interval, the risk is calculated for $(y'_i + 2)/(n_i + 4) \in \mathcal{I}^{F}(\theta_i, s_i, \varsigma)$.  In other words, the minimization of the risk is done regardless of the observed value. 

Since there is no closed-form solution, the integral in \eqref{eq:risk} can then be numerically computed given $(y_i, \mu_i, \tau_i^2)$ using algorithms like QUADPACK as implemented in the R function \texttt{integrate}. Because we are dealing with discrete data, there may be multiple values of $s_i(\cdot)$ that capture a similar set of $y_i$ and have similar risk values. To determine which value of $s_i(\cdot)$ to use, we pick the value that leads to the smallest interval with continuity corrections. In addition, by taking advantage of the prior information on $\eta_i$, we might want a shorter interval overall when $\widehat{\theta}_i$ is close to $\textrm{inv\_logit}(\mu_i)$ because there will be more probability around $\mu_i$ given that $\theta_i = \textrm{inv\_logit}(\eta_i)$ and $\eta_i \sim \textrm{N}(\cdot \mid \mu_i, \tau_i^2)$. Meanwhile, the interval might be slightly wider in one direction when $\widehat{\theta}_i$ is further away. 

In sum, the FAB intervals use the observed sample mean as the point estimate $\widehat{\theta}_i$ and leverage Bayesian inference to obtain $(\mu_i, \tau_i^2)$. While this forms the basic FAB intervals, there are two computational extensions that we also consider. First, to avoid picking $s_i$ based on the values that minimize the length of the interval, we might consider a continuous approximation of $p(y_i)$. Our simulation studies reveal that the intervals and coverage probability do not substantially change; therefore, we do not pursue this direction further. Second, when we compare the originally defined intervals against their FAB versions, the latter will be significantly wider because the Bayesian prior introduces more uncertainty. One consequence is that $\mathcal{I}^{F}(\theta, s_i, \varsigma)$ extends past the support. In other words, $\theta + \varsigma z_{(1 - s_i(\theta))\alpha} < 0$ or $\theta + \varsigma z_{1 - s_i(\theta)\alpha} > 1$. Because $y'_i / n_i$ is not defined past 0 or 1, any part of the interval that includes values of $\theta$ below 0 or above 1 does not incur any risk. As a result, a way to minimize the risk for a given $\theta$ is to have $\mathcal{I}^{F}(\theta, s_i, \varsigma)$ exceed 0 or 1 as much as possible. Hence, we wish to punish the interval for this behavior. To that end, we introduce an ``all-in" penalty for $\theta$ defined as following:
\begin{align*}
    \lambda(s_i \mid \theta, \varsigma) =
    1 + \max\left(-(\theta + \varsigma z_{(1 - s_i(\theta))\alpha}), \theta + \varsigma z_{1 - s_i(\theta)\alpha} - 1\right).
\end{align*}
Through the penalty, the maximization compares how far the interval extends beyond 0 or 1. We then add one to whichever is larger because the maximum risk for a given $\theta$ is one. Because this penalty punishes the interval for exceeding past 0 or 1, the penalization encourages intervals skewed toward the middle instead of the ends. With this, we re-define the Bayes risk as 
\begin{align}
R'(s_i \mid \varsigma) = \int_{\theta \in [0,1]} f(s_i \mid \varsigma, \theta) d \theta,
\label{eq:newrisk}
\end{align}
where
\begin{align}
f(s_i \mid \varsigma, \theta) = 
\begin{cases}
\lambda(s_i \mid \theta, \varsigma)  & 
\theta + \varsigma z_{(1 - s_i(\theta))\alpha} < 0 \textrm{ or } \theta + \varsigma z_{1 - s_i(\theta)\alpha} > 1\\
\sum_{y'_i \in \{0, 1, \ldots, n_i\}} p(y'_i) \mathbbm{1}\left(\frac{y'_i}{n_i} \in \mathcal{I}^{F}(\theta, s_i, \varsigma)\right) & \textrm{otherwise}.
\end{cases}
\label{eq:all_in_penalty}
\end{align}
Note that the penalization replaces the original risk whenever the interval for a given $\theta$ extends past the support. As before, we then aim to find $s_i(\theta)$ that minimizes this new risk $R'(s_i \mid \varsigma)$.

 This truncation issue also affects the FAB interval itself. During the numerical computation, when $\theta$ is near 0 or 1 we truncate the constructed interval at the boundaries. This truncation can alter the interval’s nominal behavior and lead to departures from 95\% coverage as before depending on $s_i$. The all-in penalty is one way to mitigate this problem by discouraging boundary-spilling intervals and favoring intervals that remain within $[0,1]$. However, boundary truncation is a general challenge for Wald, Wilson, and AC intervals, so additional methodological work is needed. We outline several promising directions in the Discussion section.

\subsection{Computation algorithms}
Our approach to computing the FAB intervals is comprised of two steps. 

\begin{algorithm}[!t]
\caption{Learn $s_i(\theta)$ for $\theta$}\label{alg:learn_s_i}
\begin{algorithmic}
\Require $\theta, \alpha, n_i, \mu_i, \tau_i^2, \varsigma$
\For{$s_i(\theta) = 0; s_i(\theta) \leq 1; s_i(\theta) \mathrel{{+}{=}} 0.01$}
\State Let $\mathcal{Y} \gets \{y'_i \in \{0, 1, 2, \ldots, n_i\} : \frac{y'_i}{n_i} \in \mathcal{I}^{F}(\theta_i, s_i, \varsigma)\}$ 
\State Set $r(s_i(\theta)) \gets \sum_{y'_i \in \mathcal{Y}} p(y'_i)$ 
\Comment{$p(y'_i)$ is defined in \eqref{eq:risk}; When using the penalty function, we replace \eqref{eq:risk} by \eqref{eq:newrisk} and \eqref{eq:all_in_penalty} defined above.}

\EndFor
\State Set $s_{\textrm{min}} \gets \arg \min_{s_i(\theta) \in [0,1]} r(s_i(\theta))$ \Comment{This minimization is over the grid}
\If{$s_{\textrm{min}} >$  0}
\State Apply a binary search and the above for-loop steps to find 
\vspace{-2mm}
\[
s_i(\theta)^{\textrm{min}} \gets \arg\min_{s_i(\theta) \in [s_{\min} - 0.01, s_{\min}]} r(s_i(\theta))
\]
\vspace{-5mm}
\Else
\State Set $s_i(\theta)^{\textrm{min}} = 0$
\EndIf
\State Set $s_{\textrm{max}} \gets \arg \max_{s_i(\theta) \in [0,1]} \min r(s_i(\theta))$ \Comment{This minimization is over the grid}
\If{$s_{\textrm{max}} <$  1}
\State Apply a binary search and the for-loop steps to find 
\vspace{-2mm}
\[
s_i(\theta)^{\textrm{max}} \gets \arg\max_{s_i(\theta) \in [s_{\max}, s_{\max} + 0.01]} \min r(s_i(\theta))
\]
\vspace{-5mm}
\Else
\State Set $s_i(\theta)^{\textrm{max}} = 1$
\EndIf
\State Return $\arg\min_{s_i(\theta) \in [s_i(\theta)^{\textrm{min}}, s_i(\theta)^{\textrm{max}}]} \abs{\mathcal{I}^{F}(\theta, s_i, \varsigma)}$ 
\end{algorithmic}
\end{algorithm}

\begin{algorithm}[!t]
\caption{FAB Interval}\label{alg:learn_fab}
\begin{algorithmic}
\Require $\alpha, y_i, n_i, \mu_i, \tau_i^2, \varsigma_{\textrm{risk}}, \varsigma_{\textrm{fab}}$
\For{$\theta = 0; \theta \leq 1; \theta \mathrel{{+}{=}} 0.01$}
\State Use Algorithm \ref{alg:learn_s_i} to learn $s_i(\theta)$ with $\varsigma_{\textrm{risk}}$
\EndFor
\State Find $t_{\textrm{min}}, t_{\textrm{max}}$ such that
\begin{itemize}
    \item $t_{\textrm{min}} = 0.01c_1$, $t_{\textrm{max}} = 0.01c_2$ for some $c_1, c_2 \in \mathbbm{N} \cup \{0\}$
    \item For all $\theta = 0.01 t, t\in \mathbbm{N} \cup \{0\}$, and $\theta \in [t_{\textrm{min}}, t_{\textrm{max}}]$, $\theta \in \mathcal{I}^{F}(\theta_i, s_i, \varsigma_{\textrm{fab}})$ 
    \item $t_{\textrm{max}} - t_{\textrm{min}}$ is maximized
\end{itemize}
\If{$t_{\textrm{min}} >$  0}
\State Apply a binary search and Algorithm \ref{alg:learn_s_i} to find 
\vspace{-2mm}
\[
\theta^{\textrm{min}} \gets \arg\min_{\theta \in [t_{\min} - 0.01, t_{\min}]} \theta \in \mathcal{I}^{F}(\theta_i, s_i, \varsigma_{\textrm{fab}})
\] 
\vspace{-5mm}
\Else
\State Set $\theta^{\textrm{min}} = 0$
\EndIf
\If{$t_{\textrm{max}} <$  1}
\State Apply a binary search and Algorithm \ref{alg:learn_s_i} to find 
\vspace{-2mm}
\[
\theta^{\textrm{max}} \gets \arg\max_{\theta \in [t_{\max}, t_{\max} + 0.01]} \theta \in \mathcal{I}^{F}(\theta_i, s_i, \varsigma_{\textrm{fab}})
\] 
\vspace{-5mm}
\Else
\State Set $\theta^{\textrm{max}} = 1$
\EndIf
\State Return $[\theta^{\min}, \theta^{\max}]$
\end{algorithmic}
\end{algorithm}

\begin{enumerate}
    \item Learn $s_i(\theta)$ by minimizing the risk for each $\theta$ on a grid of the interval $[0, 1]$ with details described in Algorithm~\ref{alg:learn_s_i}. 
    \item Use $s_i(\theta)$ to learn the FAB interval with detailed steps in Algorithm~\ref{alg:learn_fab}. 
\end{enumerate}

\subsection{Three FAB intervals}
\label{fab:three}

We now discuss how these steps apply to different confidence intervals for a proportion. We begin with the FAB Wald interval, $\mathcal{C}^F_N(\widehat{\theta}_i, n_i, \alpha, s_i)$. For the first step, we aim to minimize
\[
\int_{\theta \in [0,1]} \sum_{y'_i \in \{0, 1, \ldots, n_i\}} p(y'_i) \mathbbm{1}\left(\frac{y'_i}{n_i} \in \mathcal{I}^{F}\left(\theta, s_i, \sqrt{\frac{\textrm{var}(y'_i)}{n^2_i}}\right) \right) d \theta.
\]
Here $\textrm{var}(y'_i)$ is based on $p(y'_i)$ in \eqref{eq:risk}. Despite the variance not having a closed form, we can numerically compute the variance via Monte Carlo simulations given $(\mu_i, \tau_i)$. With this definition of $\mathcal{I}^{F}(\theta, s_i, \varsigma)$, we can calculate the risk given $\theta$ on a grid of the interval $[0,1]$. Given the lack of uniqueness, we can use a binary search to find the endpoints of the intervals. Over this new interval, we aim to minimize the length of the risk interval, $\mathbbm{1}\left(\frac{y'_i}{n_i} \in \mathcal{I}^{F}(\theta, s_i, \sqrt{\frac{\textrm{var}(y'_i)}{n^2_i}})\right)$. Once we have learned $s_i(\theta)$ for $\theta$ on some grid, we can use it to learn the FAB Wald interval. On the same grid of $\theta$ used to learn $s_i(\cdot)$ in the previous step, we check whether $\widehat{\theta}_i \in \mathcal{I}^{F}(\theta, s_i, \sqrt{\frac{\widehat{\theta}_i (1 - \widehat{\theta}_i)}{n_i}})$. We refer to this interval as the determination interval. Then, the FAB interval is the largest continuous region of $\theta$ for which this condition holds. We again use a binary search to find the exact end points. This requires repeating the first step to learn the $s_i(\theta)$ with Algorithm~\ref{alg:learn_s_i} for the various points being considered as part of the search. 

The FAB Agresti-Coull interval can also be found by repeating these steps with slight modifications. This affects the risk calculation in two ways. First, we aim to reduce the following in the first step:
\[
\int_{\theta \in [0,1]} \sum_{y'_i \in \{0, 1, \ldots, n_i\}} p(y'_i) \mathbbm{1}\left(\frac{y'_i + 2}{n_i + 4} \in \mathcal{I}^{F}\left(\theta, s_i, \sqrt{\frac{\textrm{var}(y'_i)}{(n_i + 4)^2}}\right)\right) d \theta,
\]
where we divide the standard deviation by $n_i + 4$. We use the modification on $y_i$ and calculate the risk based on $\frac{y'_i + 2}{n_i + 4} \in \mathcal{I}^{F}(\theta, s_i, \varsigma)$. Next, we used the modified observation, $\widetilde{\theta}_i = \frac{n_i \widehat{\theta}_i + 2}{n_i + 4}$, for the second step. We are interested in finding the longest continuous interval of $\theta$ such that $\widetilde{\theta}_i \in \mathcal{I}^{F}\left(\theta, s_i, \sqrt{\frac{\widetilde{\theta}_i (1 - \widetilde{\theta}_i)}{n_i + 4}}\right)$. 

While the same framework applies, the FAB Wilson interval's computation is different. First, given $\theta$, computing the risk in the first step involves calculating
\[
\int_{\theta \in [0,1]} \sum_{y'_i \in \{0, 1, \ldots, n_i\}} p(y'_i) \mathbbm{1}\left(\frac{y'_i}{n_i} \in \mathcal{I}^{F}\left(\theta, s_i, \sqrt{\textrm{var}(\theta)}\right)\right) d \theta.
\]
Here, $\textrm{var}(\theta) = \textrm{var}(\textrm{inv\_logit}(\eta))$, where $\eta \sim \textrm{N}(\cdot \mid \mu_i, \tau_i^2)$. This reflects the fact that we use $\theta$ instead of $y_i$ to define the variance of the interval. As a result, we can use numerical simulations to find the variance given the prior. Again, we aim to find $s_i(\theta)$ that minimizes this risk for every $\theta$ on a grid of the interval [0,1]. Then, with this $s_i(\cdot)$, we aim to find the longest continuous interval of $\theta$ for which $\widehat{\theta}_i \in \mathcal{I}^{F}\left(\theta, s_i, \sqrt{\frac{\theta(1 - \theta)}{n_i}}\right)$, as the determination interval. That resulting interval is the FAB Wilson interval, i.e., $\left\{\theta: \widehat{\theta}_i \in \mathcal{I}^{F}\left(\theta, s_i, \sqrt{\frac{\theta(1 - \theta)}{n_i}}\right)\right\}$.

\begin{table}[!tp]
    \centering
    \resizebox{\textwidth}{!}{
    \begin{tabular}{l | ccc}
    \hline\\[-1.8ex]
    \textbf{Type} & \textbf{Risk interval for} $\v{\theta}$ &\textbf{Determination interval for} $\v{\theta}$ &  \textbf{FAB Interval for} $\v{\widehat{\theta}}$ \\
    \\[-1.8ex]\hline\\  
     Wald& $\mathcal{I}^{F}\left(\theta, s_i, \sqrt{\frac{\textrm{var}(y'_i)}{n^2_i}}\right)$&$\mathcal{I}^{F}\left(\theta, s_i, \sqrt{\frac{\widehat{\theta}_i(1 - \widehat{\theta}_i)}{n_i}}\right)$
    &$\left\{\theta : \widehat{\theta}_i  \in \mathcal{I}^{F}\left(\theta, s_i, \sqrt{\frac{\widehat{\theta}_i(1 - \widehat{\theta}_i)}{n_i}}\right)
\right\}$\\
     Agresti-Coull$^*$& $\mathcal{I}^{F}\left(\theta, s_i, \sqrt{\frac{\textrm{var}(y'_i)}{(n_i + 4)^2}}\right)$& $\mathcal{I}^{F}\left(\theta, s_i, \sqrt{\frac{\widetilde{\theta}_i(1 - \widetilde{\theta}_i)}{n_i+4}}\right)$& $\left\{\theta : \widetilde{\theta}_i  \in \mathcal{I}^{F}\left(\theta, s_i, \sqrt{\frac{\widetilde{\theta}_i(1 - \widetilde{\theta}_i)}{n_i+4}}\right)
\right\}$\\
    Wilson$^*$ & $\mathcal{I}^{F}\left(\theta, s_i, \sqrt{\textrm{var}(\theta)}\right)$& $\mathcal{I}^{F}\left(\theta, s_i, \sqrt{\frac{\theta(1 - \theta)}{n_i}}\right)$& $\left\{\theta : \widehat{\theta}_i  \in \mathcal{I}^{F}\left(\theta, s_i, \sqrt{\frac{\theta(1 - \theta)}{n_i}}\right)\right\}$\\
    \hline
    \end{tabular}
    }
    \caption{Summary of different intervals used in the construction of the FAB intervals. Here, $\widetilde{\theta}_i = \frac{y_i + 2}{n_i + 4}$, $\textrm{var}(y_i')$ is defined in \eqref{eq:risk}, and $\textrm{var}(\theta)$ is defined to be $\textrm{var}(\theta) = \textrm{var}(\textrm{inv\_logit}(\eta))$. Note that an asterisk next to the FAB interval type indicates the use of the all-in penalty when computing the risk with the risk interval.}
    \label{tab:interval_table}
\end{table}

Table~\ref{tab:interval_table} summarizes the risk, determination, and FAB intervals for the Wald, AC, and Wilson intervals, respectively. The explicit instruction for obtaining the FAB intervals in Algorithm \ref{alg:learn_fab} and the code are available on Github.

\subsection{Prior specification and extension to MRP estimates}

\textcite{BurrisHoffExactAdaptiveConfidence2020} show that the FAB intervals have exact domain-specific coverage and are coverage-robust regardless of whether or not a linking model is misspecified. Improvements to the linking model are expected to reduce the interval width. In hierarchical Bayesian domain inference, the prior specification determines the amount of shared information across domains and the inferential uncertainty. In MRP applications, an exchangeable normal prior could lead to overshrinkage toward the overall mean and result in inferior performance \parencite{Valliant19-jssam, mrp-si20}. To weaken the strong assumption of exchangeability, \textcite{BNFP:SI15} applied a Gaussian process (GP) prior to the mean functions in poststratification cells, \textcite{prior-si2018} developed a structured prior for high-order interaction terms among covariates, \textcite{priorSAE-Tang18} applied a global-local shrinkage prior to SAE, and \textcite{WakefieldEtAlSmallAreaEstimation2020} incorporated spatio-temporal correlation into the modeling of domain estimates. We examine the impact of prior choice on the proposed FAB intervals for a proportion. We compare inferences with the GP prior and the global-local normal prior and present the prior specifications in detail in the following session.

We present the FAB intervals for each domain proportion, $\theta_i$, or a combination of the domain proportions. MRP estimates (with details given below) are a weighted average of the covered domain estimates, and we can modify the FAB intervals using poststratified point estimates and variance estimates based on the Bayesian inference for MRP under the hierarchical prior specification. We demonstrate the extension in the following sections.

\section{Simulation studies}
\label{sec:sim_results}
To assess properties of the FAB intervals for proportions, we conduct two simulation studies. The first considers data generated from a binomial distribution and the standard normal prior as \eqref{bin}. The first simulation is designed to determine which method of computing risk is most suitable for each of the three FAB intervals and understand whether and why the FAB interval procedure can improve the coverage of confidence intervals for proportions, using the observed proportion as the estimate for $\widehat{\theta}_i$ and a standard normal distribution as prior. These findings are then carried forward to the second simulation, in which the data-generating process is designed to closely reflect the application study of interest and we focus on examining the effect of hyperprior choice. The second simulation uses Bayesian inference and uses the post-stratified sample means as $\widehat{\theta}_i$. For Bayesian inference, we consider hierarchical prior specifications, including the GP and global-local priors, and use the posterior mean and variance from the hierarchical structure as the normal ``prior" distribution for $\eta_i$. We use the posterior information to inform the choice of $\mu_i$ and $\tau^2_i$. Across both studies, we evaluate the frequentist properties and compare the 95\% FAB intervals with the original confidence or credible intervals for proportions, i.e., the Wald, Wilson, and AC intervals.

\subsection{Calibrated Bayes inference for a proportion}
First, we design a simulation study to compare the Wald, Wilson and AC intervals based on their original formulations described in Section~\ref{original} and the FAB versions given in Section \ref{fab:three}, examining both the default FAB intervals and those with all-in penalty. As stated above, because we have two ways of computing risk, our goal with this study is to determine which method works better for each interval according to the coverage probability. We also want to understand how these risks induce the changes observed in the FAB intervals' behavior. To accomplish this, we assume that $y_i$ is generated from a binomial distribution and the prior for the FAB interval is the standard normal distribution given in \eqref{bin}. In other words, the prior mean for the FAB interval is 0 and the prior variance is 1. We fix $n_i = 100$. For each $y_i \in \{0,1,\ldots,100\}$, we compute the interval
$\mathcal{C}\!\left(\frac{y_i}{n_i},\, n_i,\, 0.05,\, s_i\right)$.
Rather than enumerating every original and FAB interval separately, we use $\mathcal{C}(\cdot)$ as a generic placeholder for the specific interval procedure being evaluated.

We assess the performance of different intervals over a grid of $\theta_i \in [0,1]$ by computing the coverage probability at each $\theta_i$. Specifically, for a given $\theta_i$, the coverage is the binomial likelihood weighted probability that the interval constructed from $y_i$ contains $\theta_i$, where $y_i \sim \text{Binomial}(\cdot \mid \theta_i, n_i)$:
\begin{align}
\label{def:cr}
\sum_{y_i = 0}^{100} 
\mathbbm{1}\!\left\{\theta_i \in \mathcal{C}\!\left(\frac{y_i}{n_i},\, n_i,\, 0.05,\, s_i\right)\right\}
\, \text{Binomial}(y_i \mid \theta_i, n_i).
\end{align}
That is, we sum the probabilities of those outcomes $y_i$ whose corresponding intervals contain the true parameter value $\theta_i$.




\begin{figure}[!t]
\centering
\begin{subfigure}{0.32\textwidth}
    \centering
    \includegraphics[width=0.99\textwidth]{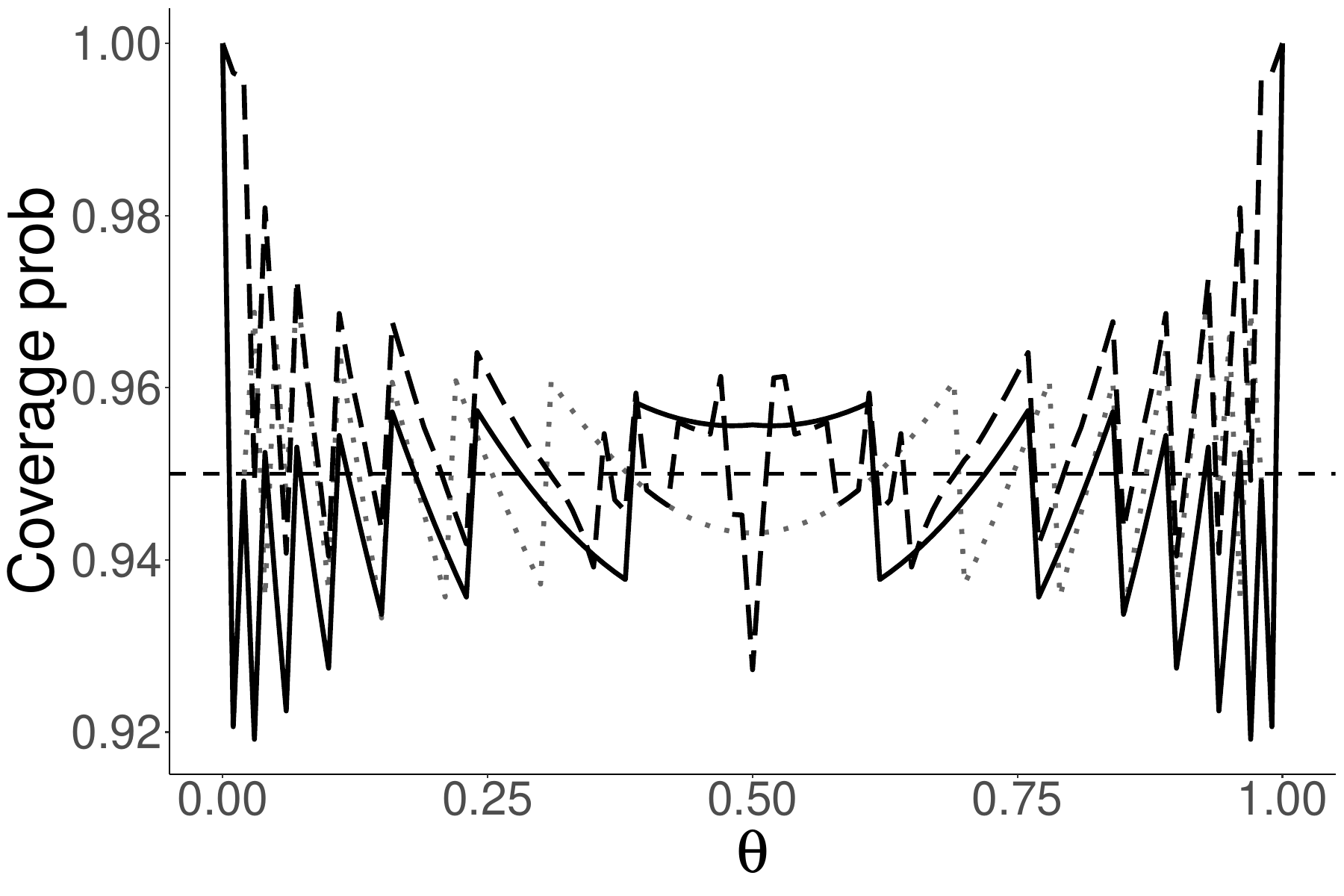} 
    \caption{Wilson}
\end{subfigure}
\begin{subfigure}{0.32\textwidth}
    \centering
    \includegraphics[width=0.99\textwidth]{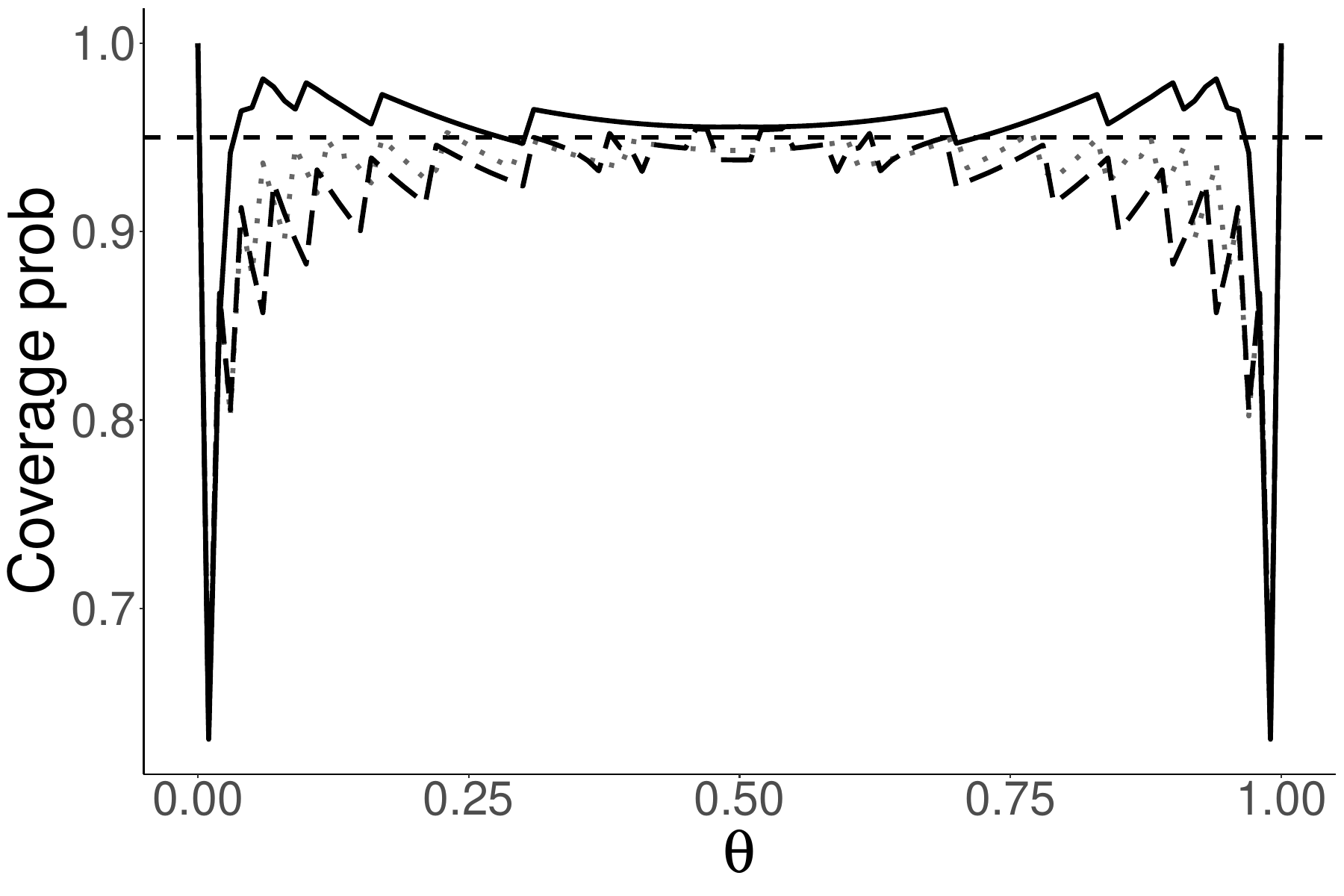}
    \caption{Wald}
\end{subfigure}
\begin{subfigure}{0.32\textwidth}
    \centering
    \includegraphics[width=0.99\textwidth]{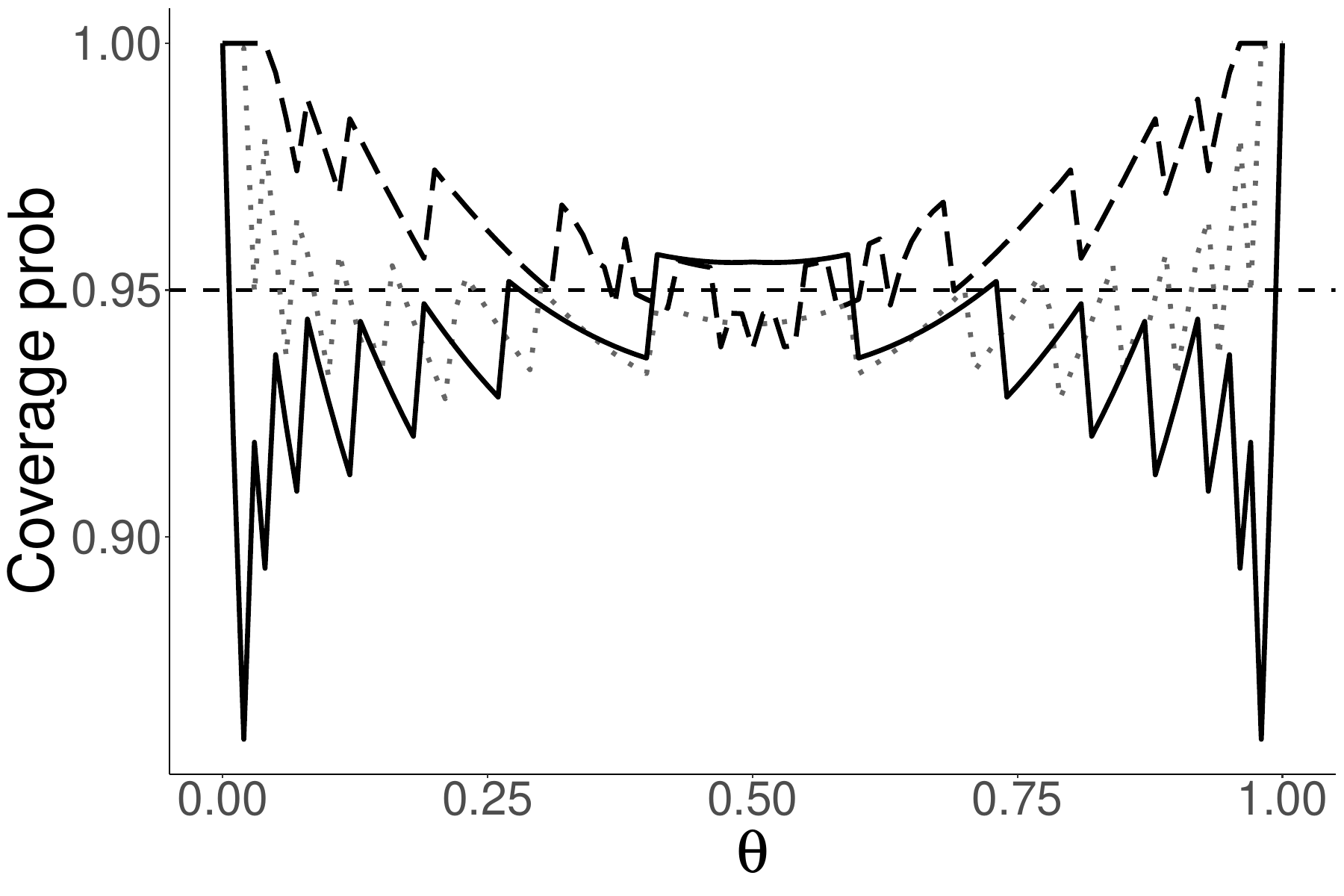}
    \caption{AC}
\end{subfigure}\\
\centering
\begin{subfigure}{0.32\textwidth}
    \centering
    \includegraphics[width=0.99\textwidth]{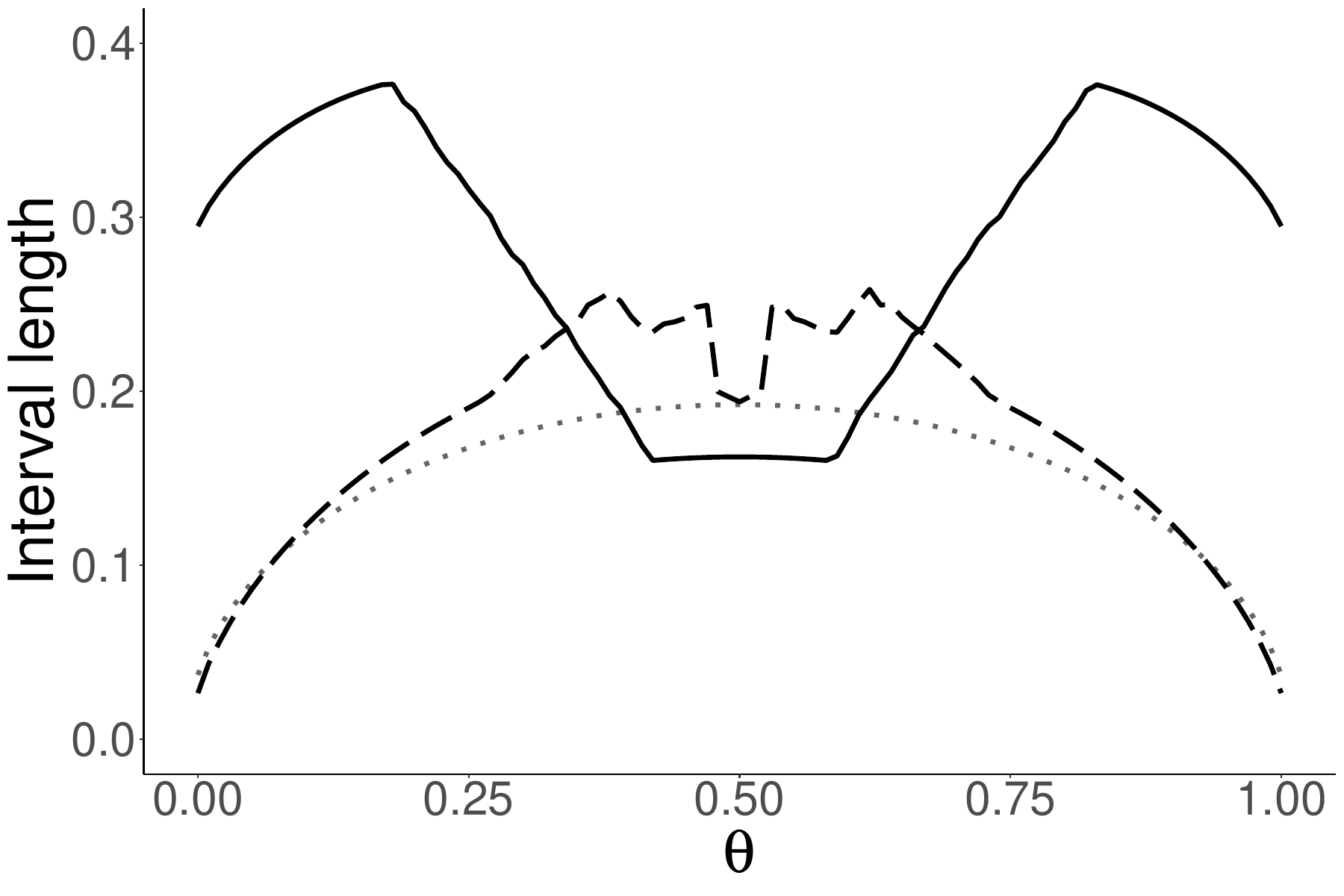} 
    \caption{Wilson}
\end{subfigure}
\begin{subfigure}{0.32\textwidth}
    \centering
    \includegraphics[width=0.99\textwidth]{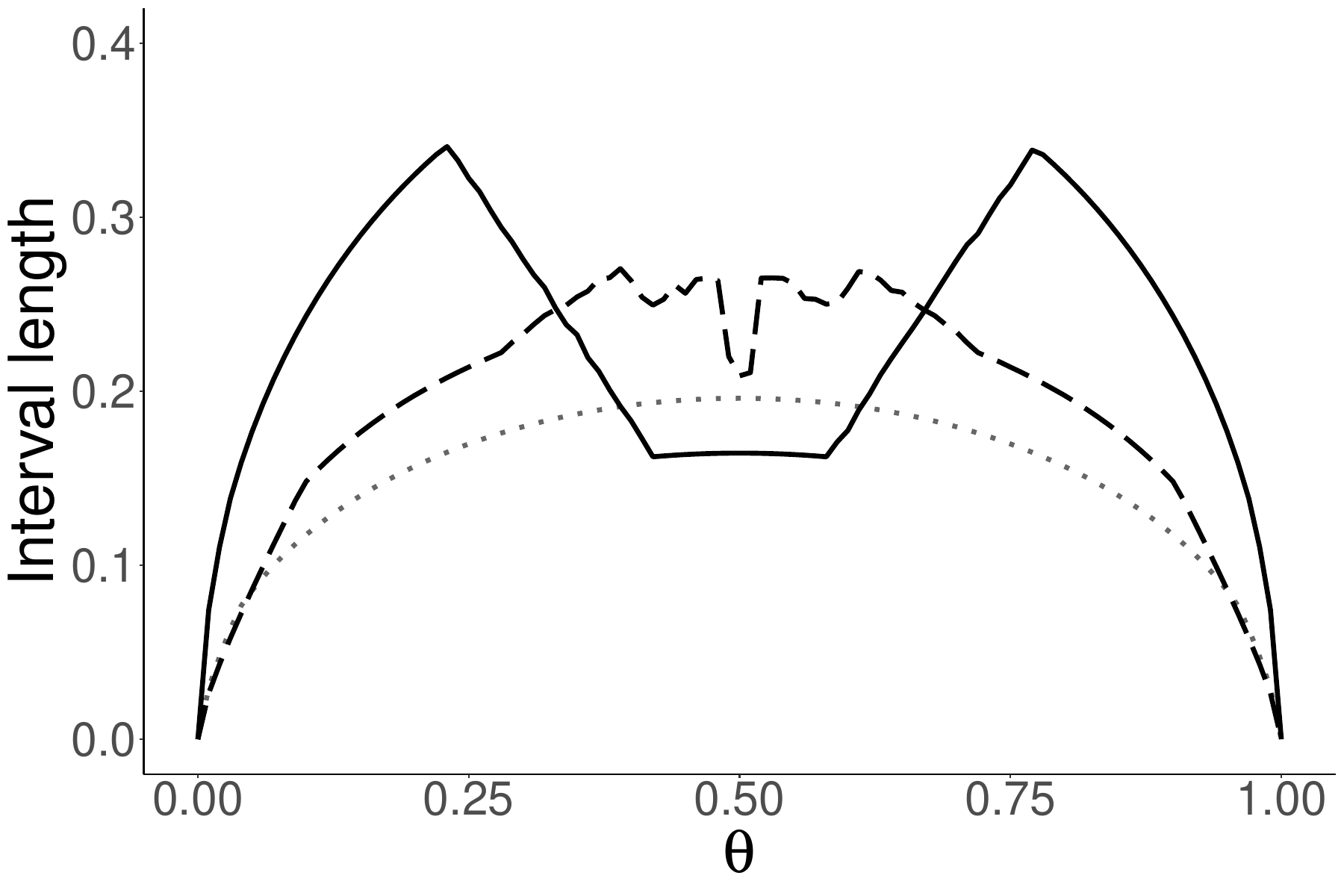}
    \caption{Wald}
\end{subfigure}
\begin{subfigure}{0.32\textwidth}
    \centering
    \includegraphics[width=0.99\textwidth]{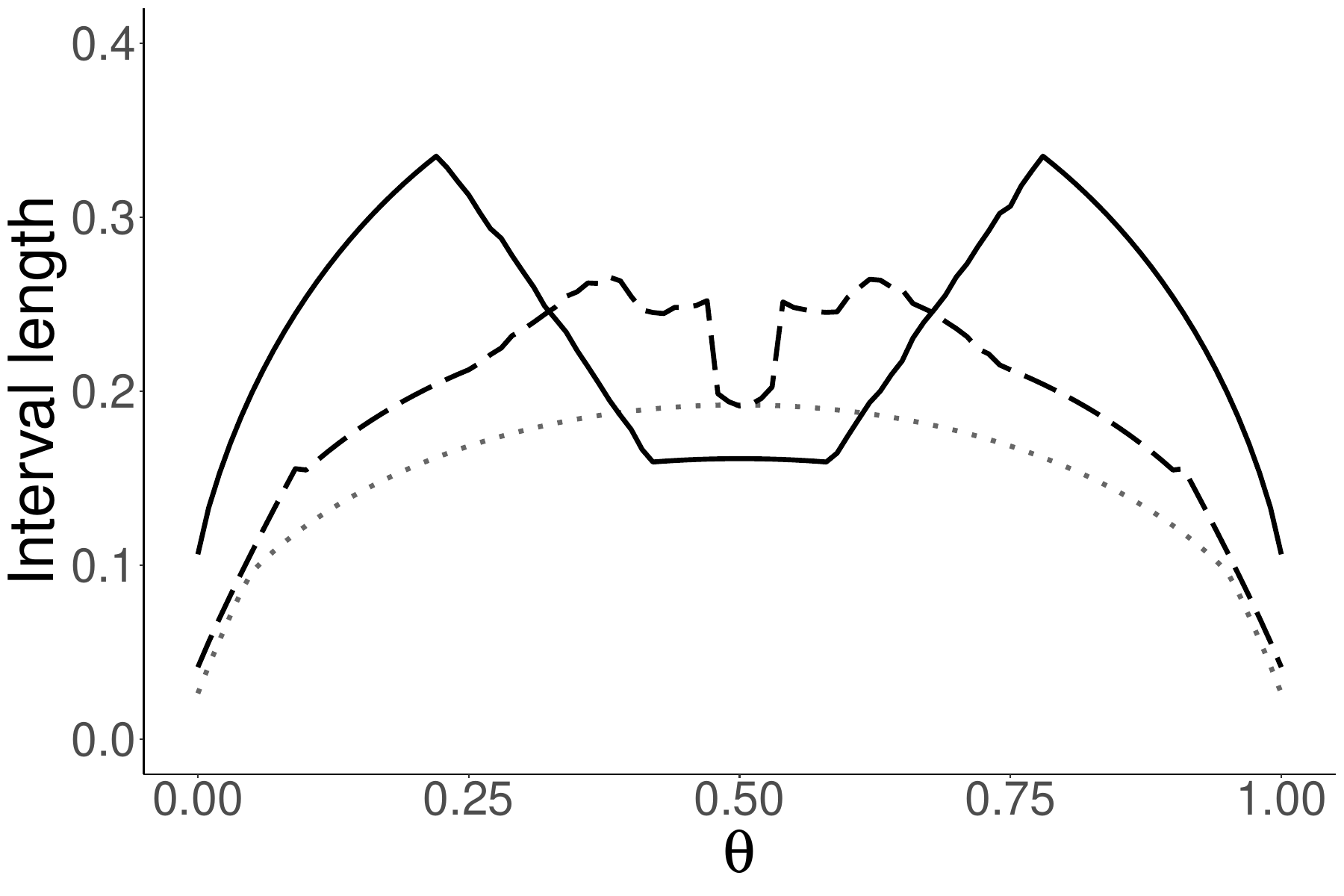}
    \caption{AC}
\end{subfigure}\\
\caption{Coverage probability (top row) and interval length (bottom row) for the original and FAB versions of the Wilson, Wald, and Agresti--Coull (AC) intervals under fixed $n_i=100$ and varying $\theta_i$. The solid black line denotes the default FAB interval, the dashed black line denotes the FAB interval with the all-in penalty, and the gray dot-dashed line denotes the originally defined interval. The dashed horizontal line in the top row indicates the nominal 95\% coverage level.}
\label{fig:coverage_prob_binary_sim}
\end{figure}

Figure \ref{fig:coverage_prob_binary_sim} compares the coverage probabilities and interval lengths compares the coverage probabilities and interval lengths of the original Wilson, Wald, and AC intervals with those of their FAB versions. The Wilson intervals are the most stable across $\theta_i$. The FAB Wilson interval with the all-in penalty improves coverage near the boundaries while avoiding the large interval lengths of the default FAB Wilson interval. The Wald intervals perform poorly near 0 and 1, reflecting the well-known degeneracy of the Wald standard error at the boundaries. Although the FAB Wald interval improves coverage for some values of $\theta_i$, it often requires much wider intervals and remains unreliable near the endpoints. For the AC intervals, the all-in penalty improves coverage away from the center and especially near the boundaries, with only a moderate increase in interval length relative to the original AC interval. Overall, the FAB Wilson interval with the all-in penalty provides the best balance between coverage and interval length, while the FAB AC interval with the all-in penalty offers a more conservative alternative.

\begin{figure}[!t]
\centering
\begin{tabular}{ccc} 
When $\theta_i=0.05$&&\\
\includegraphics[width=0.3\textwidth]{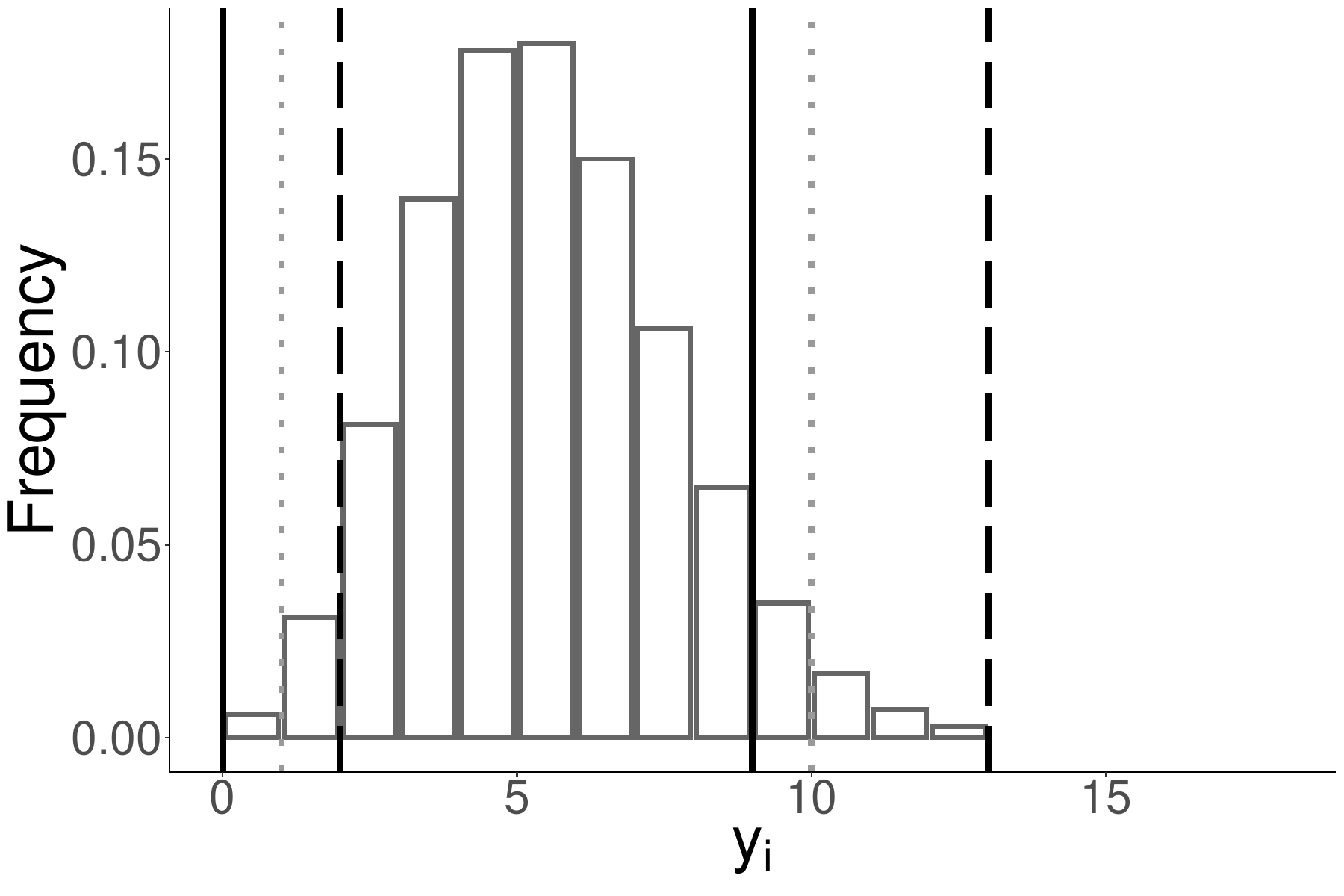}   &     
\includegraphics[width=0.3\textwidth]{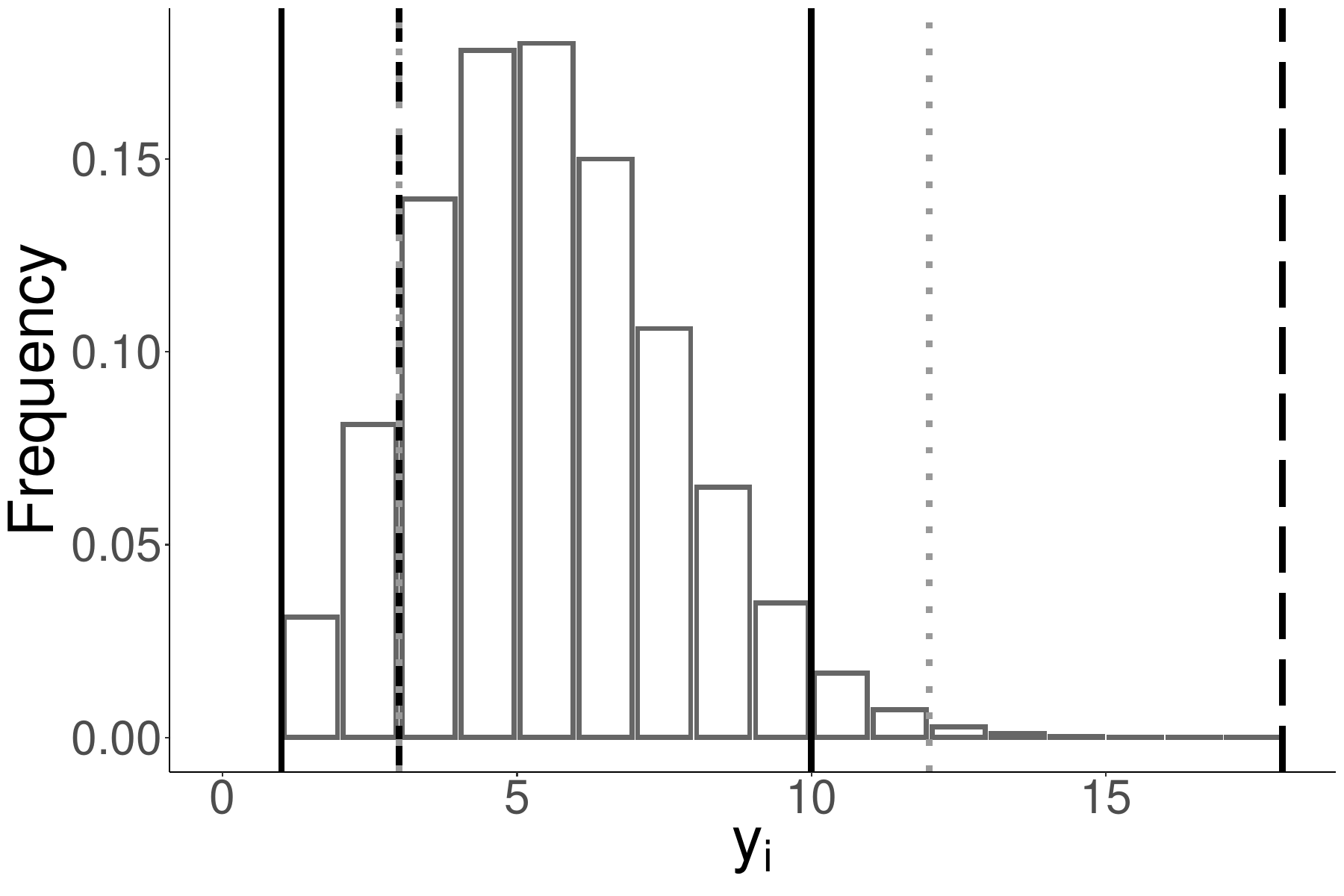} 
&
\includegraphics[width=0.3\textwidth]{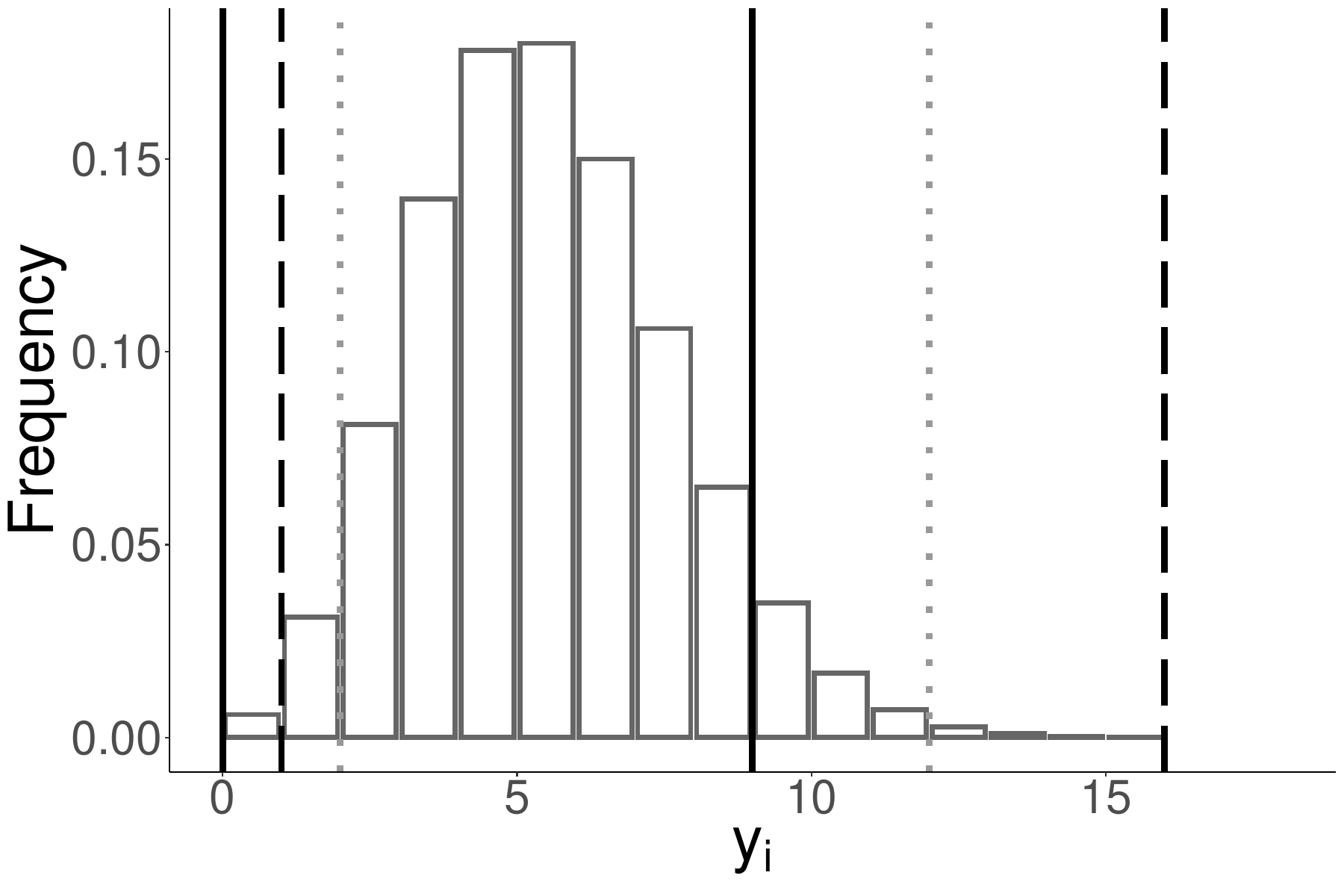} 
\\
When $\theta_i=0.1$&&\\
\includegraphics[width=0.3\textwidth]{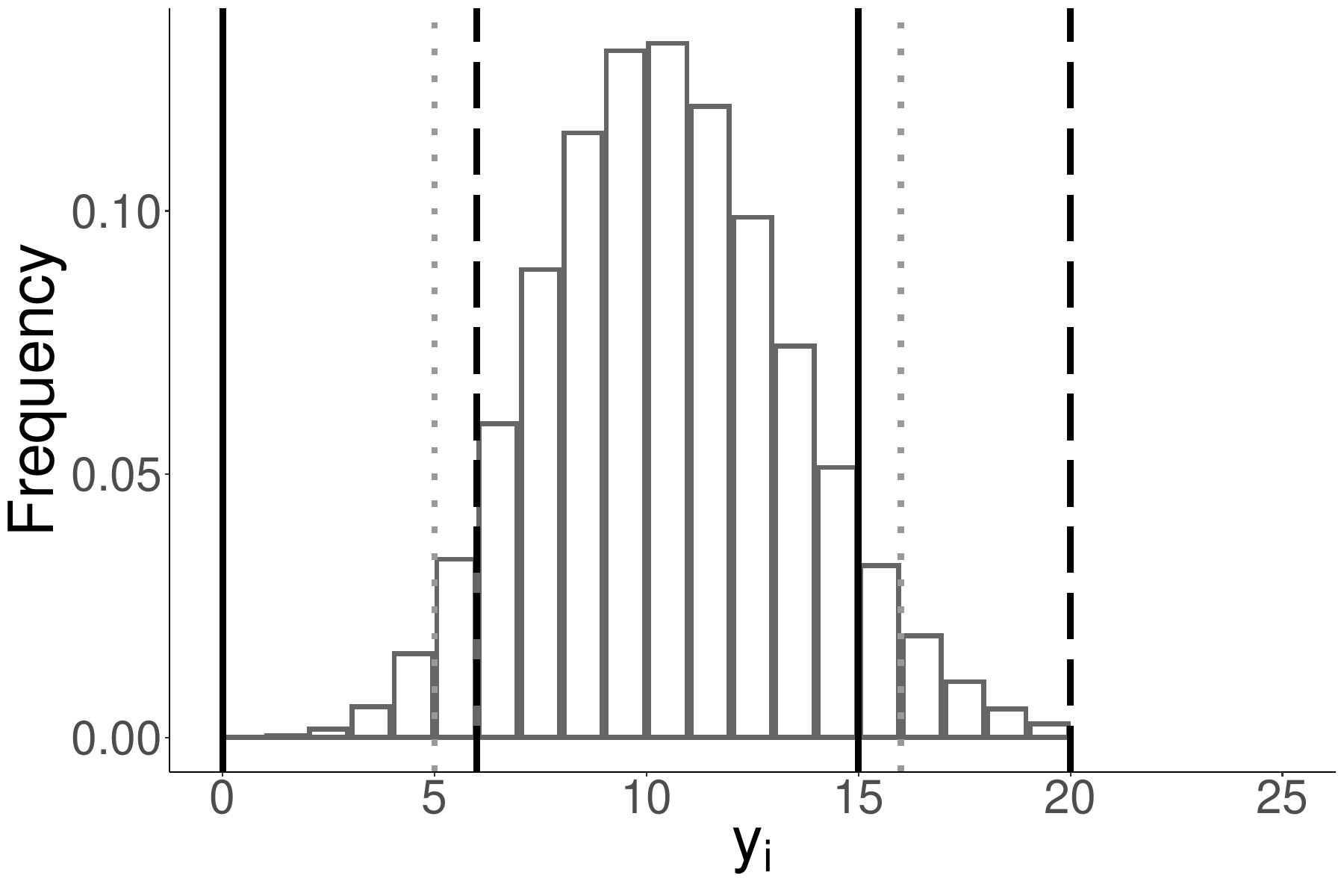}  &     
\includegraphics[width=0.3\textwidth]{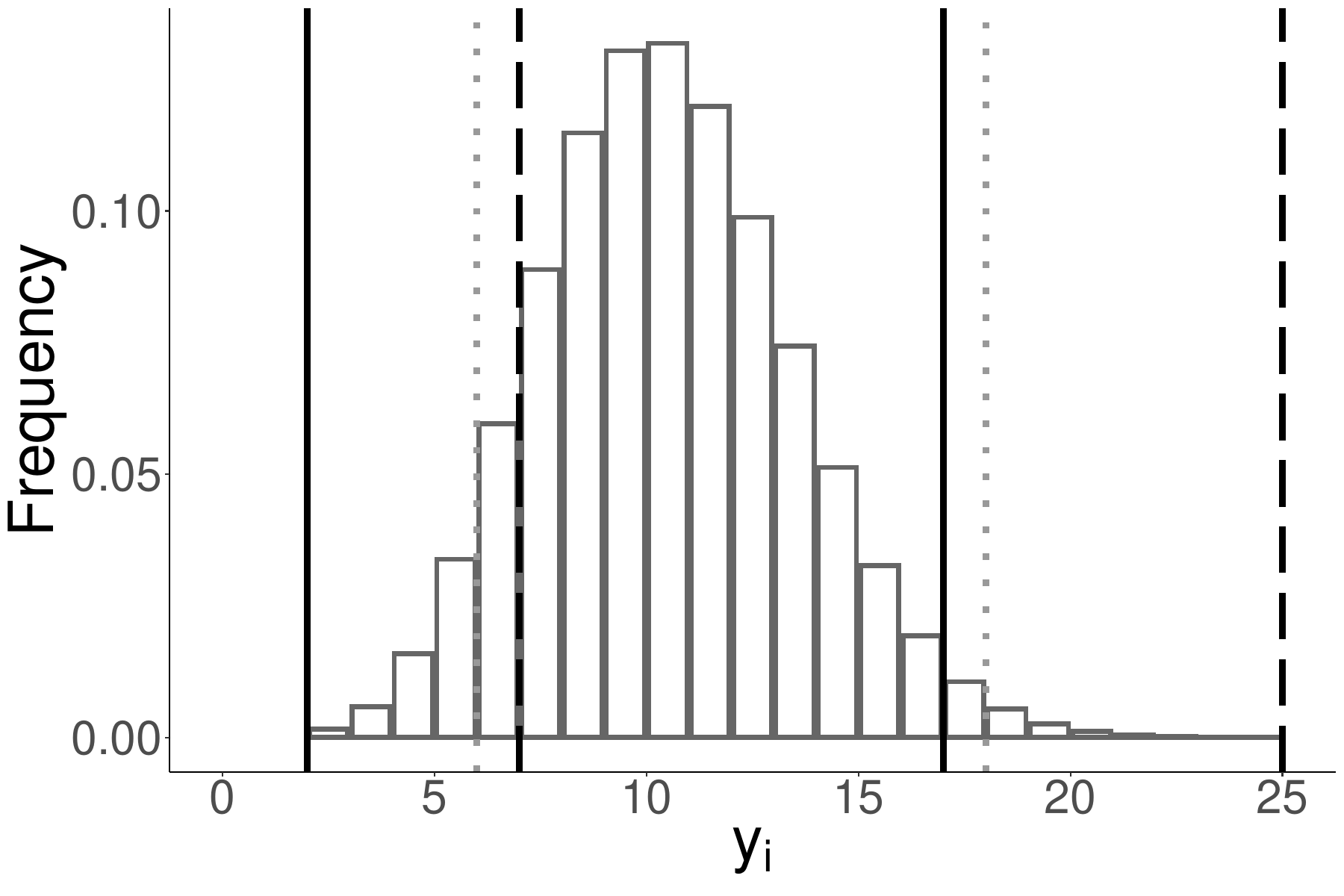}
&
\includegraphics[width=0.3\textwidth]{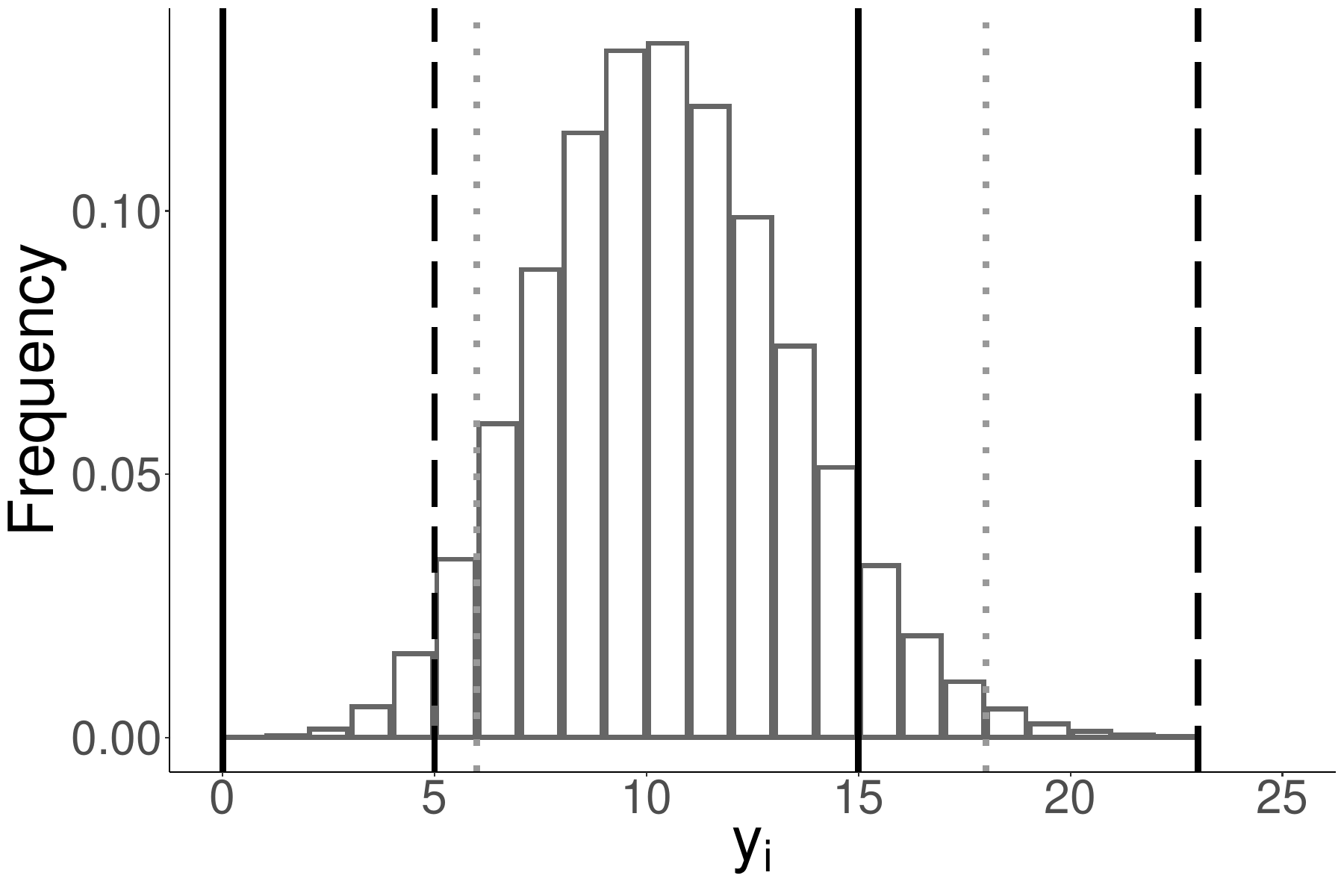}
\\
When $\theta_i=0.4$&&\\
\includegraphics[width=0.3\textwidth]{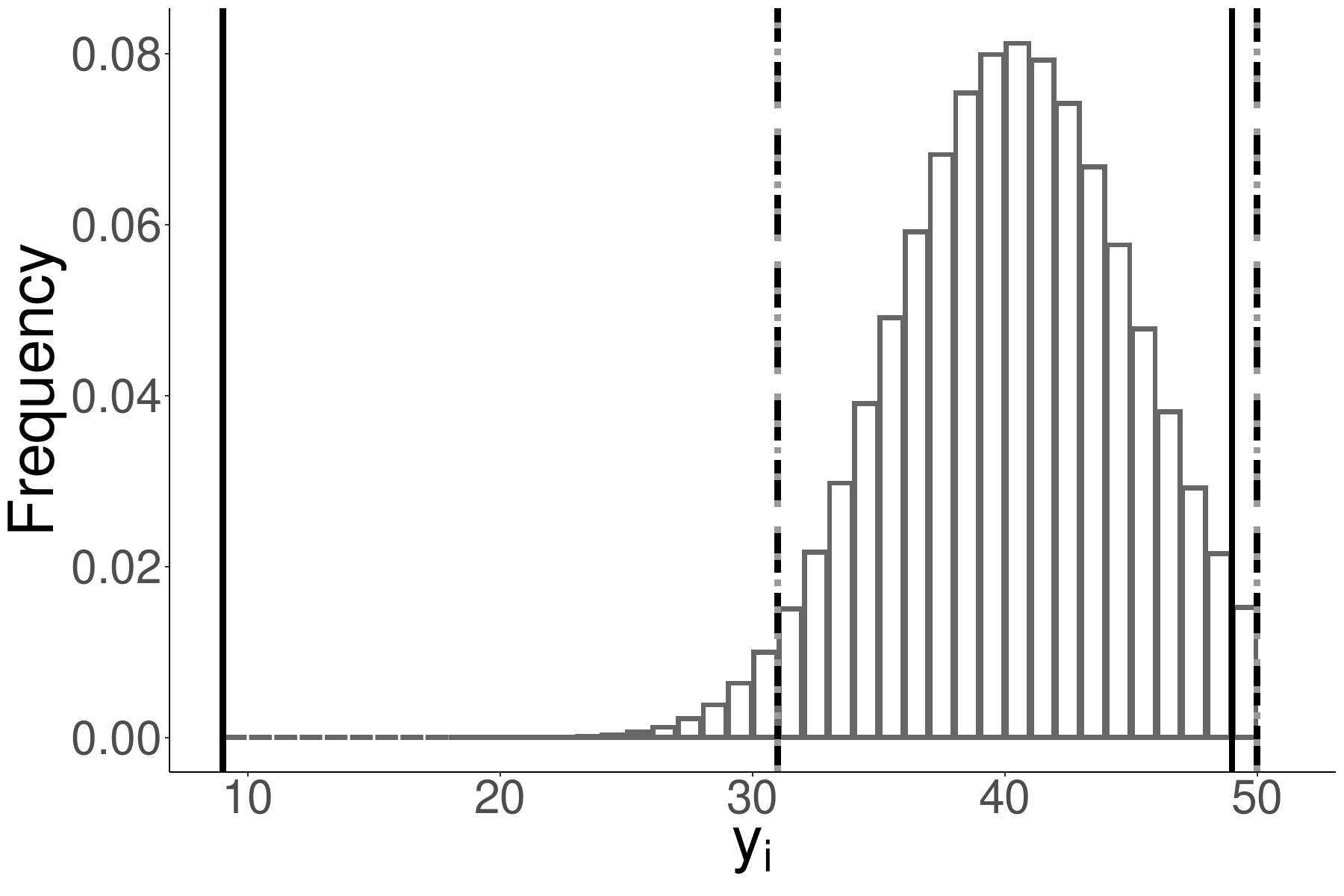}  &     
\includegraphics[width=0.3\textwidth]{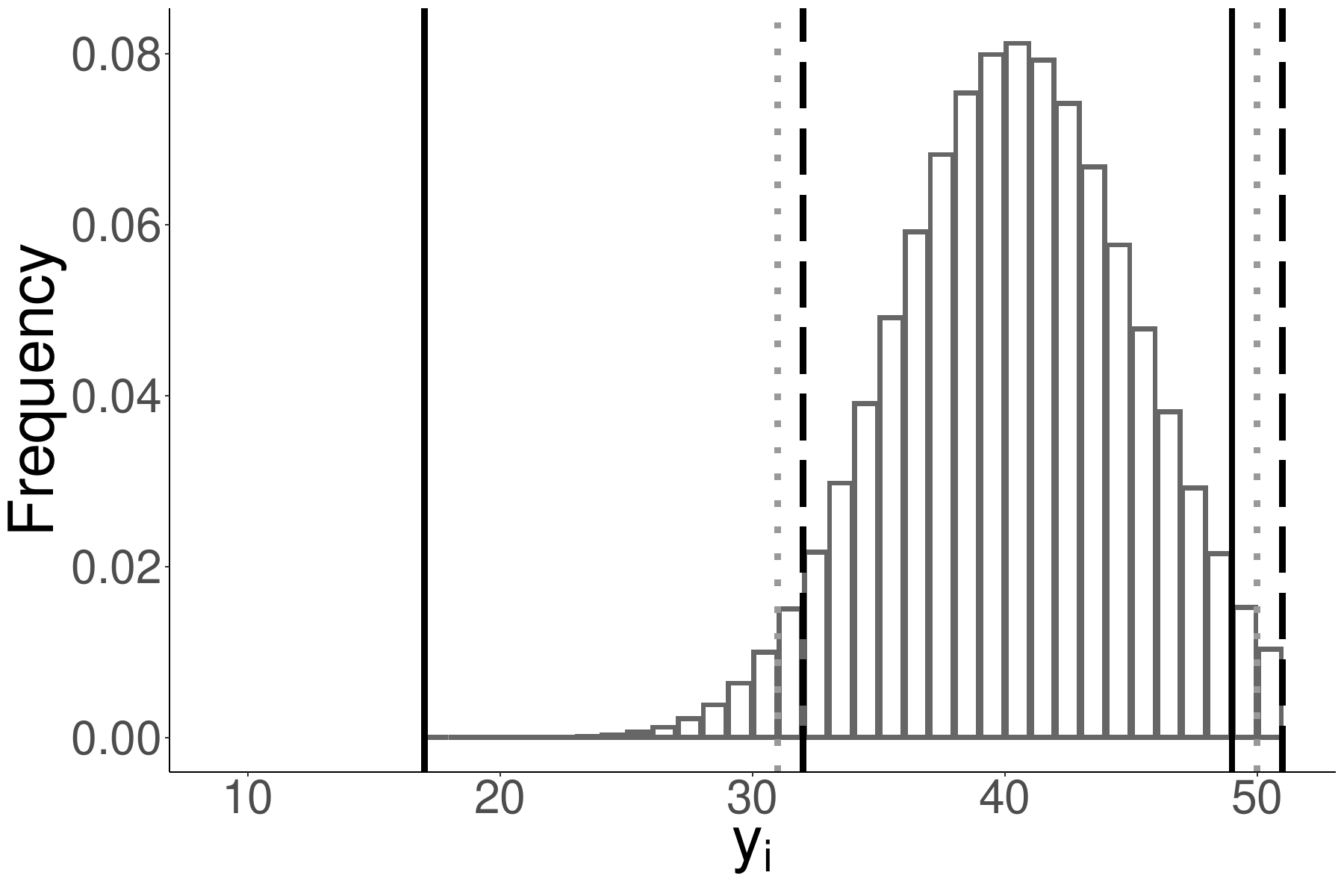}
&
\includegraphics[width=0.3\textwidth]{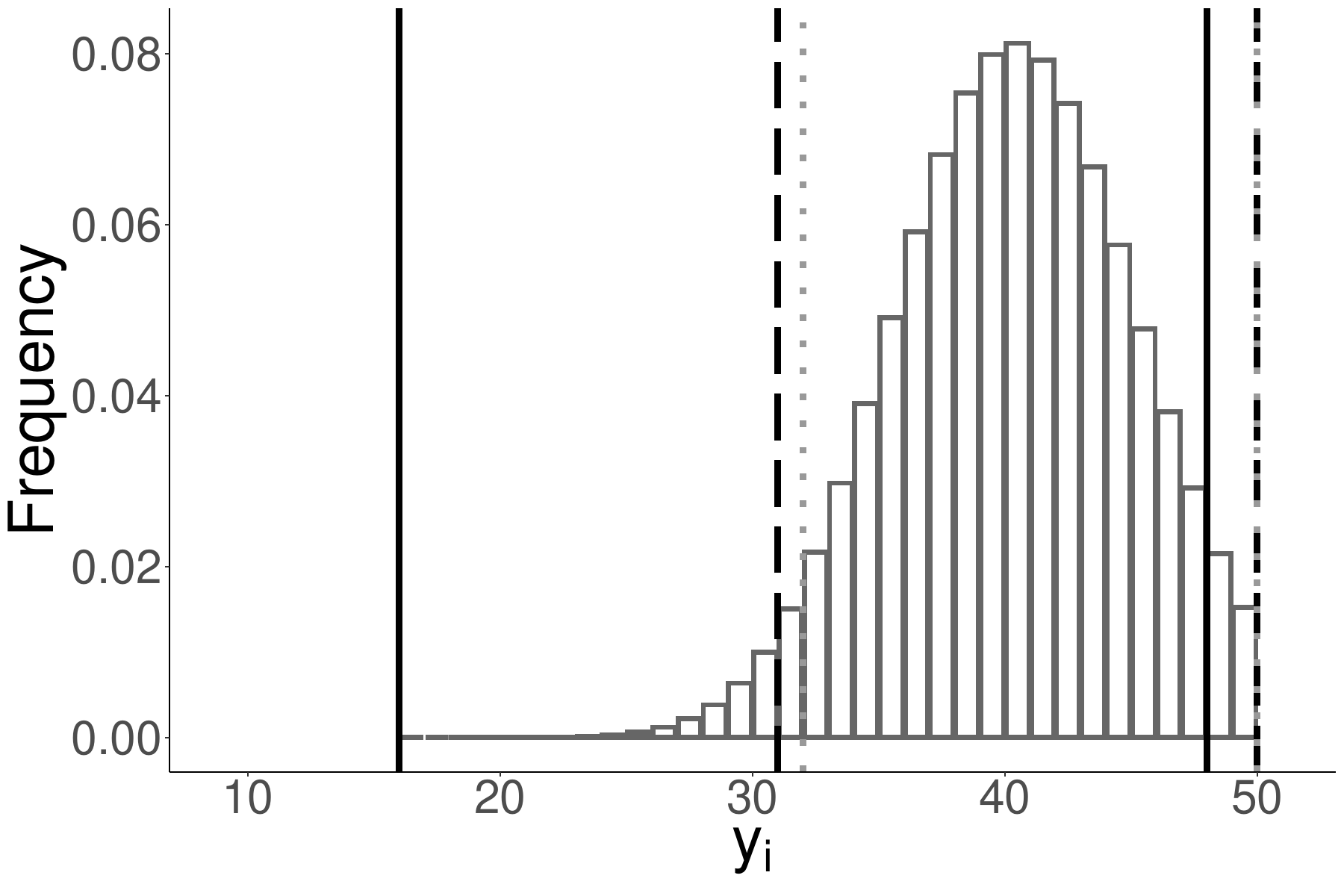}
\\
(a) Wilson interval & (b) Wald interval & (c) AC interval\\
\end{tabular}
\caption{The range of values of $y_i$ for which a Frequentist assisted Bayes (FAB) or confidence interval built on $y_i / n_i$ captures a given $\theta_i$. The FAB intervals created from $y_i$'s within the black solid lines, the FAB (all-in penalty) intervals created from $y_i$'s within the black dashed lines, and the original intervals created from $y_i$'s within the gray dotted lines include $\theta_i$. The histogram frequency displays the probability $\text{Binomial}(y_i \mid \theta_i, n_i)$, i.e., the associated probability of the $y_i$ value that yields the FAB or original intervals covering $\theta_i$.}
\label{fig:binary_sim_ci_comp}
\end{figure}

To understand why the FAB intervals have these coverage probabilities, we consider three different values of $\theta_i$, (0.05, 0.1, 0.4), and assess which values of $y_i$ lead to the intervals that contain these values of $\theta_i$. Based on \eqref{def:cr}, the coverage for a given interval type is obtained by summing the likelihood $\text{Binomial}(y_i \mid \theta_i, n_i)$ over all outcomes $y_i$ whose intervals contain $\theta_i$. Consequently, the coverage can differ across interval types because the set of values of $y_i$ for which $\theta_i$ is covered and the probabilities of those outcomes under $\theta_i$ are different. Figure 3 illustrates that for $\theta_i \in \{0.05, 0.1, 0.4\}$, the binomial probabilities associated with each $y_i$ value that yields FAB or original intervals covering $\theta_i$. 
One trend across all values of $\theta_i$ is that compared to the regular intervals, the FAB intervals shift the range of $y_i$ values whose intervals cover the $\theta_i$'s to the ends because there may be a long tail of $y_i$ values close to $0$ whose intervals include the corresponding $\theta_i$'s. For instance, for $\theta_i = 0.1$, the FAB intervals associated with $y_i \leq 5$ include 0.1, which is not the case for the originally defined and the FAB intervals (all-in penalty). In other words, the FAB intervals are shifting the range to the ends. Meanwhile, the FAB intervals (all-in penalty) also change which $y_i$'s intervals cover the $\theta_i$'s in that there may be a shift toward the right. For example, when we examine $\theta_i = 0.1$, the FAB intervals (all-in penalty) based on larger values of $y_i$ now include the $\theta_i$. This is the opposite behavior compared to the FAB intervals. Overall, there is a shift in the range of $y_i$ values toward the middle.

Because of these dynamics with the FAB intervals, it is crucial to identify the range of $y_i$ values included by these intervals and the corresponding coverage probabilities. We begin with the Wald interval. When we examine the originally defined Wald interval, the intervals associated with $y_i$ between 3 and 11 include $\theta_i = 0.05$. If we then use the FAB intervals (all-in penalty), the range of $y_i$ whose intervals cover $\theta_i = 0.05$ expands to the left and intervals based on any $y_i$ value between 3 and 18 cover the $\theta_i$. However, the probability of $y_i \geq 12$ given $\theta_i = 0.05$ is close to zero. Hence, while the FAB intervals (all-in penalty) expand the range of $y_i$, the increase in coverage probability is small. Using the default FAB interval means that intervals associated with $y_i$ between 1 and 9 include $\theta_i = 0.05$. The default FAB interval extends the range of $y_i$ on the left at the cost of losing some large $y_i$ values. Given that $\theta_i = 0.05$, the probability of $y_i \in \{1, 2\}$ is greater than that of $y_i \in \{10, 11\}$.  Hence, by extending the range of $y_i$ whose intervals cover $\theta_i$ in this way, the default FAB interval improves the coverage probability for the Wald interval. A similar trend applies to various Wald intervals with $\theta_i = 0.1$ and $\theta_i = 0.4$.

Nevertheless, shifting the range of $y_i$ to the left does not benefit the AC interval or Wilson interval. With $\theta_i = 0.05$, the originally defined Wilson interval applied to $y_i$ between 1 and 9 includes $\theta_i$. Meanwhile, the originally defined AC interval includes $\theta_i$ for $y_i$ between 2 and 11. Because the minimum of $y_i$ is 0, there is not much room to shift the range to the left. Not only that, given $\theta_i = 0.05$, the total probability of $y_i = 0, 1$ is much smaller compared to the total probability of $y_i \in \{9, 10, 11\}$, making the shift toward the ends detrimental. As a result, the default FAB interval does not improve the coverage probability. Meanwhile, expanding the range of $y_i$ whose intervals cover $\theta_i$ to the right helps improve the coverage. In the case of the Wilson intervals for $\theta_i = 0.05$, this overcomes the fact that the FAB interval with the all-in penalty based on $y_i = 1$ no longer includes $\theta_i$. For the AC intervals, the FAB interval with the all-in penalty also expands the range of $y_i$ on both sides with most of the expansion on the right side. Hence, the changes to the range of $y_i$ induced by the FAB intervals (all-in penalty) improve the coverage probability of the AC and Wilson intervals. Examining the range of $y_i$ whose interval includes $\theta_i$ and the probabilities associated with those $y_i$ explains the improvement in coverage probability. Further discussion of why these shifts occur due to the construction of the FAB intervals can be found in the supplementary material.

In summary, these differences explain the performance of the intervals. For the Wilson and AC intervals, we are functionally adjusting the estimates to be more central whereas for the Wald interval, we want to shift it closer to the boundary. The results of these experiments also support the use of the all-in penalty for the FAB AC and Wilson intervals. However, for the FAB Wald intervals, the default FAB interval can be utilized. To simplify notation when discussing these intervals in the rest of the paper, the FAB intervals for the AC and Wilson intervals refer to those intervals with the all-in penalty whereas the FAB Wald intervals have no penalty.

\subsection{Simulation study based on COVID-19 test data}
\label{sim-2}

We next look at an example motivated by real-world data. Our goal is to see whether using the results from Bayesian inference to build the FAB intervals improves on the coverage probability of domain estimates compared to the Bayesian credible intervals or the original confidence intervals. Using the MRP estimates that also adjust for sample selection bias in the motivating application study, we are interested in understanding how the FAB intervals perform for small area estimates after poststratification. 

\subsubsection{Data generation}
We use a subset of COVID-19 test records collected from a Midwest hospital system, which was also used by \textcite{mrp-covid21, mrp-covid22,si2024multilevel} to implement MRP. The dataset includes COVID-19 test results of asymptomatic patients who have elective surgery appointments and their sex (male or female), age in years (0-17, 18-34, 35-64, 65-74, or 75+), race (White, Black, or other), and residence ZIP code. We choose a subset of the test records restricted to a pre-selected time period (January 2022) and catchment area covering 30 ZIP codes. \textcite{si2024multilevel} construct poststratification cells based on the cross-tabulation of sex, age, race, and ZIP code and obtain MRP estimates by using the population cell counts from the American Community Survey (ACS). In the simulation, we use the same poststratification structure as \textcite{si2024multilevel}, with cell index $j=1,\dots, J$, where $J$ is the total number of cells in the contingency table of sex, age, race, and ZIP code. We ignore time and model the cross-sectional test results for simulation purposes. We first fit a model to the observed data and then use the model with estimated parameters to generate synthetic data. In particular, we fit the following logistic regression model:
\begin{align}
\label{eq:logit_model}
y_j  \sim \textrm{Binomial}(\cdot \mid \pi_j, n_j),\,\,\, 
\pi_j = \textrm{inv\_logit}\left(\alpha_0 + \v{X}_j \beta_{sex} + \alpha^{\textrm{age}}_{j} + \alpha^{\textrm{race}}_{j} + \alpha^{\textrm{ZIP}}_{j}\right).        
\end{align}
Here, the cell-wise positivity rate $\pi_j$ is a function of the intercept ($\alpha_0$) and varying effects of age ($\alpha^{\textrm{age}}_{j}$), race ($\alpha^{\textrm{race}}_{j}$) and ZIP code ($\alpha^{\textrm{ZIP}}_{j}$). We use $\v{X}_j$ to indicate sex where $X_j$ = 1 for men and $X_j$ = 0 for women. We introduce the following prior distributions: $\alpha_0 \sim t(3, 0, 2.5)$, a Student t distribution with three degrees of freedom, mean of zero, and a standard deviation of 2.5; and $\alpha^{\textrm{var}}_{j} \sim \textrm{N}(0, {\sigma^{\textrm{var}}}^2)$, with an unknown standard deviation parameter $\sigma^{\textrm{var}} \sim t^+(3, 0, 2.5)$, a half-t distribution restricted to positive values with three degrees of freedom, mean of zero, and a scale of 2.5, where $\textrm{var} \in \{ \textrm{age}, \textrm{race}\}$. Note that we use the same prior for $\beta_{sex}$ as that for the intercept $\alpha_0$. We consider two prior distributions for the varying effect across ZIP codes: 1) assuming independence across locations, i.e., using the same prior used as those for other varying effects, $\alpha^{\textrm{ZIP}}_{j} \sim N(0, {\sigma^{\textrm{ZIP}}}^2)$, $\sigma^{\textrm{ZIP}} \sim t^+(3, 0, 2.5)$; and 2) accounting for spatial correlation across locations via a GP spatial prior, $\alpha^{\textrm{ZIP}} \sim \textrm{GP}(0, \Sigma_{x})$, with a squared exponential kernel: 
\[
k(x_j, x_k) = \sigma^2 \exp\left(-\frac{d_H(x_j, x_k)^2}{2 \ell^2}\right),
\]
where $k(x_j, x_k)$ is the $j$th row and $k$th column element of the covariance matrix $\Sigma_{x}$, $(x_j, x_k)$ denote the longitude and latitude of locations $j$ and $k$, defined as the centroids of ZIP code $j$ and $k$, respectively, $d_H(\cdot, \cdot)$ is the Haversine distance, $\sigma^2$ is the variance of the GP, and $\ell$ represents the bandwidth of the kernel. We set $\sigma = 1$ and $\ell = 5$ as the hyperparameter values in the GP spatial prior for the convenience of data generation. 

We fit the model with the specified priors and draw one set of parameters and varying effects from their posterior distributions to generate synthetic outcome values. Based on the simulated dataset, we generate 100 repeated samples by bootstrapping the cell-wise observations so that different cells are included across repetitions. During each repetition, we draw $503$ cells from the 503 original cells across all 30 ZIP codes uniformly with replacement. The original sample cell size varies between 1 and 67 with a median cell count of 3. If a cell is included multiple times, we keep one copy of the cell and change the counts of the cell by multiplying the simulated number of positive tests and total number of tests by the number of times the index was drawn. Thus, the overall sample sizes are different across repeated samples. For instance, the spatial data set sample cell size varies from 1 to 268 observations whereas the independent data set sample cell size varies from 1 to 402 observations. Still, the median cell size is four for both. The average observed positivity rate is 0.094 for the independent data set and 0.032 for the spatial data set. The rates in both data sets are small, with more than half of the values being zero.
 
\subsubsection{Model estimation}
For Bayesian estimation, we fit the logistic regression model in \eqref{eq:logit_model} with the specified prior distributions using the R package, \texttt{cmdstanr}~\parencite{cmdstanr}. To understand the effect of the prior choice on our FAB intervals, we consider two prior choices for $\alpha^{\textrm{ZIP}}_{j}$ in the estimation: the GP and global-local normal priors. The global-local normal prior is only for estimation, but the data generation process uses either the GP or independent prior as described above. The global-local normal prior assumes $\alpha^{\textrm{ZIP}}_{j} \sim \textrm{N}(0, \sigma_j^2\tau^2)$, the global scale $\tau \sim \textrm{N}^+(0,3^2)$, and the local scale $ \sigma_j \sim \textrm{N}^+(0, 1)$, where $\textrm{N}^+()$ is a half-normal distribution restricted to positive values. We use the half-normal prior distribution for the scale parameters to improve MCMC convergence across repeated simulation studies, which provides regularization and can be replaced by half-Cauchy prior distributions if the scale informed by the data has large values~\parencite{gelman06-prior}. We run four MCMC chains with 2000 iterations and keep the last 1000 iterations each. Using the posterior draws of cell-wise incidence $\hat{\pi}_j$, we obtain the MRP estimates of ZIP code-level infection rates expressed as
\begin{align}
\label{mrp-ZIP}
\tilde{\theta}_i = \frac{\sum_{\textrm{cell }j \in \textrm{ZIP }i} N_j\hat{\pi}_j}{\sum_{\textrm{cell }j \in \textrm{ZIP }i} N_j},
\end{align}
which is an aggregated estimate of cell-wise incidence weighted by the population cell size $N_j$, among all cells belonging to the ZIP code $i$. In the MRP application, \textcite{mrp-covid21, si2024multilevel} obtain the population cell counts $N_j$ using the ACS for the catchment area covered by the ZIP codes, the linking of which is based on the U.S. Department of Housing and Urban Development crosswalk from ZIP codes to census tracts that involves approximation since the mapping is not one-to-one. Here, we construct the FAB intervals for the ZIP-level incidence rates in \eqref{mrp-ZIP} by using the poststratified sample mean estimate from all observed cells for a ZIP code as the center and the posterior mean and standard deviation of $\textrm{logit}(\tilde{\theta}_i)$ as the mean and standard deviation of the prior normal distribution mentioned in \eqref{bin} to obtain $(\varsigma, s_i)$. In other words, we use the posterior information to build the FAB intervals around the poststratified sample means. We vary the prior assigned to the ZIP-varying intercept, which affects the prior mean and prior standard deviation of $\text{logit}(\theta_i)$ used to build the FAB intervals, and examine the effect of prior choices. In our simulation, only one prior choice matches what is used to generate the data (GP prior for estimation and data creation).

\subsubsection{Outputs and comparison}

\begin{figure}[!t]
\begin{tabular}{cc} \includegraphics[width=0.49\textwidth]{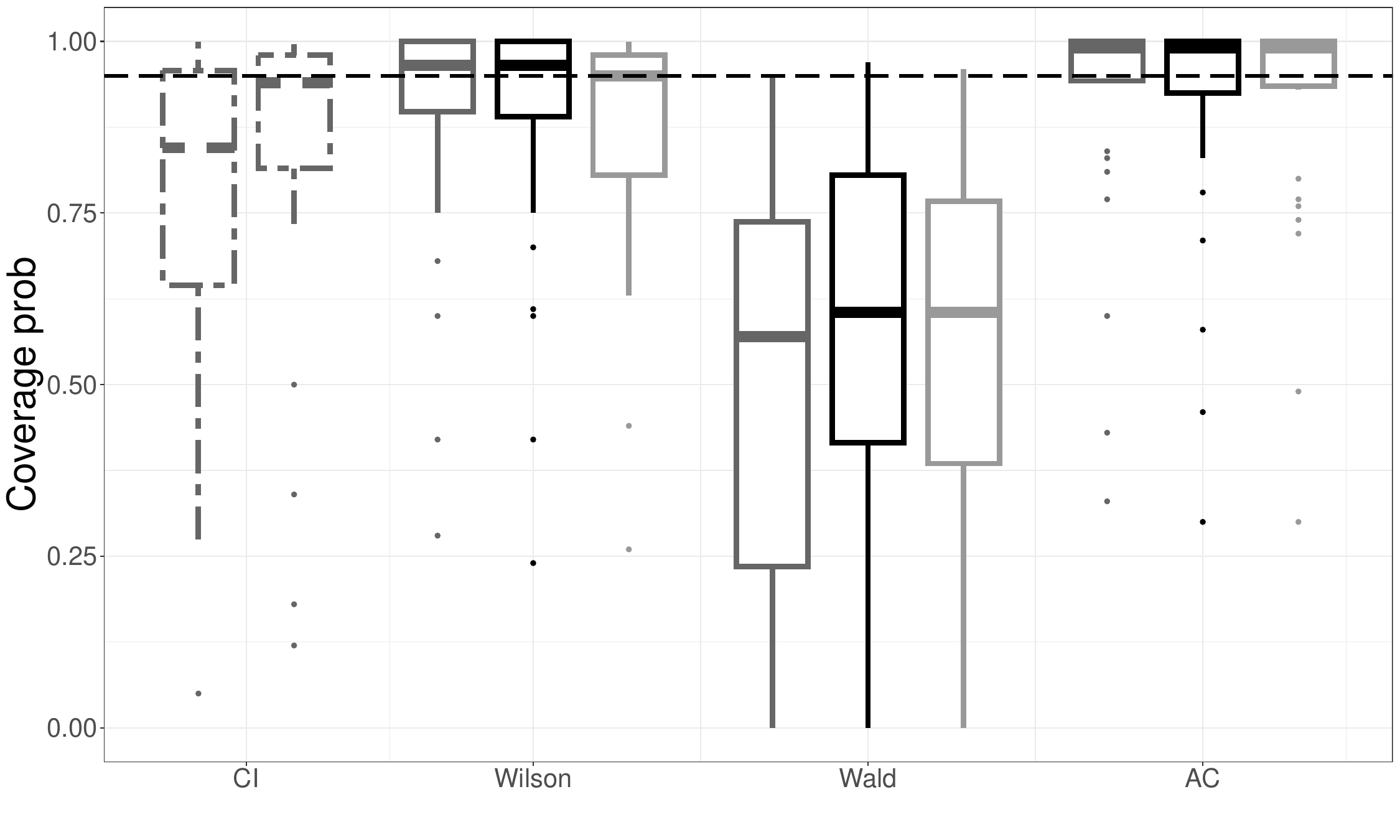}   &     \includegraphics[width=0.49\textwidth]{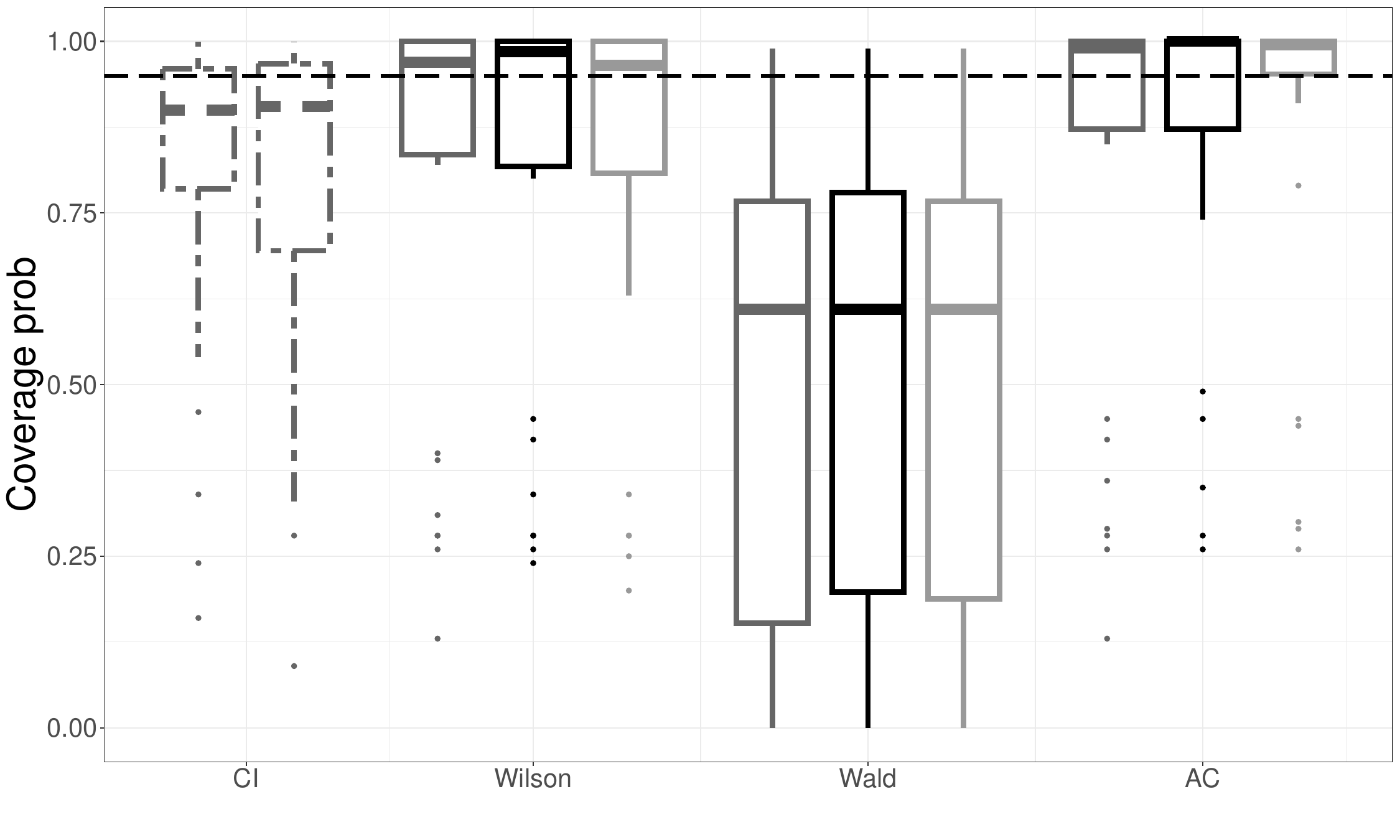}\\
\includegraphics[width=0.49\textwidth]{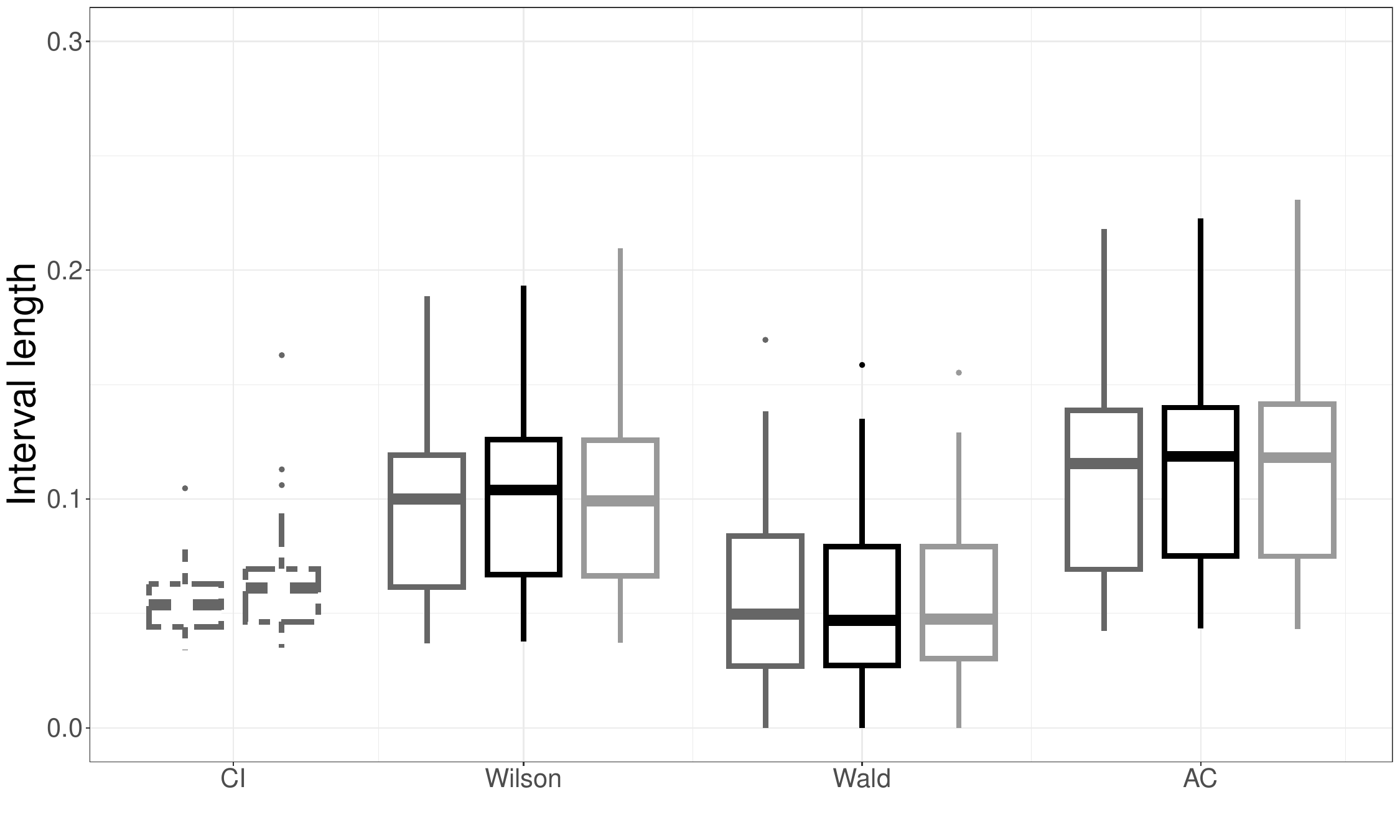} &   
\includegraphics[width=0.49\textwidth]{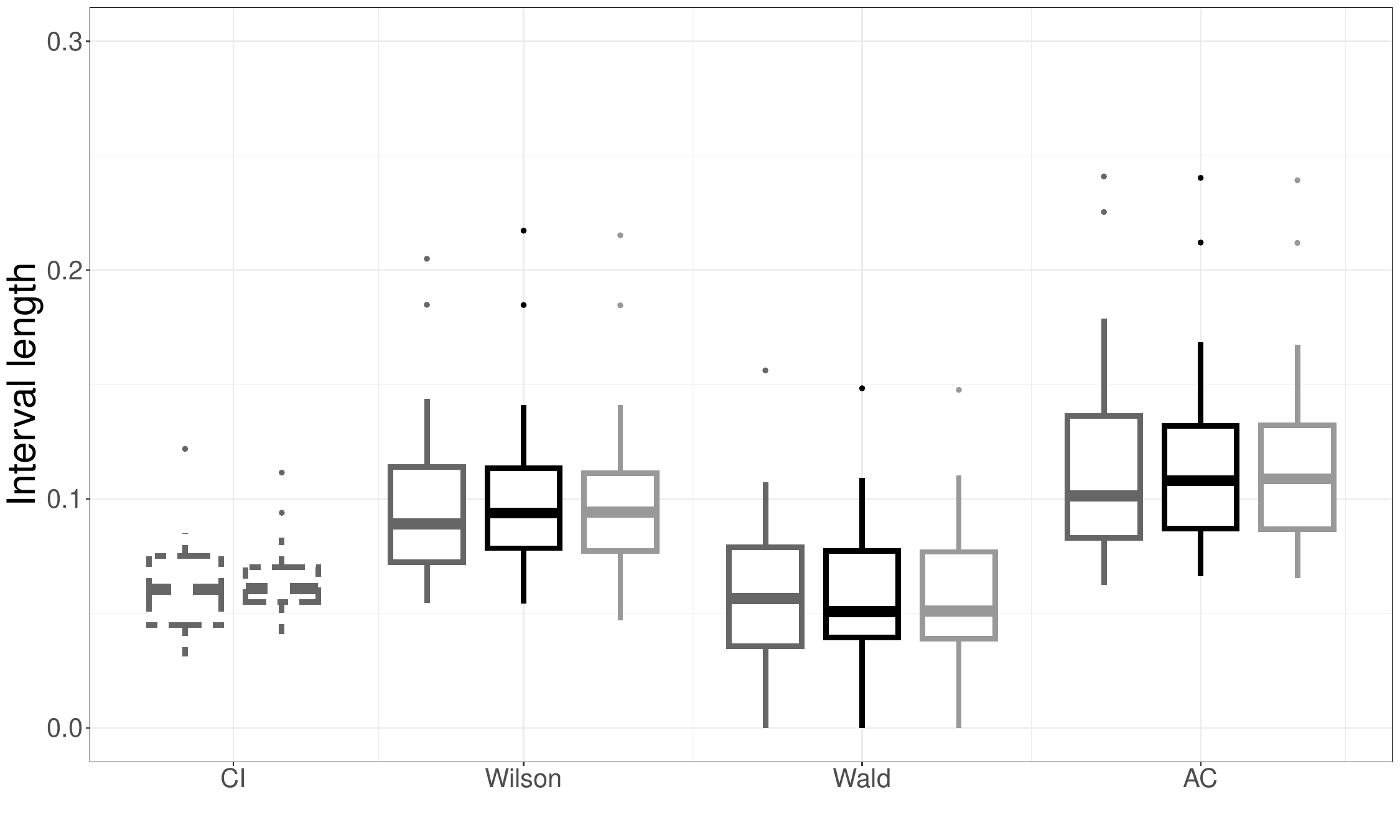}\\
(a) Simulated with independent prior & (b) Simulated with spatial prior
\end{tabular}
\caption{The coverage probabilities and average interval lengths of Bayesian credible intervals (CIs) and Frequentist assisted Bayes (FAB) intervals of the ZIP-level estimates with different prior specifications. The left panel is based on data simulated under the independent prior, and the right panel is for data simulated under the spatial prior. For the Wilson, Wald, and Agresti-Coull (AC) interval types, the FAB intervals based on the spatial prior are displayed in black on the left, whereas the FAB intervals based on the global-local normal prior are colored in light gray on the right. The credible intervals are colored in semi-dark gray and displayed with long-short dashed lines on the left of the plot, with the left interval corresponding to the spatial prior and the right one corresponding the global-local prior. Meanwhile, the interval as originally defined appears to the left of the corresponding FAB intervals colored in semi-dark gray, but with solid lines. The dashed line represents 95\% coverage probability.
}
\label{fig:sim_coverage_prior_comp_info_ZIP}
\end{figure}

\begin{table}[!th]
\centering
\resizebox{\textwidth}{!}{
\begin{tabular}{lcccc}
\toprule
& \multicolumn{2}{c}{\textbf{Coverage probability ratio}} 
& \multicolumn{2}{c}{\textbf{Interval length ratio}} \\
\cmidrule(lr){2-3} \cmidrule(lr){4-5}
\textbf{Interval type} 
& \textbf{Ind. scenario} 
& \textbf{GP scenario} 
& \textbf{Ind. scenario} 
& \textbf{GP scenario} \\
\midrule

\multicolumn{5}{l}{\textit{Credible interval}} \\
\quad Global-local prior 
& 1.21 (0.44, 3.60) 
& 0.97 (0.56, 1.42) 
& 1.18 (0.84, 2.68) 
& 1.10 (0.86, 1.97) \\

\addlinespace
\multicolumn{5}{l}{\textit{Wilson interval}} \\
\quad Original interval 
& 1.80 (0.54, 19.40) 
& 1.12 (0.28, 4.17) 
& 1.72 (0.96, 3.10) 
& 1.71 (0.94, 4.02) \\

\quad FAB interval (GP prior) 
& 1.77 (0.54, 19.00) 
& 1.13 (0.26, 4.17) 
& 1.76 (0.99, 3.18) 
& 1.75 (0.99, 4.15) \\

\quad FAB interval (global-local prior) 
& 1.74 (0.56, 18.80) 
& 1.13 (0.27, 4.13) 
& 1.74 (0.88, 3.45) 
& 1.73 (0.94, 4.24) \\

\addlinespace
\multicolumn{5}{l}{\textit{Wald interval}} \\
\quad Original interval 
& 1.00 (0.00, 10.00) 
& 0.62 (0.00, 1.24) 
& 1.01 (0.00, 2.79) 
& 1.02 (0.00, 2.83) \\

\quad FAB interval (GP prior) 
& 1.03 (0.00, 9.00) 
& 0.62 (0.00, 1.24) 
& 1.00 (0.00, 2.72) 
& 0.99 (0.00, 2.71) \\

\quad FAB interval (global-local prior) 
& 1.05 (0.00, 9.60) 
& 0.62 (0.00, 1.24) 
& 0.99 (0.00, 2.60) 
& 0.99 (0.00, 2.70) \\

\addlinespace
\multicolumn{5}{l}{\textit{Agresti--Coull interval}} \\
\quad Original interval 
& 1.85 (0.55, 19.80) 
& 1.15 (0.28, 4.17) 
& 1.99 (1.07, 3.88) 
& 1.98 (0.98, 4.69) \\

\quad FAB interval (GP prior) 
& 1.84 (0.59, 19.80) 
& 1.21 (0.28, 4.17) 
& 2.02 (1.16, 3.66) 
& 2.03 (1.19, 4.73) \\

\quad FAB interval (global-local prior) 
& 1.85 (0.63, 19.80) 
& 1.22 (0.28, 4.17) 
& 2.03 (1.17, 3.79) 
& 2.03 (1.18, 4.73) \\

\bottomrule
\end{tabular}
}
\caption{Mean coverage probability ratios and interval length ratios, with minimum and maximum values shown in parentheses, under the independent and GP data-generating scenarios. Ratios are computed relative to the Bayesian credible intervals from the model fit using the GP prior. The ZIP-code sample sizes have a mean of 94.5, with an average range 23.9--263.6, under the independent scenario and mean 94.6, with average range 23.9--255.8, under the GP scenario.}
\label{tab:interval-ratios}
\end{table}

We calculate the coverage probabilities and average lengths of the Bayesian 95\% credible intervals and three FAB intervals (Wald, Wilson, and AC) of the ZIP-level estimates. To find the ``true" proportion at the ZIP code-level, we use the MRP estimate in \eqref{mrp-ZIP} with the simulated $\pi_j$ for each cell and the corresponding poststratification cell counts.

Figure \ref{fig:sim_coverage_prior_comp_info_ZIP} displays the results of our experiment comparing prior specifications. When applied to data generated in the independent scenario, the Bayesian credible intervals for the global-local normal distribution outperform those based on the GP prior with higher coverage but slightly larger interval lengths. However, when data are simulated under the spatial prior, both prior specifications result in Bayesian credible intervals with undercoverage. On the other hand, the median coverage probabilities of the FAB Wilson and AC intervals are above the nominal rate. As the FAB AC intervals are wider, the coverage probabilities are higher. Meanwhile, the FAB Wald interval has lower coverage than the credible intervals and the other FAB intervals. This could be due to the narrowness of the FAB Wald interval, which is caused by the narrowness of the credible interval and the smaller $\varsigma$ based on the poststratified proportions. These proportions are small. Finally, according to the coverage probability, the FAB intervals perform slightly better on the data simulated under the spatial prior versus the independent prior.

\begin{figure}[!t]
\begin{tabular}{cccc} 
\centering
\includegraphics[width=0.23\textwidth, height=1.7in]{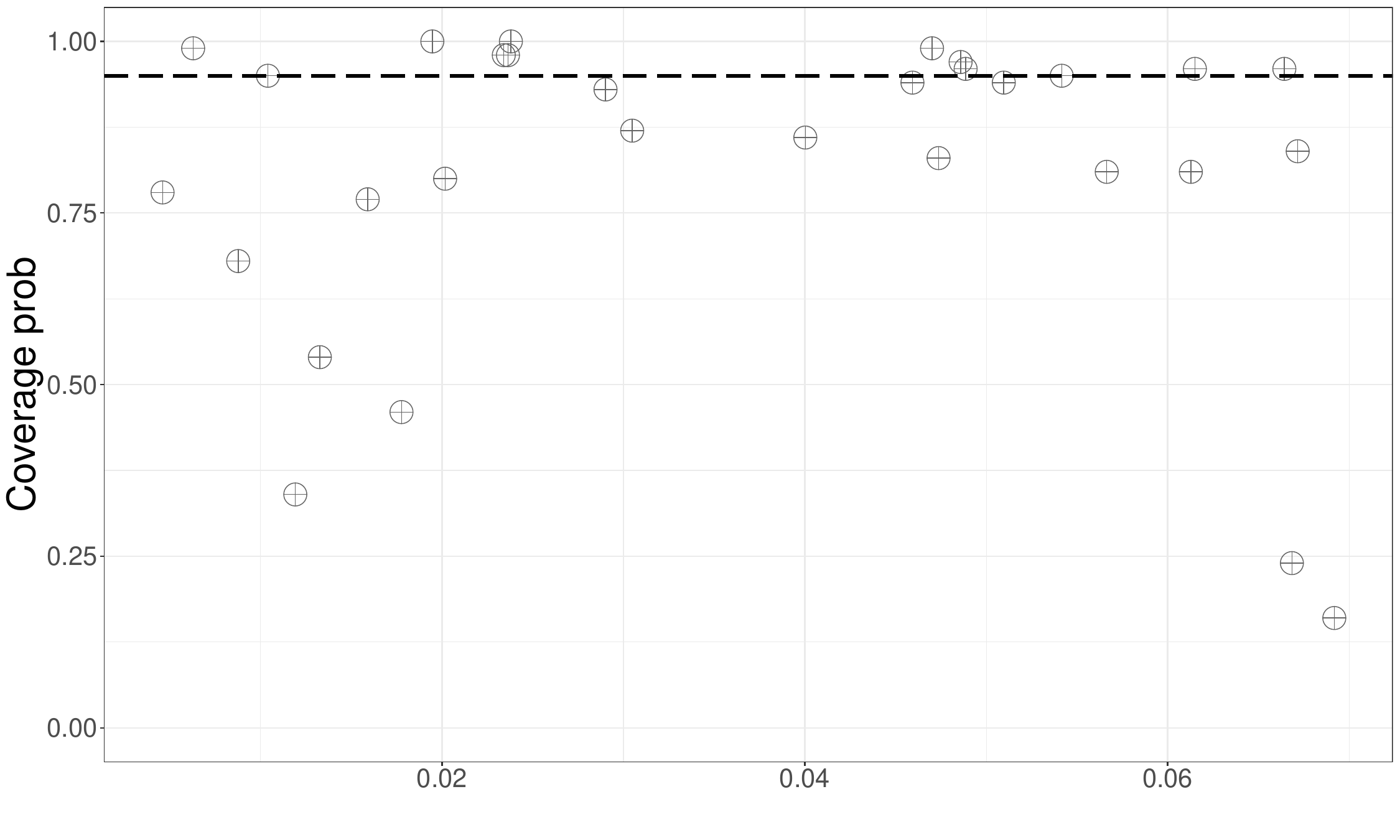}   &     
\includegraphics[width=0.23\textwidth, height=1.7in]{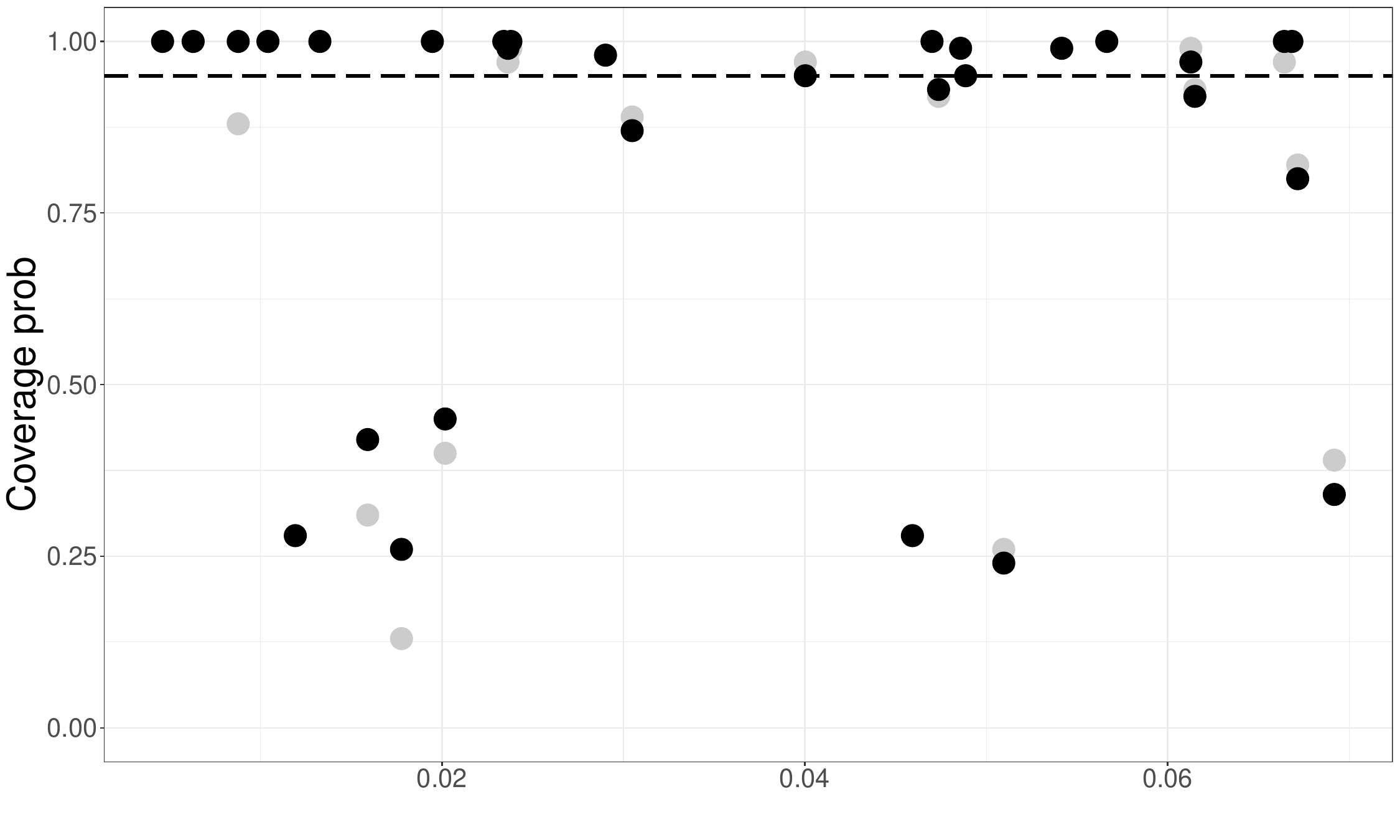}
&
\includegraphics[width=0.23\textwidth, height=1.7in]{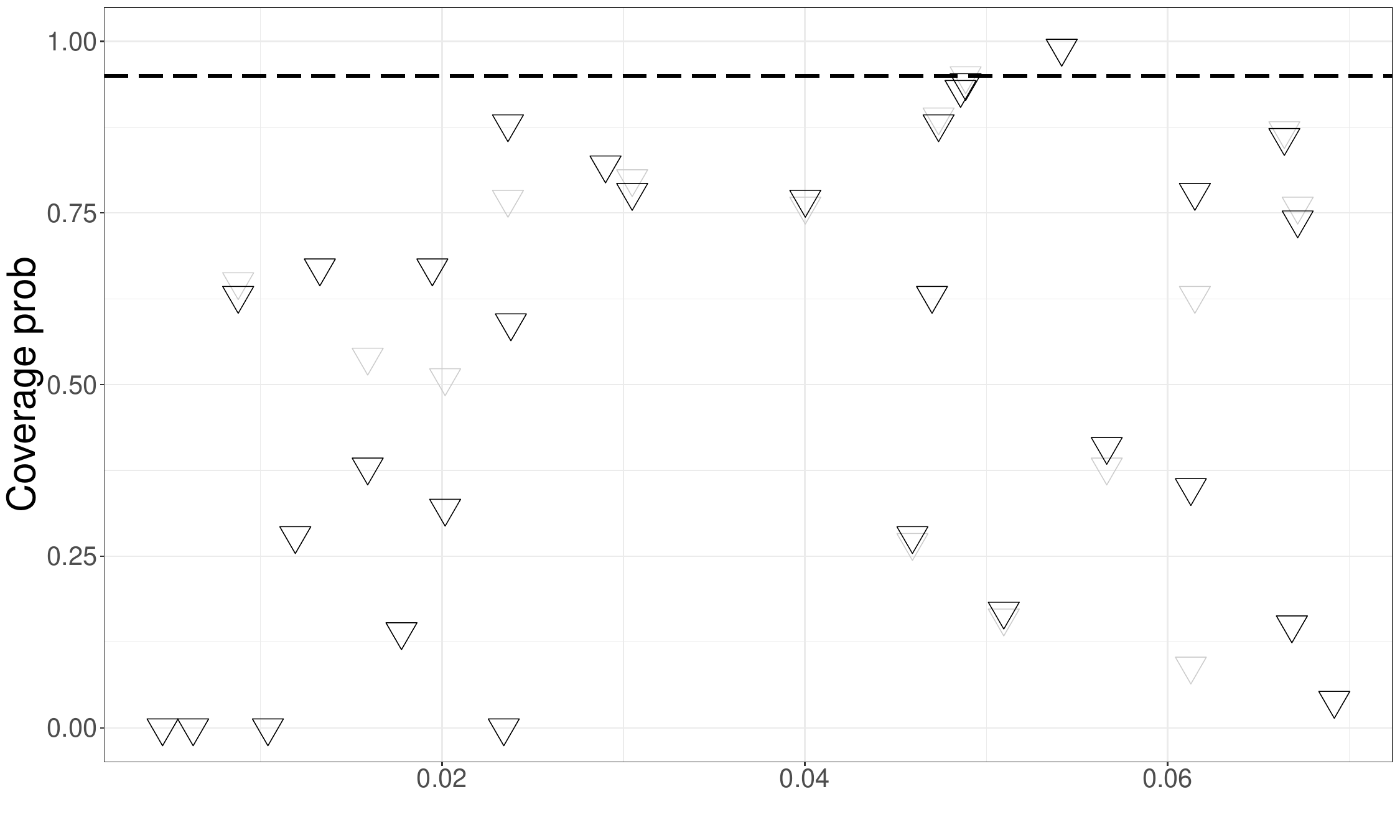}
&
\includegraphics[width=0.23\textwidth, height=1.7in]{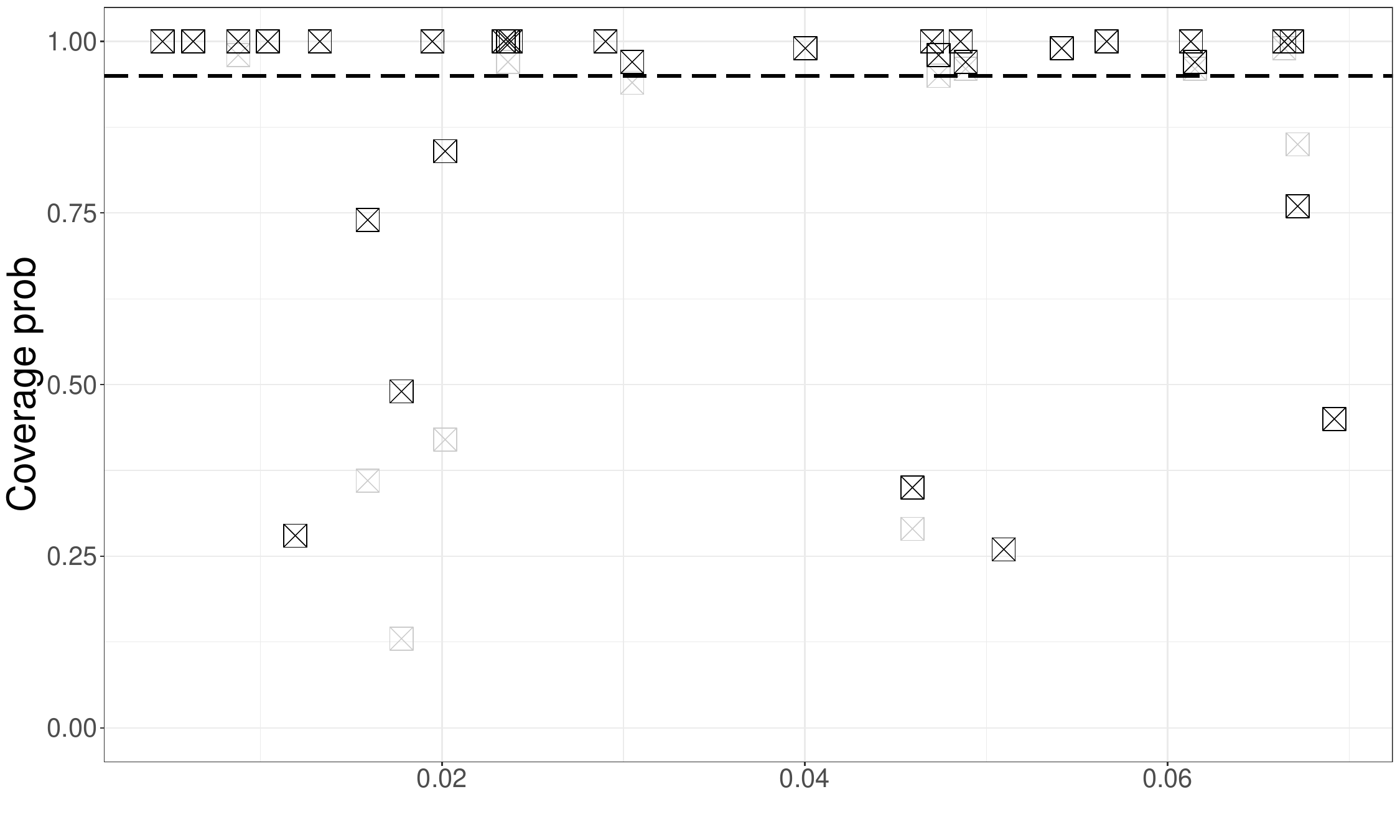}\\
\includegraphics[width=0.23\textwidth, height=1.7in]{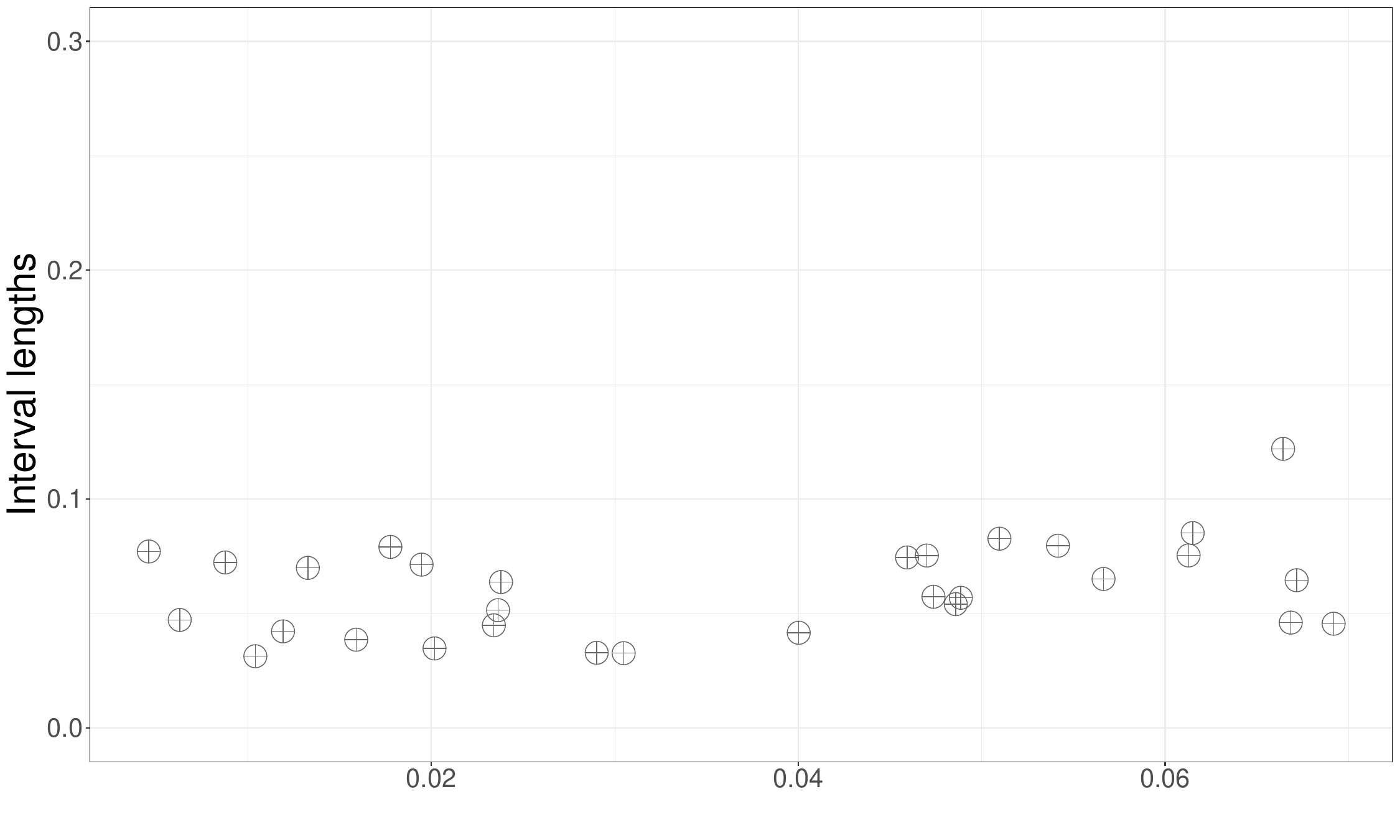} &     
\includegraphics[width=0.23\textwidth, height=1.7in]{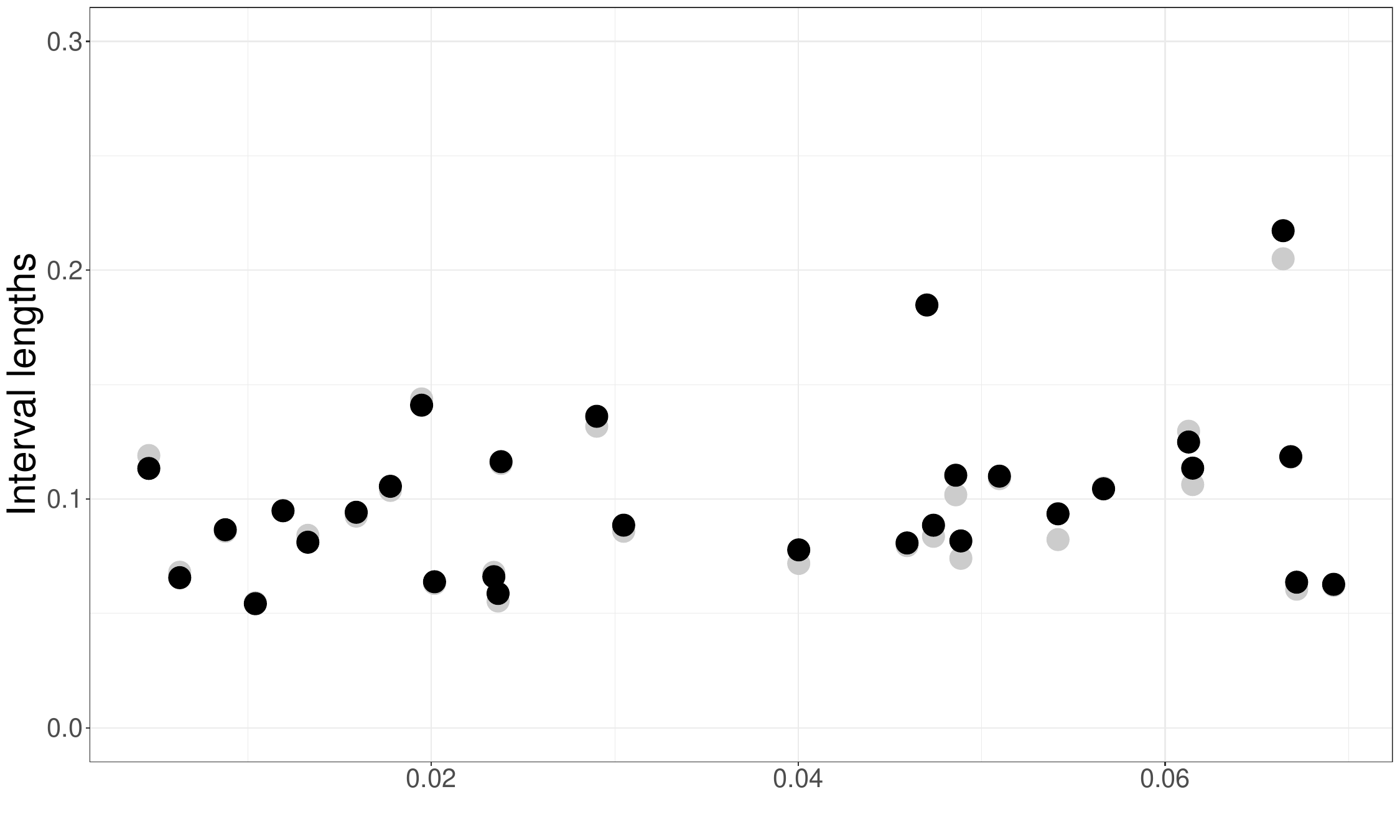} &
\includegraphics[width=0.23\textwidth, height=1.7in]{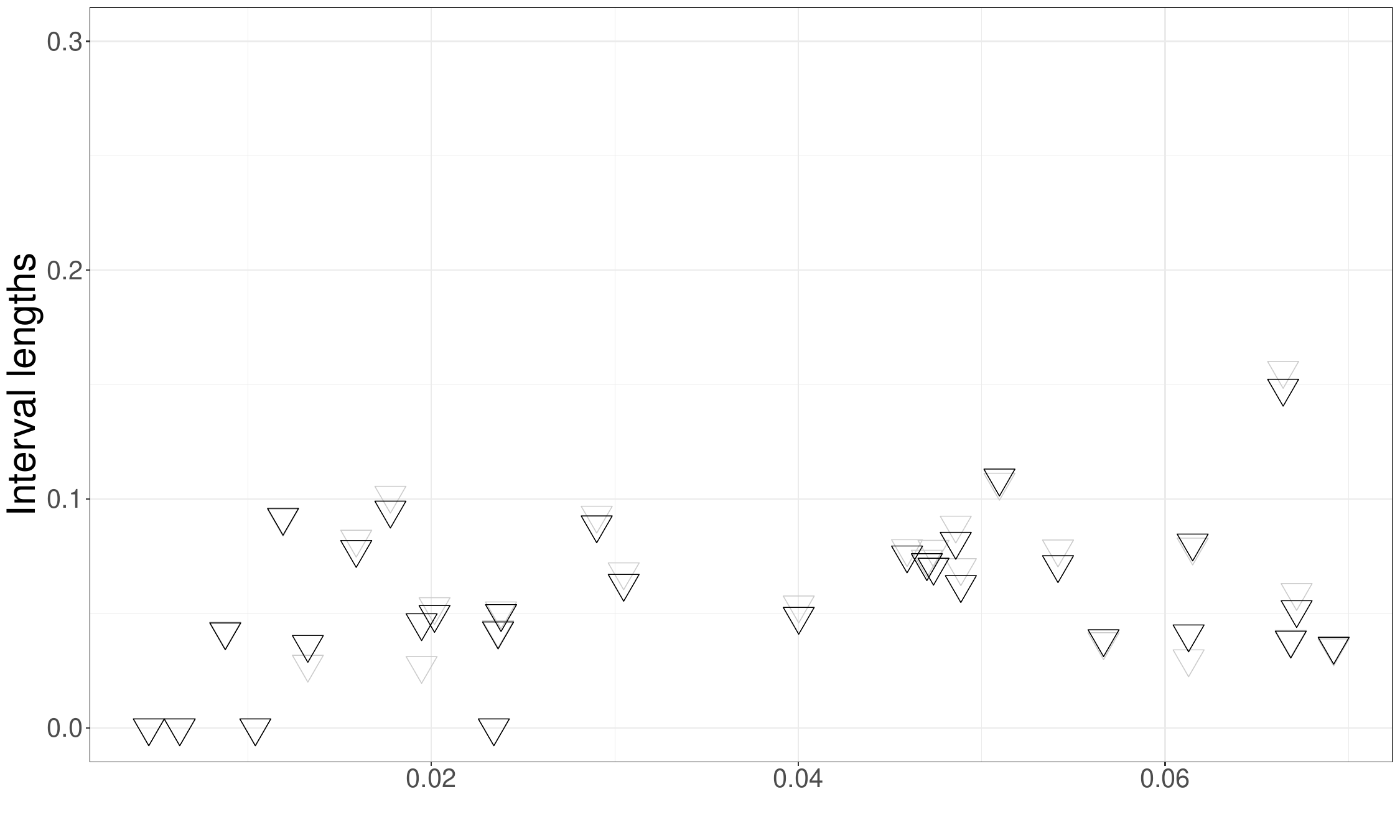} &
\includegraphics[width=0.23\textwidth, height=1.7in]{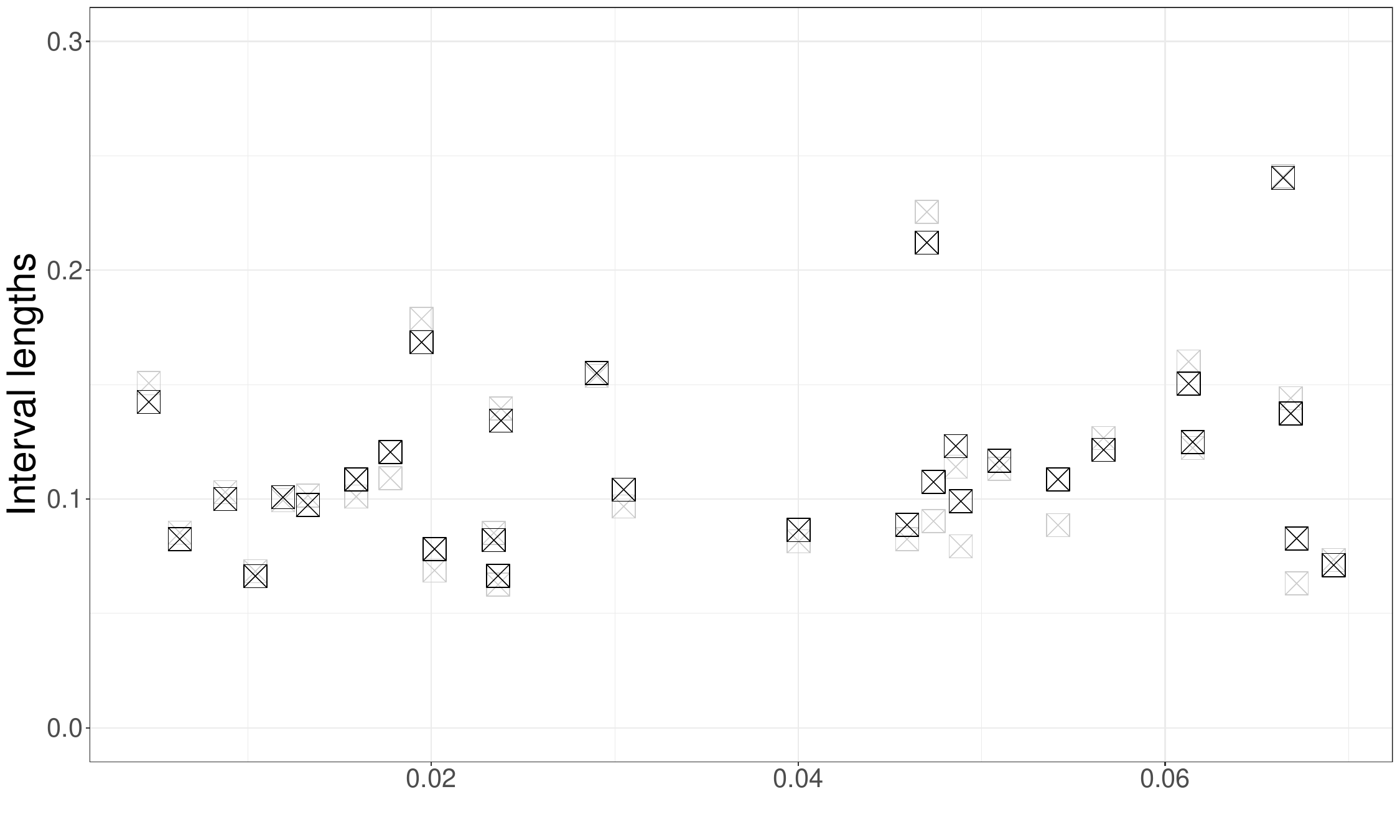} \\
(a) Credible intervals & (b) Wilson & (c) Wald & (d) Agresti-Coull\\
\end{tabular}
\caption{The coverage probabilities and average interval lengths of Bayesian credible intervals, the originally defined confidence intervals, and the Frequentist assisted Bayes (FAB) intervals of the ZIP-level estimates, with the x-axis representing the true ZIP-level proportions and data generated from the spatial prior scenario. The FAB intervals are displayed in black, whereas the standard intervals are colored in light gray.}
\label{fig:sim_coverage_prior_comp_info_prop}
\end{figure}

Next, we compare these FAB intervals against their original counterparts. For the originally defined AC, Wilson, and Wald intervals, we again use the poststratified sample mean estimate. Note that the intervals as originally defined use the total number of observations associated with the cells that belong to a ZIP code. We see that the coverage probabilities of the FAB intervals tend to be similar or higher than those of the originally defined intervals across both scenarios. The tradeoffs are that FAB intervals may be wider and have slightly higher variation in the coverage probabilities. However, as the ZIP code sample size increases, the FAB intervals become narrower. One exception is the Wald interval. Across all these cases, we see that utilizing the right prior can improve the coverage probability. This is confirmed by the mean, maximum, and minimum ratios of the coverage probability and interval length displayed in Table \ref{tab:interval-ratios}. While the minimum ratio is below 1, the average coverage probabilities and interval lengths for the FAB Wilson and AC intervals are greater than 1.

We investigate this further with the results using data generated from the spatial prior case. Here, we use the Bayesian credible interval from the GP prior as a point of comparison in order to match the prior. Figure \ref{fig:sim_coverage_prior_comp_info_prop} displays these results by ordering the information according to the poststratified true cell proportions. We notice that the FAB Wilson and AC intervals are wider than the FAB Wald intervals, because Wald intervals become degenerate when $\widehat{\theta}_i = 0$. Meanwhile, the FAB AC intervals appear to be the widest, achieve the highest coverage probabilities, and outperform the standard intervals. On the other hand, the FAB Wald intervals are inferior. However, we see that while the FAB interval improves the coverage probability compared to the original interval, the FAB interval still can struggle if the original posterior estimates are not close to the true proportions at the ZIP-level. For instance, there are two ZIP codes whose credible intervals have consistently lower coverage probabilities. For these two ZIP codes, the FAB intervals have similar coverage probabilities as the original intervals, but still fail to achieve the nominal coverage rate. When we examine these ZIP codes, we find that the poststratified sample mean at the ZIP code level, $\widehat{\theta}_i$, is above the third quantile in the range of observed proportions from the data.
This helps explain the failure because the FAB intervals contain $\theta$ such that $\widehat{\theta}_i \in \mathcal{I}^{F}(\theta, s_i, \varsigma)$. Since the sample sizes are around 162.56 and 134.33 and $\varsigma$ is based on the sample size, the $\mathcal{I}^{F}(\theta, s_i, \varsigma)$ that includes $\widehat{\theta}_i$ is not wide enough to incorporate the poststratified proportions.


Overall, we recommend the FAB Wilson interval with the all-in penalty when the target proportions are away from the center. When the proportions are closer to the center, the FAB Wilson interval without the penalty is a reasonable option. The FAB Wald intervals tend to undercover. The FAB AC interval attains higher coverage probabilities but is often the widest. If one prefers a shorter interval at the cost of undercoverage while maintaining a median coverage probability of 95\%, the FAB Wilson interval with the all-in penalty offers a practical compromise for proportions away from the center; near the center, the FAB Wilson interval without penalty is preferable.

\section{Application to the COVID-19 transmission estimation}
\label{sec:real_data_results}
We apply the calibrated Bayes inference methods to the routine hospital COVID-19 testing records from a Midwest hospital system from May 11th, 2020 to June 13th, 2022. We fit a logistic regression to estimate the incidence in cell $j$, based on the cross-tabulation of sex, age, race and ZIP code, as the following:
\begin{align}
\label{eq:logit_model_real}
y_j  \sim \textrm{Binomial}(\cdot \mid \pi_j, n_j),\,\,\, 
\pi_j = \textrm{inv\_logit}\left(\alpha_0 + \v{X}_j \beta + \alpha^{\textrm{age}}_{j} + \alpha^{\textrm{race}}_{j} + \alpha^{\textrm{ZIP}}_{j}\right),      
\end{align}
for $j=1, \dots, J$. There are 399 ZIP codes with test results, and, following ~\textcite{si2024multilevel}, we include ZIP-level predictors that could affect viral transmission, such as measures on the urbanicity, percentage of college-educated individuals, employment rate, poverty rate, average income, and area deprivation index. This information is encoded in $\v{X}$, in addition to sex. For the regression coefficient $\beta$, we apply $\textrm{N}(0,3^2)$ as the prior. For the varying effects of age and race, the prior is $\alpha^{\textrm{var}}_{j} \sim \textrm{N}(0, {\sigma^{\textrm{var}}}^2)$ and $\sigma^{\textrm{var}} \sim \textrm{N}^+(0, 3^2)$, a half-normal distribution restricted to positive values with mean 0 and variance 9, where var $\in$ \{age, race\}. Finally, we use a GP prior for $\alpha^{\textrm{ZIP}}_{j}$ based on the Haversine distance between the longitude and latitude of the ZIP code's centroids. We assign a half-Cauchy distribution with mean 0 and standard deviation 1 as the prior for the variance in the GP prior. In terms of the bandwidth $\ell$, we compared model fitting based on three different values, $(5, 7.5, 10)$, using the Leave One Out Cross-Validation in the R package \texttt{loo}~\parencite{loo-R} and chose $\ell=7.5$ for the inference.

We present county-level estimates here, extending from the ZIP-level estimates in the simulation study to demonstrate the performance of calibrated intervals for domains with different ranges of sample sizes. We use a U.S. Department of Housing and Urban Development crosswalk file to obtain the list of ZIP codes in the catchment areas of each county. The MRP estimates of the county-level infection incidence are expressed as the following. 

\begin{align}
\label{mrp-county}
\tilde{\theta}_c = \frac{\sum_{\textrm{cell }j \in \textrm{county }c} N_j\hat{\pi}_j}{\sum_{\textrm{cell }j \in \textrm{county }c} N_j}.
\end{align}
Here, $\hat{\pi}_j$ is the cell-wise estimate for a particular combination of sex, age, race and ZIP code, and $N_j$ is the population cell count obtained from the ACS for cell $j$. 

We fit the model with the R package \texttt{cmdstanr}, running four MCMC chains with 2000 iterations and keeping the last 1000 iterations from each chain.  We assess the convergence based on the Gelman-Rubin diagnostics, $\widehat{R}$, most of which are around 1.

\begin{figure}[!th]
\centering
\begin{subfigure}{0.48\textwidth}
    \centering
    \caption{Wilson intervals}   
    \includegraphics[width=0.9\textwidth]{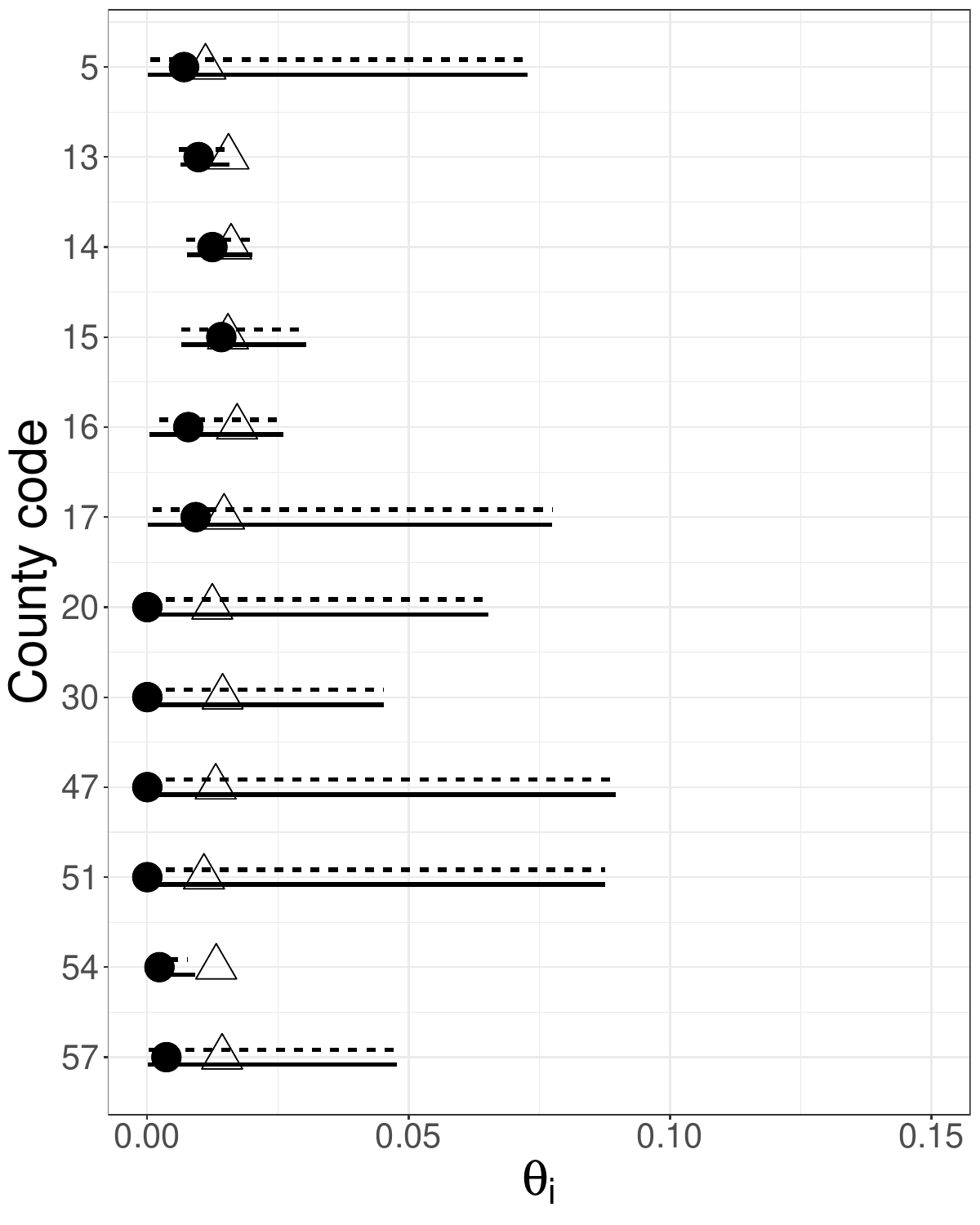} 
\end{subfigure}
\begin{subfigure}{0.48\textwidth}
    \centering  
    \caption{Wald intervals}
    \includegraphics[width=0.9\textwidth]{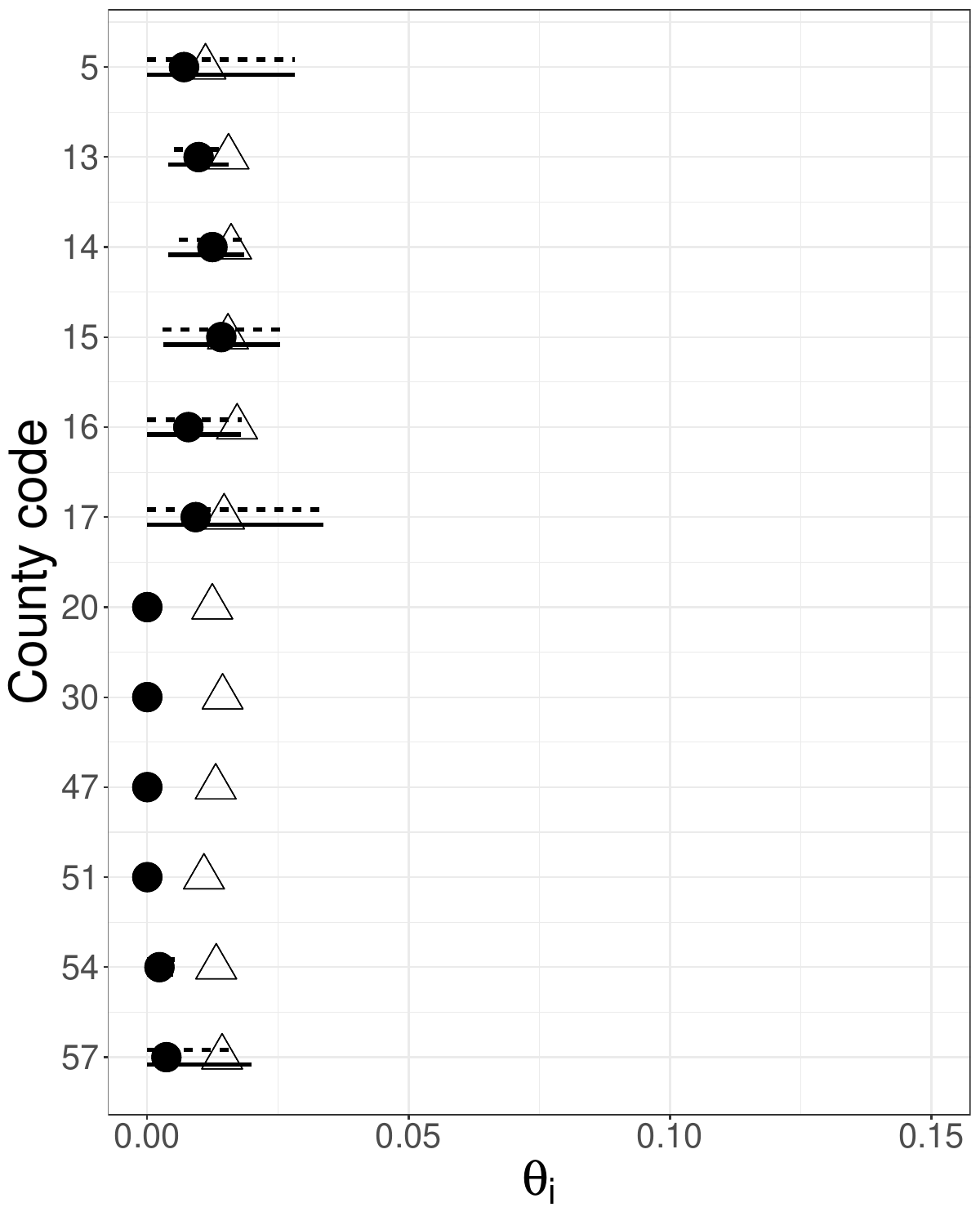}  
\end{subfigure}\\
\begin{subfigure}{0.48\textwidth}
    \centering
    \caption{Agresti-Coull intervals} 
    \includegraphics[width=0.9\textwidth]{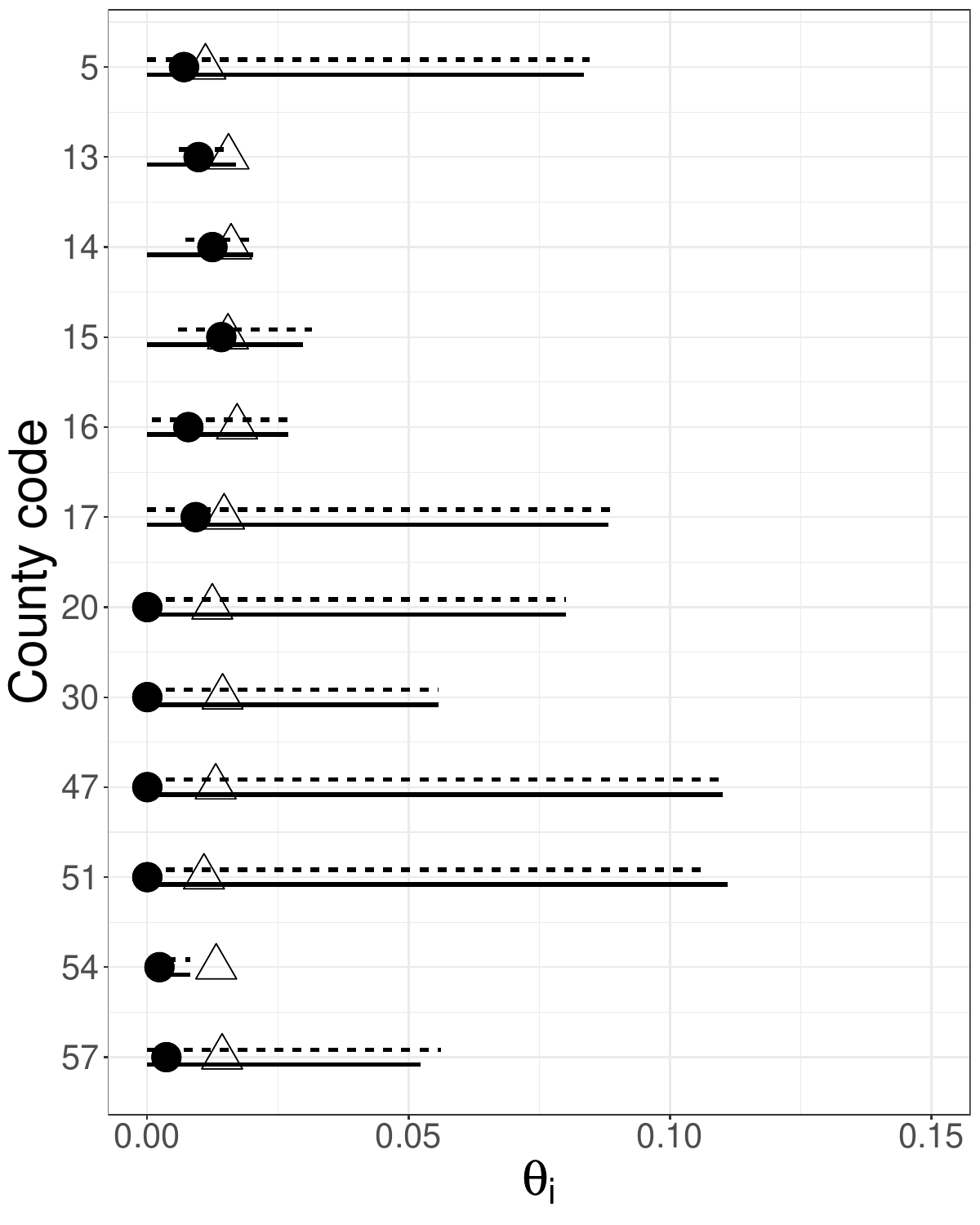}
\end{subfigure}
\begin{subfigure}{0.48\textwidth}
    \centering
    \caption{Credible intervals}
    \includegraphics[width=0.9\textwidth]{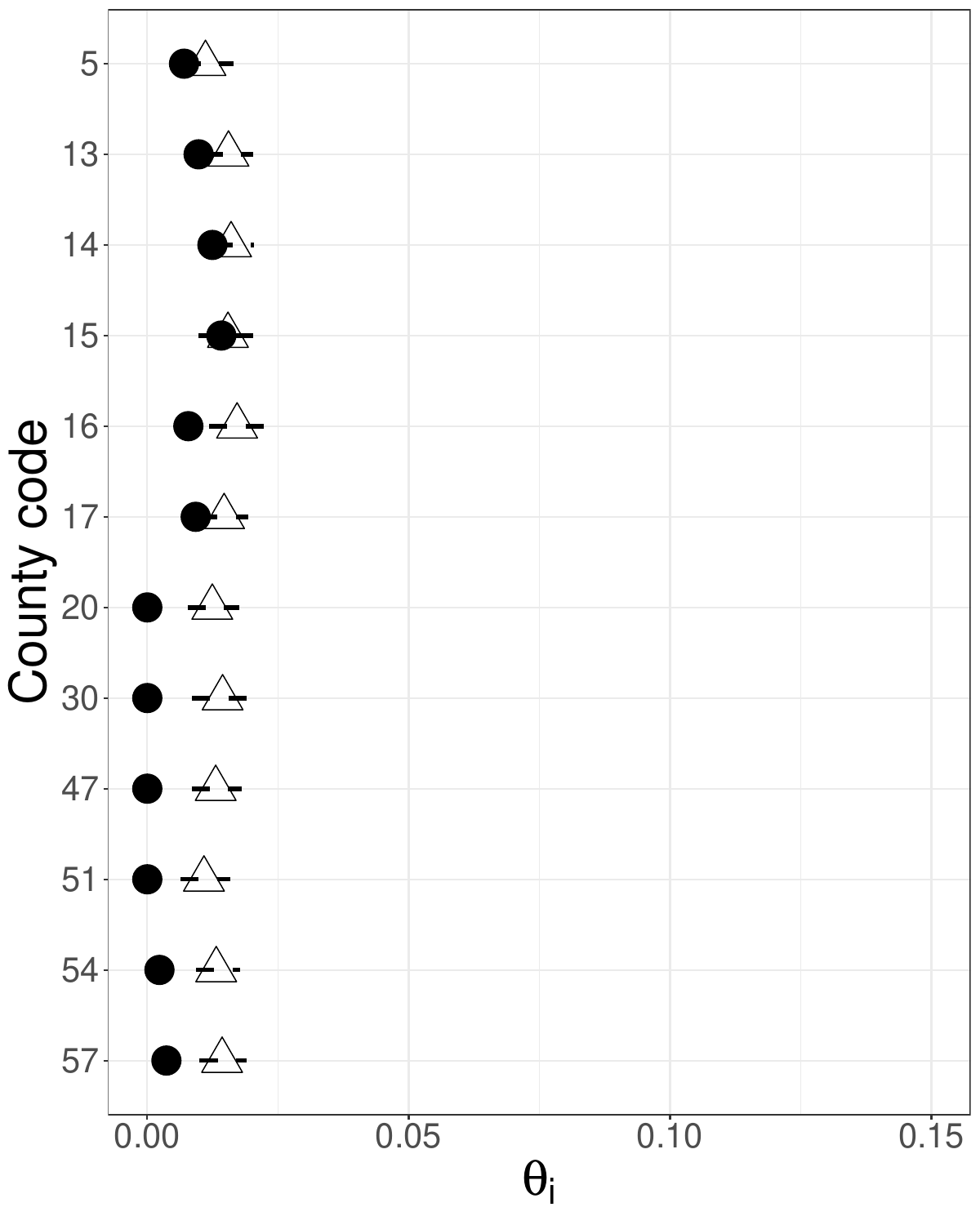}
\end{subfigure}
\caption{County-level incidence inference with the post-stratified observed positivity rates shown as circles and the posterior mean estimates under multilevel regression and poststratification shown as open triangles. For the Wilson, Wald, and Agresti-Coull intervals, the dotted lines represent the originally defined intervals and the solid lines represent the Frequentist assisted Bayes (FAB) intervals.}
\label{fig:real_data_results_forest_plot_county_two}
\end{figure}

 \FloatBarrier

\begin{table}[!th]
    \centering
    \begin{minipage}{.45\linewidth}
    \centering
    \resizebox{0.94\textwidth}{!}{
    \begin{tabular}{c c c c c}
    \\[-1.8ex]\hline 
    \multirow{2}{*}{\textbf{County}} & 
    \multirow{2}{*}{\shortstack{\textbf{Sample}\\ \textbf{Size}}} &
    \multirow{2}{*}{\shortstack{\textbf{Observed}\\ \textbf{Proportions}}} & \multirow{2}{*}{\shortstack{\textbf{Posterior}\\ \textbf{Means}}} & \multirow{2}{*}{\textbf{Posterior CI}}\\
    & \\
    \hline \\[-1.8ex] 
    5 & 60 & 0.007 & 0.011 & (0.007, 0.017)\\
    &\\
    13 & 1644 & 0.010 & 0.016 & (0.011, 0.020)\\
     &\\
    14 & 1121 & 0.012 & 0.016 & (0.013, 0.020)\\
     &\\
    15 & 426 & 0.014 & 0.015 & (0.010, 0.022)\\
     &\\
    16 & 285 & 0.008 & 0.017 & (0.012, 0.024)\\
     &\\
    17 & 59 & 0.009 & 0.015 & (0.010, 0.019)\\
     &\\
     20 & 55 & 0.000 & 0.012 & (0.008, 0.018)\\
     &\\
     30 & 81 & 0.000 & 0.014 & (0.009, 0.021)\\
     &\\
     47 & 39 & 0.000 & 0.013 & (0.008, 0.018)\\
     &\\
     51 & 40 & 0.000 & 0.011 & (0.006, 0.016)\\
     &\\
     54 & 1027 & 0.002 & 0.013 & (0.009, 0.018) \\
     &\\
     57 & 90 & 0.004 & 0.014 & (0.010, 0.019)\\
     &\\
     \hline \\[-1.8ex]
    \end{tabular}
    }
    \end{minipage}
    \quad
    \begin{minipage}{.45\linewidth}
    \centering
    \resizebox{.68\textwidth}{!}{
    \begin{tabular}{c c c}
    \\[-1.8ex]\hline 
    \multirow{2}{*}{\textbf{County}} & 
    \multirow{2}{*}{\textbf{Wilson CI}}
    & \multirow{2}{*}{\textbf{FAB Wilson CI}}\\
    & \\
    \hline \\[-1.8ex] 
    5 & (0.001, 0.073) & (0.000, 0.073) \\
 & \\
 13 & (0.006, 0.016) & (0.006, 0.016) \\
 & \\
 14 & (0.007, 0.021) & (0.008, 0.020) \\
 & \\
 15 & (0.007, 0.030) & (0.007, 0.030) \\
 & \\
 16 & (0.002, 0.026) & (0.000, 0.026) \\
 & \\
 17 & (0.001, 0.077) & (0.000, 0.077) \\
 & \\
 20 & (0.000, 0.065) & (0.000, 0.065) \\
 & \\
 30 & (0.000, 0.045) & (0.000, 0.045) \\
 & \\
 47 & (0.000, 0.090) & (0.000, 0.090) \\
 & \\
 51 & (0.000, 0.088) & (0.000, 0.088) \\
 & \\
 54 & (0.001, 0.008) & (0.000, 0.009) \\
 & \\
 57 & (0.000, 0.048) & (0.000, 0.048) \\
 & \\
     \hline \\[-1.8ex]
    \end{tabular}
    }
    \end{minipage}\\
    \begin{minipage}{.45\linewidth}
    \centering
    \resizebox{.68\textwidth}{!}{
    \begin{tabular}{c c c}
    \\[-1.8ex]\hline 
    \multirow{2}{*}{\textbf{County}} & 
    \multirow{2}{*}{\textbf{Wald CI}}
    & \multirow{2}{*}{\textbf{FAB Wald CI}}\\
    & \\
    \hline \\[-1.8ex] 
     5 & (0.000, 0.028) & (0.000, 0.028) \\
     & \\
     13 & (0.005, 0.015) & (0.004, 0.016) \\
     & \\
     14 & (0.006, 0.019) & (0.004, 0.018) \\
     & \\
     15 & (0.003, 0.025) & (0.003, 0.025) \\
     & \\
     16 & (0.000, 0.018) & (0.000, 0.018) \\
     & \\
     17 & (0.000, 0.034) & (0.000, 0.034) \\
     & \\
     20 & (0.000, 0.000) & (0.000, 0.000) \\
     & \\
     30 & (0.000, 0.000) & (0.000, 0.000) \\
     & \\
     47 & (0.000, 0.000) & (0.000, 0.000) \\
     & \\
     51 & (0.000, 0.000) & (0.000, 0.000) \\
     & \\
     54 & (0.000, 0.005) & (0.000, 0.005) \\
     & \\
     57 & (0.000, 0.016) & (0.000, 0.020) \\
     & \\
     \hline \\[-1.8ex]
    \end{tabular}
    }
    \end{minipage}
    \begin{minipage}{.45\linewidth}
    \centering
    \resizebox{.68\textwidth}{!}{
    \begin{tabular}{c c c}
    \\[-1.8ex]\hline 
    \multirow{2}{*}{\textbf{County}} & 
    \multirow{2}{*}{\textbf{AC CI}}
    & \multirow{2}{*}{\textbf{FAB AC CI}}\\
    & \\
    \hline \\[-1.8ex] 
     5 & (0.000, 0.085) & (0.000, 0.083) \\
     & \\
     13 & (0.006, 0.016) & (0.000, 0.017) \\
     & \\
     14 & (0.007, 0.021) & (0.000, 0.020) \\
     & \\
     15 & (0.006, 0.031) & (0.000, 0.030) \\
     & \\
     16 & (0.001, 0.029) & (0.000, 0.027) \\
     & \\
     17 & (0.000, 0.089) & (0.000, 0.088) \\
     & \\
     20 & (0.000, 0.080) & (0.000, 0.080) \\
     & \\
     30 & (0.000, 0.056) & (0.000, 0.056) \\
     & \\
     47 & (0.000, 0.109) & (0.000, 0.110) \\
     & \\
     51 & (0.000, 0.107) & (0.000, 0.111) \\
     & \\
     54 & (0.000, 0.008) & (0.000, 0.008) \\
     & \\
     57 & (0.000, 0.056) & (0.000, 0.052) \\
     & \\
     \hline \\[-1.8ex]
    \end{tabular}
    }
    \end{minipage}\\
    \caption{Comparison of different intervals for county-level incidence estimates.}
    \label{table:county_vals}
\end{table} 

 \FloatBarrier

While the 399 ZIP codes cover 76 counties, we focus on incidence estimates of 12 counties in which the number of tests is at least 20 and fewer than 2000. Further, the observed proportions are very close to zero, allowing us to examine the performance examination of FAB intervals with a range of sample sizes and observed proportions near zero. Figure \ref{fig:real_data_results_forest_plot_county_two} displays the forest plots comparing the various 95\% confidence intervals. Similar to Section~\ref{sim-2}, we use the posterior mean and the posterior standard deviation of the logit transformed MRP estimates as the mean and standard deviation of the normal (prior) distribution used in the FAB interval construction. Meanwhile, the estimate used as the center of the FAB interval is a poststratified sample mean, aggregated from the cell level to the county level. 

Shown in Figure \ref{fig:real_data_results_forest_plot_county_two}, the posterior mean estimates (ranging from 0.011 to 0.016) are different from the observed positivity rates (between 0 and 0.031). The hierarchical model shrinks the cell-wise and thus county-level estimates, especially for counties with small sample sizes. Counties with sample sizes larger than 1500 have narrower FAB intervals or original confidence intervals than those with sample sizes smaller than 20. 

Table~\ref{table:county_vals} gives the detailed values. The Bayesian credible intervals are narrower than the FAB intervals. Our FAB procedure does not help the Wald interval overcome its degeneracy at zero or one, where Wald intervals for counties with an estimated incidence near zero are narrow. As a result, the FAB Wald and Wald intervals are very similar in length across our counties of interest. However, the FAB Wald interval can be wider, likely because of the behavior of $s_i$ discussed previously that favors intervals pushed to the boundary. Consequently, the FAB Wald interval will include larger $\theta$ that the FAB Wilson or AC interval may not include because the interval used to determine whether to include $\theta$ in the FAB Wald interval can still contain the county sample means for these larger values of $\theta$.

The FAB Wilson intervals provide results similar to, or slightly longer than, the Wilson interval. However, the FAB Wilson intervals generally decrease in length as sample size increases. Outside of county 54, the exception to this trend is that when the FAB Wilson intervals are wider, they extend to zero due to the all-in penalty. Finally, the AC intervals are the widest among all intervals, which is consistent with the results above. Generally, the FAB AC intervals become shorter on the right-hand side as the sample size increases. For intervals in which the proportion is zero and the posterior means are close to zero, the lengths are equivalent when the number of tests in a county is 81 and the FAB AC interval is shorter when the number is at least 90. However, the FAB intervals tend to be longer on the left-hand side compared to the AC intervals as originally defined. This characteristic might be useful because the original AC interval might pull the interval too much toward the middle by adding two to $y_i$ and four to $n_i$. The FAB interval appears to address this by extending toward the ends. 

Meanwhile, the shrinkage on the right-hand side reflects the fact that because the risk peaks for small $y_i$, the risk is similar for various $s_i(\theta)$. As a result, our tie-breaking procedure defaults to picking the shortest determination interval. Due to our assumption of normality for that interval, $s_i(\theta)$ is chosen to be 0.5, reverting our interval back to the default interval. However, as sample size increases, there are more $y_i$ to distinguish the different choices of $s_i(\theta)$ via the risk. For instance, suppose there are two candidate risk intervals: $[0.15, 0.25]$ and $[0.14, 0.24]$. For $n_i = 100$, only $y' = 2$ is captured in both intervals. However, for $n_i = 200$, $y' = 3, 4, 5$ are in the first interval, whereas only $y' = 3, 4$ are in the second interval. As a result, this leads to smaller FAB intervals as the sample size increases. Overall, the posterior credible intervals are generally much narrower than the Wilson, FAB Wilson, AC, and FAB AC intervals, especially for counties with small sample sizes or observed proportions near zero. The Wilson and FAB Wilson intervals have very similar lengths across counties. The AC and FAB AC intervals are typically the widest. The Wald and FAB Wald intervals are often shorter, but this is misleading when observed proportions are zero. Therefore, the FAB Wilson interval provides the best practical balance: it avoids Wald degeneracy, is more conservative than AC, and remains close in length to the original Wilson interval.

\section{Discussion}
\label{conclusion}

In this paper, we discuss how to calibrate Bayesian domain inference and compute the Frequentist, Assisted by Bayes intervals for a proportion. This work extends the approach of \textcite{BurrisHoffExactAdaptiveConfidence2020} from a normally distributed variable to a binary outcome. These intervals have to be numerically calculated because there is no closed-form solution when proportions are constrained to lie in $[0,1]$ and the prior for the logit transformation of the probability of a successful event is normal. Our simulation studies show that these FAB intervals of the proportion can achieve better coverage rates than the Bayesian credible intervals and the originally defined confidence intervals. The improvement is most evident for the FAB Wilson and FAB AC intervals. The FAB Wald interval is less reliable, especially when the observed proportions are close to 0 or 1, because the Wald interval can become too narrow or even degenerate at the boundaries. Our simulation studies also show that the prior choice mainly affects the Bayesian credible intervals, and FAB Wilson and FAB AC intervals are comparatively robust to the prior specification. However, the prior still matters because it affects the risk calculation and the amount of borrowing across domains. A well-specified linking model, such as a model that includes relevant domain-level covariates, can reduce interval length. Thus, the FAB framework can improve calibration, but careful modeling remains important for efficiency.

However, one disadvantage is that the intervals used to compute the Bayes risk are too wide. When these intervals extend beyond $[0,1]$, parts of the interval outside the parameter space do not contribute to the original risk calculation. Without correction, this can favor intervals that spill beyond the boundary, which in turn can produce unintuitive FAB intervals and may still lead to undercoverage in some settings. Our paper addresses this problem by introducing the all-in penalty, which discourages risk intervals from extending outside $[0,1]$ and therefore produces intervals that are more stable near the boundaries. In our simulations, the all-in penalty improves the performance of the FAB Wilson and FAB AC intervals, particularly when the true proportions are far from 0.5.

In practice, users should first examine the distribution of domain sample sizes and observed proportions. In our simulations and application, the FAB intervals are useful for domains with sample sizes ranging from a few dozen to several hundred observations, although intervals for very small domains were necessarily wide. Second, when the prior mean and variance are estimated from a hierarchical Bayesian model, including domain-level covariates and having a moderate or large number of domains are helpful. Third, the choice among FAB intervals should depend on the distribution of the observed proportions. When proportions are close to the boundaries, especially when many observed proportions are near 0, we recommend the FAB Wilson interval with the all-in penalty as a practical default. Finally, practitioners should compare several intervals as part of a sensitivity analysis.

There are a few interesting directions to consider moving forward. A key hurdle to overcome is that the proportions must lie in the interval $[0,1]$. One solution is to derive a FAB framework based on the logit scale. Because the logit scale is unbounded, the risk and determination intervals would always lie in the support, enabling both to be computed properly. However, it is problematic if any sample proportions are 0 or 1 because the determination interval on the logit scale would have to include $\pm \infty$. Alternatively, we might define the Wald, Wilson, and AC intervals according to a normal distribution truncated to the interval $[0,1]$ instead of a normal distribution. This again would ensure that the risk and determination intervals would always lie in the support. Finally, along those lines, we can generalize which prior distribution to use. When we perform full Bayesian inference, we have information about the posterior distribution and may not need to approximate it with a normal distribution. Instead, we can use the draws from the posterior distribution to define the various intervals, enabling us to properly compute the risk. Doing this exercise might also provide further guidance on how to inflate the variance of the hierarchical model's parameters and thus increase the credible interval length and coverage. This might better incorporate the posterior estimates in the intervals. 

The second direction could include different link functions and transformations. Our approach is also applicable for a probit link, and there may be additional computational shortcuts available with that type of link \parencite{AlbertChibBayesianAnalysisBinary1993}. Exploring that link also would let us understand the robustness of the FAB intervals under various links. 

Third, the current MRP assumes that poststratification adequately adjusts for differences between the sample and the target population. In some applications, selection into the sample may be nonignorable, meaning that inclusion depends on unobserved factors related to the outcome. Extending FAB intervals to handle informative sampling designs would improve their generalizability~\parencite{asaf065}. 

Finally, conformal methods may provide another route for constructing calibrated intervals for proportions. Conformal inference offers exact coverage guarantees under exchangeability assumptions. Combining conformal ideas with the FAB framework could produce intervals that borrow strength across domains while retaining coverage guarantees.

\section*{Data availability statement}
The data used in the simulation will be shared on reasonable request to the corresponding author. On the other hand, the data analyzed in the COVID application cannot be shared publicly due to Health Insurance Portability and Accountability Act (HIPAA) privacy requirements associated with medical records.

\printbibliography

\setcounter{section}{0}
\setcounter{figure}{0}
\section*{\Large Supplementary Material}

\section{Discussion on the shifts observed in calibrated Bayes inference}

Since the shift in which outcomes $y_i$ yield intervals that cover $\theta_i$ drives the coverage improvement and may appear counterintuitive, we explain why the default FAB Wilson interval and the all-in-penalty version produce different shifts. This involves explaining how the penalty's effect trickles across the following steps and intervals:
\begin{enumerate}
    \item For $\theta \in [0, 1]$, choose $s_i(\theta)$ by minimizing the risk from the $y_i' / n_i'$ captured by the risk intervals.
    \item Construct the determination interval for $\theta \in [0, 1]$ based on the optimal $s_i(\theta)$ learned in the previous step.
    \item Build the FAB intervals based on the determination intervals for a given $y_i$.
    \item Find the set of $y_i$ whose FAB intervals cover a given $\theta_i \in [0, 1]$.
\end{enumerate}

\begin{figure}[!t]
\begin{subfigure}{0.49\textwidth}
    \centering
    \includegraphics[width=0.99\textwidth]{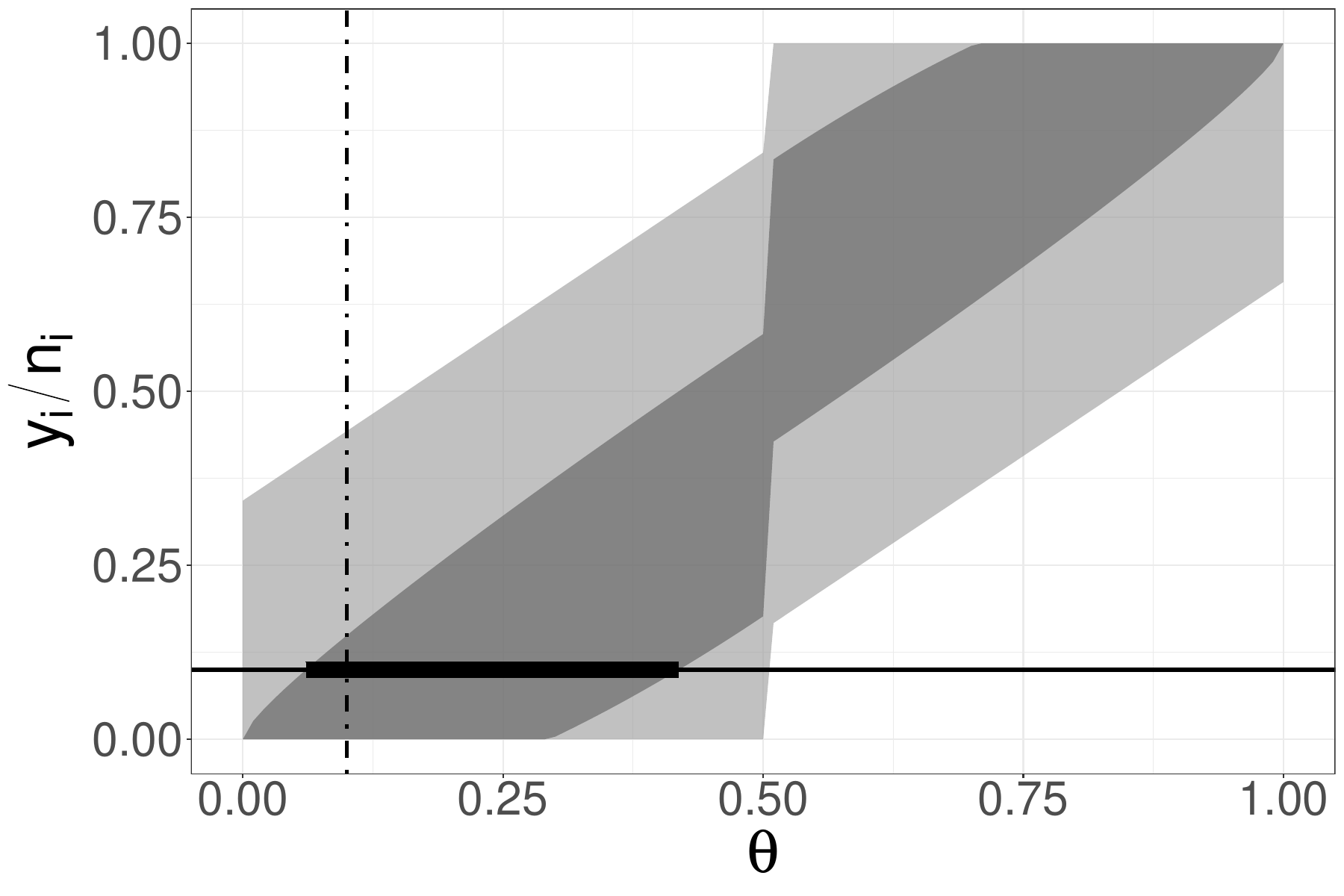} 
    \caption{Wilson FAB}
\end{subfigure}
\begin{subfigure}{0.49\textwidth}
    \centering
    \includegraphics[width=0.99\textwidth]{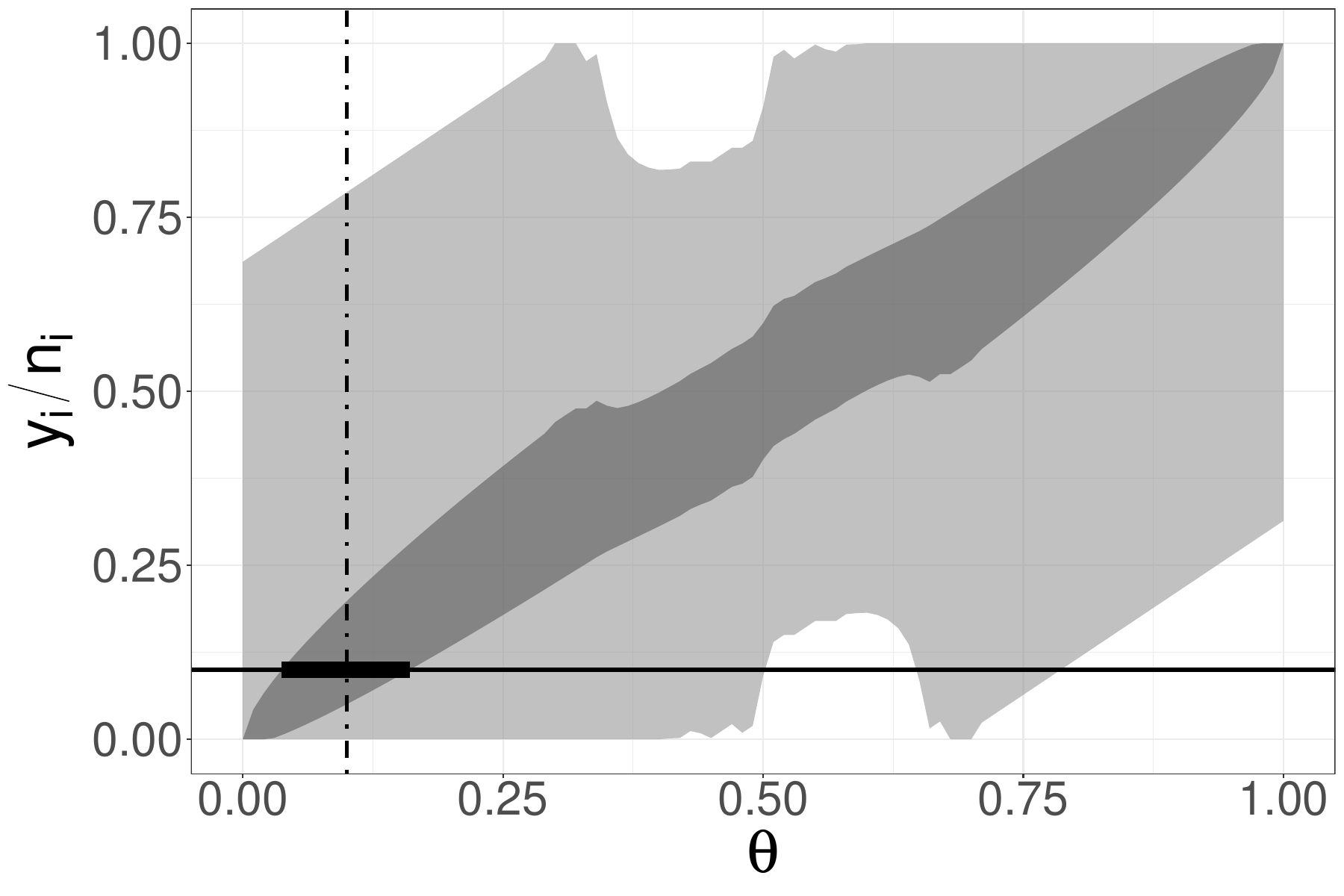}
    \caption{Wilson FAB (all-in penalty)}
\end{subfigure}
\caption{Plots showing the lower and upper bounds of the risk interval $\mathcal{I}^{F}\left(\theta, s_i, \textrm{var}(\theta)\right)$ in gray region and the determination interval $\mathcal{I}^{F}\left(\theta, s_i, \sqrt{\frac{\theta(1 - \theta)}{n_i}}\right)$ in dark gray for a grid of $\theta \in [0, 1]$ on the x-axis for the Wilson Frequentist assisted Bayes (FAB) interval and Wilson FAB interval with all-in penalty. The y-axis shows the space of $y'_i / n_i$ for the risk interval and $y_i / n_i$ for the determination interval. Both intervals are plotted based on the optimum $s_i$. The black horizontal line is $y_i / n_i$ = 0.1, i.e., the value we want to build an interval around, and the bounds of the thicker black line represent the FAB intervals for this $\theta_i$. The dot-dashed line represents $\theta = 0.1$.}
\label{fig:binary_underlying_interval_comp}
\end{figure}

\begin{figure}[!h]
\begin{subfigure}{0.49\textwidth}
    \centering
    \includegraphics[width=0.99\textwidth]{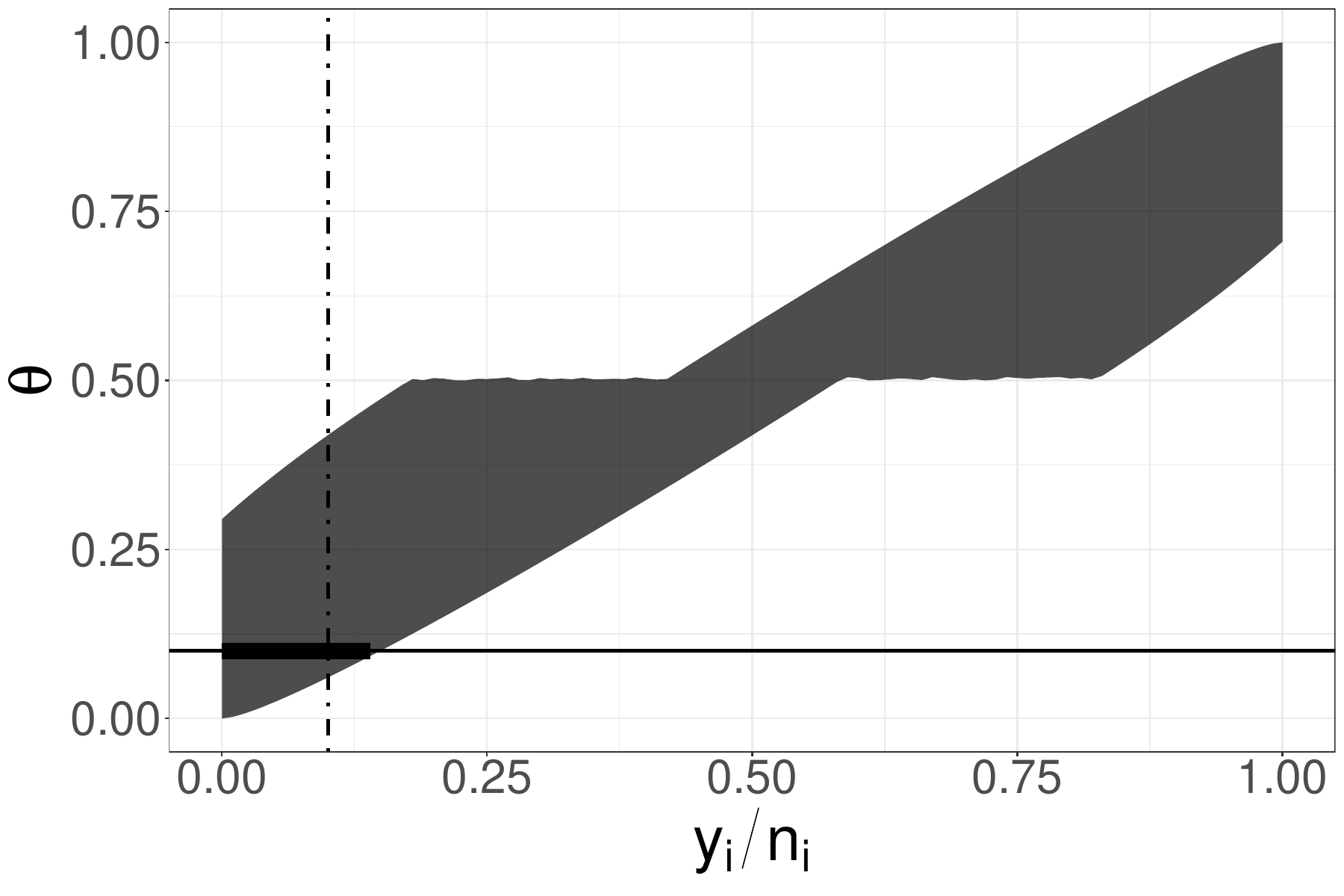} 
    \caption{Wilson FAB}
\end{subfigure}
\begin{subfigure}{0.49\textwidth}
    \centering
    \includegraphics[width=0.99\textwidth]{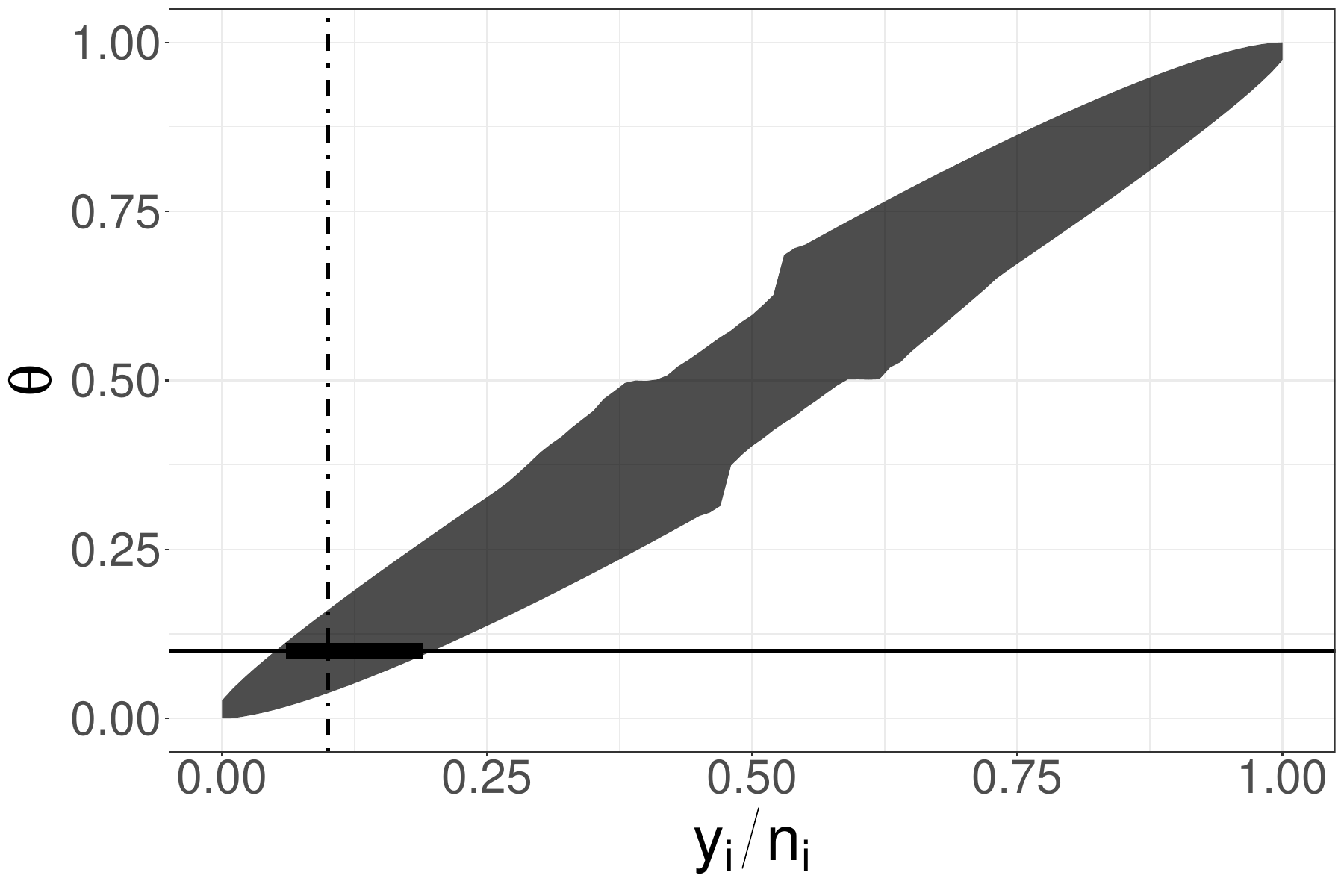}
    \caption{Wilson FAB (all-in penalty)}
\end{subfigure}
\caption{Plots showing the upper and lower bound of the Wilson Frequentist assisted Bayes (FAB) interval and Wilson FAB interval with all-in penalty for a grid of $y_i / n_i \in [0, 1]$. Both intervals are plotted based on the optimum $s_i$. The black horizontal line is $\theta$ = 0.1, i.e., the value we want to build an interval around, and the bounds of the thicker black line represent the FAB intervals for this $\theta_i$. The dot-dashed line represents $y_i / n_i = 0.1$.}
\label{fig:binary_underlying_interval_comp_theta}
\end{figure}

The effect of the penalty is clear on the first two steps.
Under the default specification (e.g., $n_i=100$, $\mu_i=0$, $\tau_i^2=1$), the risk interval can extend beyond the parameter support $[0,1]$, even for moderate $\theta$. Because values of $\hat\theta_i\notin[0,1]$ are not penalized under the default risk, the subsequent truncation to $[0,1]$ induces an asymmetric boundary effect: among candidate risk intervals of a given length, truncation favors those that extend beyond the support predominantly on one side (so that, after truncation, the retained portion within $[0,1]$ is shortest). Consequently, for $\theta<0.5$ the risk-minimizing intervals tend to allocate more of their mass below 0, whereas for $\theta\ge 0.5$ they tend to allocate more of their mass above 1. As a result, the risk-minimizing $s_i(\theta)$ aims to extend the risk interval past the boundary as much as possible in one direction. Meanwhile, the penalty assigns risk to interval mass outside $[0,1]$ so overshooting the boundary is no longer cost-free. The risk increases with the amount by which $\mathcal{I}^{F}(\theta,s_i,\mathrm{var}(\theta))$ extends past 0 or 1. This changes the risk-minimizing $s_i(\theta)$ and favors risk intervals that remain within $[0,1]$ as much as possible, reducing the boundary-driven asymmetry. Then, because the determination interval for a given $\theta$ uses the risk-minimizing $s_i(\theta)$, the determination intervals tilt toward the middle and away from the boundary when the penalty is applied. This is in contrast to the default specification in which the risk-minimizing $s_i(\theta)$ favors the risk intervals extending past the boundary. This then leads to determination intervals extending past the boundaries as well. This can be seen in Figure~\ref{fig:binary_underlying_interval_comp} because the light-gray region shows $\mathcal{I}^{F}(\theta,s_i,\mathrm{var}(\theta))$ and the dark-gray region displays $\mathcal{I}^{F}\!\left(\theta,s_i,\sqrt{\theta(1-\theta)/n_i}\right)$. Both are evaluated at the risk-minimizing $s_i(\theta)$ (minimizing $R(s_i\mid\varsigma)$ without the penalty and $R'(s_i\mid\varsigma)$ with the penalty).

The third step is challenging because there is an inversion. In our previous two steps, we examined a given $\theta$ and the corresponding intervals. Now, because the FAB interval is the set of $\theta$ whose determination interval contains the observed $y_i / n_i$, we have to compare the sets of $\theta$ that result from using and not using the penalty. In the no-penalty panel of Figure~\ref{fig:binary_underlying_interval_comp}(a), the dark-gray acceptance regions corresponding to $\theta>y_i/n_i$ still intersect the horizontal line at $y_i/n_i=0.1$, meaning many $\theta>y_i/n_i$ values “accept” this observation. In contrast, for $\theta\le 0.1$ the acceptance regions intersect the horizontal line much less, so those $\theta$ values are less often included after inversion. Consequently, the inverted interval $\mathcal{C}_{\mathrm{FAB}}(y_i/n_i)$ for $y_i/n_i=0.1$ places more mass above $\theta=0.1$ and is shifted away from the boundary. This is visible in Figure~\ref{fig:binary_underlying_interval_comp}: the thick black segment lies predominantly to the right of the vertical $\theta=0.1$ line. Interpreted in terms of inclusion, the interval based on $y_i/n_i$ near the lower end more readily covers $\theta_i >y_i/n_i$ than $\theta_i<y_i/n_i$; an analogous pattern holds near $y_i/n_i\approx 1$ by symmetry. We can then repeat this analysis for the FAB interval with the penalty. In Figure~\ref{fig:binary_underlying_interval_comp}(b), for $y_i/n_i=0.1$ the dark-gray acceptance regions for $\theta<y_i/n_i$ intersect the horizontal line more often than in the no-penalty case, so smaller $\theta$ values are now more likely to be included after inversion. Visually, using the vertical $\theta=0.1$ reference line, the set of $\theta$ values whose acceptance region contains $y_i/n_i=0.1$ expands on the left relative to the no-penalty panel. This produces a FAB interval for $y_i/n_i=0.1$ that is less one-sided and, relative to the no-penalty construction, is tilted toward the boundary. 

Finally, the fourth step again features an inversion. Given a $\theta_i \in [0, 1]$, we want to understand which $y_i$'s FAB intervals contain this $\theta_i$ because as discussed previously, the set of $y_i$'s determines the coverage probability. Again, we will need to compare the sets of $y_i$ under the penalty and the default configuration. To do so, we will examine Figure~\ref{fig:binary_underlying_interval_comp_theta}. Here, we display both configurations of the FAB interval for $y_i / n_i$. Now, the dot-dashed line represents $y_i / n_i = 0.1$ and the corresponding sets of $y_i / n_i$ for which $\theta_i = 0.1$ are displayed in a thick bold line. In Figure~\ref{fig:binary_underlying_interval_comp_theta}(a), we notice that as discussed previously, the FAB intervals without the penalty are heavily tilted toward the middle. In other words, for much of the FAB interval without the penalty, $\theta > y_i / n_i$ when $y_i / n_i$ is close to zero and $\theta < y_i / n_i$ when $y_i / n_i$ is close to one. For instance, for $y_i / n_i = 0.1$, the FAB interval is (0.061, 0.419). Due to this, when we look at which $y_i$'s FAB intervals contain $\theta_i$, the intervals for $y_i < \theta_i$ capture $\theta_i$ when $y_i / n_i$ is close to zero. In our example of $\theta_i = 0.1$ in Figure~\ref{fig:binary_underlying_interval_comp_theta}(a), this can be seen by the fact that much of the bold line is to the left of $y_i / n_i = 0.1$. When we do use the penalty to build our FAB intervals, Figure~\ref{fig:binary_underlying_interval_comp_theta}(b) displays this change at the outcome level. With the all-in penalty, outcomes $y_i/n_i$ closer to the middle are more likely to yield intervals that still contain $\theta_i=0.1$. That is, the set of $y_i$ values contributing to coverage shifts toward the center. This can be seen by the fact that much of the bold line for intervals that can capture $\theta_i = 0.1$ is to the right of $y_i / n_i = 0.1$ in Figure~\ref{fig:binary_underlying_interval_comp_theta}(b).

\end{document}